\documentclass[manuscript,screen,acmtog]{acmart}
\AtBeginDocument{%
  }

\usepackage{algorithm}
\usepackage{algpseudocode}
\usepackage{xcolor}
\usepackage{mathtools}
\usepackage{tabularx}
\usepackage[normalem]{ulem}

\newcommand{\estimatorbox}[2]{%
  \medskip
  \noindent
  \begin{minipage}{\linewidth}
    \fcolorbox{white}{gray!20}{%
      \begin{minipage}{\dimexpr\linewidth-2\fboxsep-2\fboxrule\relax}
      \textbf{#1}
      \end{minipage}%
    }
    \fcolorbox{white}{gray!10}{%
      \begin{minipage}{\dimexpr\linewidth-2\fboxsep-2\fboxrule\relax}
      #2
      \end{minipage}%
    }
  \end{minipage}
  \medskip
}

\setcopyright{acmlicensed}
\copyrightyear{2026}
\acmYear{2026}
\acmDOI{XXXXXXX.XXXXXXX}
\acmConference[Conference acronym 'XX]{Make sure to enter the correct
  conference title from your rights confirmation email}{June 03--05,
  2018}{Woodstock, NY}
\acmISBN{978-1-4503-XXXX-X/2018/06}

\begin{document}

\title{Walking on Heat Stars for Parabolic Heat Equations with Neumann Boundary Conditions}

\def\WOHSCLEAN{}

\ifdefined\WOHSCLEAN
  \newcommand{\oldtext}[1]{}%
  \newcommand{\changedtext}[1]{#1}%
\else
\makeatletter
\DeclareRobustCommand{\oldtext}[1]{{%
  \color{orange}%
  \let\submittedFloatReset\@floatboxreset
  \def\@floatboxreset{\submittedFloatReset\color{orange}%
    \captionsetup{font+={color=orange}}}%
  \let\submittedLabel\label
  \renewcommand{\label}[1]{\submittedLabel{submitted:##1}}%
  \let\submittedDisplayLabel\ltx@label
  \renewcommand{\ltx@label}[1]{\submittedDisplayLabel{submitted:##1}}%
  \let\submittedRef\ref
  \renewcommand{\ref}[1]{%
    \ifcsname r@submitted:##1\endcsname
      \submittedRef{submitted:##1}%
    \else
      \submittedRef{##1}%
    \fi}%
  #1}}
\makeatother
\newcommand{\changedtext}[1]{{\color{blue} #1}}
\fi
\newcommand{\citeTODO}[1]{[\textcolor{red}{\textsc{cite}: #1}]}

\author{Anchang Bao}
\affiliation{%
  \institution{School of Software, BNRist, Tsinghua University}
  \city{Beijing}
  \country{China}
}
\email{baoanchang02@gmail.com}

\author{Enya Shen}
\affiliation{%
  \institution{School of Software, BNRist, Tsinghua University}
  \city{Beijing}
  \country{China}
  }
\affiliation{%
  \institution{Haihe Lab of ITAI}
  \city{Tianjin}
  \country{China}
  }

\email{shenenya@tsinghua.edu.cn}

\author{Yongjun Zhang}
\affiliation{%
  \institution{Yunji Intelligent Engineering Co. Ltd}
  \city{Shenzhen}
  \country{China}
  }
\email{zhangyongjun@szewec.com}  

\author{Zhongwei Liu}
\affiliation{%
  \institution{Yunji Intelligent Engineering Co. Ltd}
  \city{Shenzhen}
  \country{China}
  }
\email{liuzhongwei@szewec.com}  

\author{Jianmin Wang}
\affiliation{%
  \institution{School of Software, BNRist, Tsinghua University}
  \city{Beijing}
  \country{China}
  }
\email{jimwang@tsinghua.edu.cn}

\renewcommand{\shortauthors}{Bao et al.}

\begin{abstract}
  Monte Carlo methods have proven highly effective for elliptic partial differential equations through algorithms such as Walk on Spheres and Walk on Stars, which evaluate solutions at individual points without volumetric meshing or global linear solves. 
  \oldtext{Extending these methods to the transient regime has remained an open challenge: parabolic equations couple space and time through an anisotropic scaling, requiring joint sampling of spatial displacements and backward time steps whose distribution was not previously available in a unified, exact form.}\,\changedtext{Parabolic heat equations couple space and time, so extending the local star-shaped construction of Walk on Stars requires joint sampling of spatial displacements and backward time steps.} 

  We present Walk on Heat Stars, a grid-free Monte Carlo solver \oldtext{that closes this gap by extending}\,\changedtext{that extends} the boundary integral framework of Walk on Stars to the parabolic setting.
  Our method uses a non-cylindrical boundary integral formulation to accommodate the evolving auxiliary heat balls within a fixed physical domain.
  The heat ball geometry is parameterized by a logarithmic time coordinate and a spatial direction, revealing that the double-layer kernel factorizes into \oldtext{independent Gamma and uniform components}\changedtext{a uniform directional density and a piecewise Gamma density conditional on the direction, both of which can be sampled directly}.
  This parameterization enables \oldtext{exact directional importance sampling of}\,\changedtext{importance-sampling strategies for} the recursive next walk position, the Neumann flux contribution, and the volumetric source term, yielding unbiased Monte Carlo estimators for all three components.

  We additionally derive a preliminary gradient estimator that expresses spatial derivatives as weighted \oldtext{boundary}\,\changedtext{surface and volume} integrals of the solution, requiring no recursion on the gradient, and adapt a heteroscedastic regression-based denoiser to the space-time domain for variance reduction. We validate our method on analytical solutions across a range of geometries and spatial frequencies, \oldtext{confirm convergence at the expected Monte Carlo rate}\,\changedtext{observe approximately inverse-time decay of raw MSE}, and demonstrate practical applicability on heat sink and cooling scenes with mixed or pure Neumann boundary conditions.
\end{abstract}

\ccsdesc[500]{Computing methodologies~Modeling and simulation}
\ccsdesc[500]{Computing methodologies~Shape analysis}

\keywords{Monte Carlo methods, heat equation, Walk on Spheres, boundary integral equations, Neumann boundary conditions, transient heat conduction}


\begin{teaserfigure}
  \changedtext{\includegraphics[width=\textwidth]{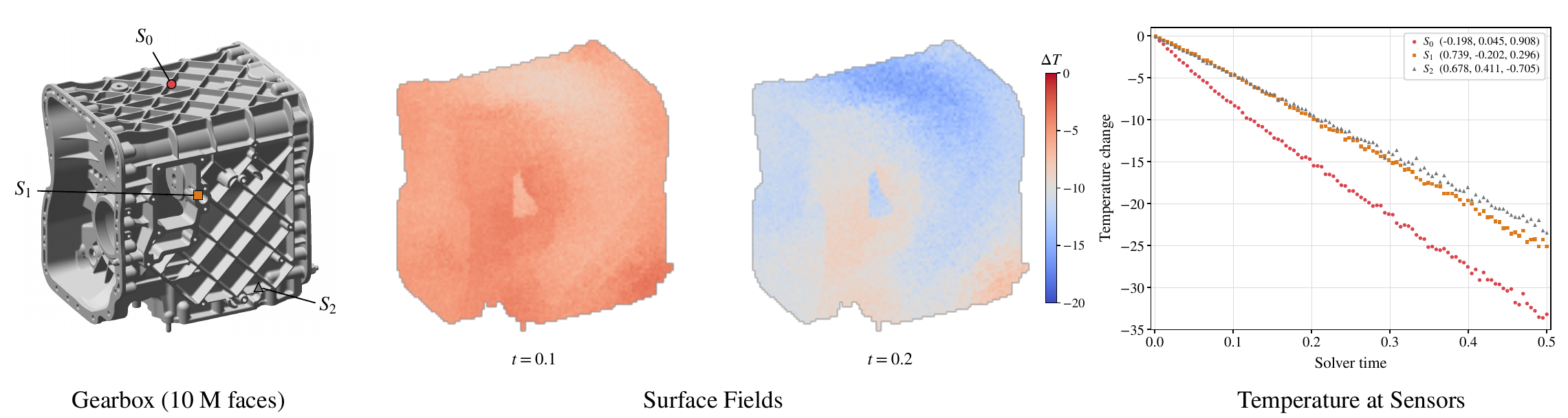}}
  \caption{\oldtext{Monte Carlo estimation of transient heat conduction in a turbine housing geometry with pure Neumann boundary conditions. A uniform initial temperature cools under a constant surface heat flux, with no Dirichlet absorption. \textbf{Left:} The model geometry. \textbf{Middle:} Estimated temperature distribution on a fixed 2D spatial slice at three time instants (top to bottom), each computed with only 16 random walks per pixel. \textbf{Right:} Cooling curves at three fixed sensor locations, demonstrating the temporal accuracy of the method. The complex geometry and pure-Neumann setup highlight the flexibility and robustness of the proposed approach.}\changedtext{WoHSt simulation of a gearbox cooling from a uniform initial temperature under a constant outward heat flux on its entire surface. \textbf{Left:} The gearbox model has nearly 10 million triangles, and the markers $S_0$, $S_1$, and $S_2$ are fixed surface sensors. \textbf{Middle:} Surface temperature changes at $t=0.1$ and $t=0.2$ share a color scale in solver units, and each visible pixel uses 16 random walks at each time. \textbf{Right:} Cooling curves at the three sensors, estimated with 128 walks per sensor at each of 100 equally spaced times from $t=0.001$ to $0.5$. Each legend entry gives the sensor's $(x,y,z)$ position in model coordinates.}}
  \label{fig:housing}\label{fig:gearbox_teaser}
\end{teaserfigure}

\maketitle

\section{Introduction}

Monte Carlo methods have become increasingly popular for the numerical solution of elliptic partial differential equations. Algorithms such as Walk on Spheres (WoS)~\cite{muller1956some,sawhney2020monte}, Walk on Boundary~\cite{sugimoto2023practical}, and Walk on Stars (WoSt)~\cite{sawhney2023walk,miller2024walkin} evaluate solutions at individual points by recursive random sampling, avoiding volumetric meshing, global linear solves, and the curse of dimensionality that affects finite element methods. These algorithms now handle Dirichlet, Neumann, and Robin conditions on geometries of arbitrary complexity, and have been extended to fluid simulation, differentiable inverse problems, and nonlinear boundary conditions. \oldtext{Yet despite this rapid progress in the elliptic regime, the \emph{transient} case has remained largely open.}\changedtext{Building on the well-developed elliptic setting, we extend the Walk on Stars algorithm to the parabolic heat equation.}

\oldtext{The central obstacle is the anisotropic coupling of space and time in parabolic equations: the fundamental solution $\Phi(x,t) = (4\pi t)^{-n/2}\exp(-|x|^2/(4t))$ depends on the ratio $|x|^2/t$, so spatial and temporal coordinates are coupled through the parabolic scaling and must be treated jointly. The mean value property that underpins elliptic walk methods becomes a \emph{space-time} statement defined over \emph{heat balls}: spindle-shaped regions whose bounding \emph{heat spheres} are level sets of $\Phi$. Sampling from this space-time measure requires jointly drawing a spatial direction and a backward time step from a distribution that reflects the anisotropic scaling, and doing so exactly even when the heat ball is truncated by domain boundaries.}

\changedtext{The heat equation has a structure that permits direct space--time Monte Carlo estimation. Its explicit heat kernel determines the coupled spatial and temporal sampling measure. The analogue of a sphere is a \emph{heat ball}, a spindle-shaped region in space--time bounded by a level set of the heat kernel (Figure~\ref{fig:heat_ball}). A walk within a heat ball advances in space and backward in time simultaneously, terminating at the Dirichlet boundary or the initial time plane. For prescribed source and Neumann data, our estimator follows one recursive continuation per step and accumulates the source and flux contributions without recursive branching, time discretization, or operator splitting. We exploit this structure to extend the local star-shaped construction of WoSt to the heat equation.}

\oldtext{Prior work on transient Monte Carlo methods is limited. The foundational series by \citet{deaconu2013hitting,deaconu_walk_2017,deaconu2018initial} introduced the Walk on Moving Spheres (WOMS) algorithm and its reinterpretation via heat balls, enabling joint simulation of exit times and positions for Brownian motion and the heat equation with Dirichlet conditions. In a different direction, \citet{Sabelfeld2019Global} developed a global Random Walk on Spheres (gRWS) method for the heat equation that computes solutions at many points simultaneously by series-based first-passage time sampling. Neither line of work handles mixed Dirichlet-Neumann boundary conditions or supports the truncated ``heat star'' geometry needed for non-convex domains.}

\changedtext{Previous work on transient Monte Carlo methods includes heat-ball walks~\cite{deaconu2013hitting,deaconu_walk_2017,deaconu2018initial}, series-based sphere walks~\cite{Sabelfeld2019Global}, and boundary-integral approaches~\cite{sabelfeld2016stochastic,sugimoto2025toward}. The series by \citet{deaconu2013hitting,deaconu_walk_2017,deaconu2018initial} introduced the Walk on Moving Spheres (WoMS) algorithm, which performs the random walk in space--time instead of the largest inscribed ball used by classical WoS. At every step, the next spatial position and the time increment are drawn jointly from a heat ball, so both coordinates advance together. Walk-on-Boundary (WoB) instead constructs walks from boundary integral representations. \citet[Chapter 4]{sabelfeld2016stochastic} develop heat-equation formulations, and \citet[Chapter 4]{sugimoto2025toward} present diffusion experiments.

Our construction starts from the local heat-ball geometry and extends it to mixed Dirichlet--Neumann conditions. Following the paradigm of the Walk on Stars algorithm, we estimate the solution of transient heat equations by walking on the \emph{heat star}, which is a part of the heat ball that remains inside the domain. In the standard Walk-on-Stars formulation~\cite{sawhney2023walk}, the next position is drawn by directional sampling in the star region, which importance-samples the Poisson kernel exactly and yields unit weight. To importance-sample the boundary measure exactly in the transient case, we introduce a radial trace parameterization of the heat ball, which reveals that the space--time analogue of directional sampling is tracing along a parabolic ray.}

In this paper we introduce \emph{Walk on Heat Stars} (WoHSt), a mesh-free Monte Carlo solver that extends WoSt to the transient regime while supporting mixed Dirichlet--Neumann boundary conditions.\oldtext{ The method rests on three foundations. First, we formulate the transient problem as a \emph{non-cylindrical} boundary integral equation whose domain evolves in time, generalizing the stationary-domain BIE used in elliptic WoSt (Section~\ref{sec:problem_setup}). Second, we derive the heat ball geometry explicitly, parameterizing the heat ball by a logarithmic time coordinate $u$, a spatial direction $\omega$, and a radial coordinate $\rho$, and show that the double-layer kernel factorizes into independent Gamma and uniform components (Section~\ref{sec:heat_ball}). Third, we develop a family of exact importance-sampling strategies that exploit this parameterization:}
\changedtext{ 
Concretely, to extend Walk on Heat Spheres~\cite{deaconu2018initial} to mixed Dirichlet--Neumann conditions, we introduce a Walk-on-Stars formulation in coupled space--time, in which the straight radial rays of the Euclidean star are replaced by parabolic radial traces.
The method rests on three foundations.
Firstly, we use a \emph{non-cylindrical} boundary integral equation (BIE) for the evolving auxiliary heat-ball slices within a fixed physical domain, extending the stationary-domain representation used in elliptic WoSt (Section~\ref{sec:problem_setup}). Then we derive the heat ball geometry explicitly, parameterizing the heat ball by a logarithmic time coordinate $\mu$, a spatial direction $\omega$, and a radial coordinate $\rho$, and show that the double-layer kernel factorizes into a uniform directional factor and a direction-conditioned piecewise Gamma factor (Section~\ref{sec:heat_ball}). Finally, we develop a family of exact importance-sampling strategies that exploit this parameterization:} directional sampling on the heat-star boundary, flux contribution from Neumann data, and source term sampling via a heat-spherical change of variables (Section~\ref{sec:sampling}).

Building on this core, we derive a gradient estimator that expresses spatial derivatives $\nabla u$ as weighted \oldtext{boundary}\,\changedtext{surface and volume} integrals of $u$, requiring no recursion on the gradient (Section~\ref{sec:gradient}), and adapt a heteroscedastic regression-based denoiser to the space-time domain (Section~\ref{sec:variance}). We validate WoHSt on analytical solutions across different geometries and spatial frequencies, \oldtext{confirming convergence at}\,\changedtext{with fitted raw-MSE slopes close to} the expected Monte Carlo rate. We further demonstrate its practical utility on heat transport problems with mixed or pure Neumann boundary conditions, where WoHSt estimates match finite element reference solutions across multiple time instants.

\oldtext{Our key contributions are:
\begin{itemize}
    \item A Walk-on-Stars formulation for transient heat equations supporting mixed Dirichlet--Neumann boundary conditions, source terms, and complex geometries (Sections~\ref{sec:problem_setup}--\ref{sec:sampling}).
    \item A geometric factorization of heat-sphere sampling into independent Gamma-distributed temporal and uniform directional components, yielding exact unit-weight importance sampling (Section~\ref{sec:heat_ball}).
    \item A recursive heat-star estimator that generalizes classical heat-ball walks: the latter are recovered as a special case for pure Dirichlet problems with vanishing source (Section~\ref{sec:wohs}).
    \item Applications to transient heat transport in complex engineering geometries, including a quantitative comparison with finite element solutions (Section~\ref{sec:experiments}).
\end{itemize}
We additionally contribute a gradient estimator (Section~\ref{sec:gradient}) and a space-time heteroscedastic denoiser (Section~\ref{sec:variance}).}\changedtext{}

\section{Related Work}

\oldtext{}\changedtext{We first review Monte Carlo methods for elliptic PDEs, then discuss transient Monte Carlo solvers and their relation to our heat-star construction. Finally, we discuss the variance-reduction methods for Monte Carlo PDE solvers.}

\subsection{Monte Carlo PDE Solvers}

Monte Carlo PDE solvers originated in the 1950s with the Walk-on-Spheres (WoS) algorithm~\cite{muller1956some}, which exploits the mean value property of harmonic functions to recursively sample exit points from inscribed spheres. WoS is based on the Kakutani theorem~\cite{kakutani1944143}, which characterizes the solution of the Dirichlet problem for Laplace equation as the expected value of the boundary data at the exit point of a Brownian motion:
\begin{equation}
    u(x) = \mathbb{E}^x[g(X_\tau)],
\end{equation}
where $X_\tau$ is the first exit point of a Brownian path starting at $x$. This result provides a direct probabilistic interpretation of the PDE solution, and motivates the recursive sampling of spheres until absorption at the boundary.

In the computer graphics and geometry processing community, the pioneering work of \citet{sawhney2020monte} revitalized interest in WoS for efficiently solving elliptic PDEs on complex geometries with the spatially accelerated closest-point queries. Subsequent works extended WoS to Neumann boundary conditions~\cite{sawhney2023walk}, Robin boundary conditions~\cite{miller2024walkin} and spatially varying coefficients~\cite{sawhney2022grid}. Specifically, Walk-on-Stars (WoSt)~\cite{sawhney2023walk} generalizes WoS by sampling from the largest star-shaped region contained in the domain, rather than a sphere, achieving significant speedups on non-convex geometries with Neumann boundary conditions. \citet{sugimoto2023practical} described an alternative Walk-on-Boundary (WoB) method for Neumann and Robin problems. \citet{huang2025geometric} extends the Walk-on-Stars to implicit geometries using interval arithmetic.
Monte Carlo PDE solvers have been extended to fluid simulation~\cite{rioux2022monte,sugimoto2024velocity}, participating media~\cite{miller2025solving}, shape deformation~\cite{de2024stochastic}, heat radiation~\cite{bao2026monte}, differentiable formulations for inverse problems~\cite{miller2024differential,yu2024differential,yilmazer2024solving}, and gradient estimation~\cite{yu2025robust}. \oldtext{}\changedtext{For elliptic problems on exterior domains, a walk can escape to infinity without reaching a boundary. \citet{nabizadeh2021kelvin} address this through a Kelvin transformation to a bounded domain, while \citet{zhou1994brownian} use analytical escape and return probabilities for an enclosing sphere.}

\oldtext{}\changedtext{Recent work also estimates or learns boundary operators. \citet{maesumi2026monte} estimate Dirichlet-to-Neumann operators and their Steklov eigenmodes using Monte Carlo methods. \citet{he2026neural} learn the harmonic-measure density from WoS exit samples, allowing a trained boundary kernel to solve Laplace problems with different boundary values without retraining.}

\subsection{Transient Problems}

\oldtext{Extending Monte Carlo PDE solvers to the transient regime has received comparatively little attention.
The most closely related contributions}\,\changedtext{The most closely related heat-ball methods} are the series of papers by \citet{deaconu2013hitting,deaconu_walk_2017,deaconu2018initial}.

\citet{deaconu2013hitting} developed the Walk on Moving Spheres (WoMS) algorithm to simulate the first time a Bessel process hits a constant level. The key idea is to recursively sample exit times from time-dependent spheres (moving spheres) whose radii shrink as the process approaches the target. They proved that the \oldtext{first hitting time from a specific family of time-varying value $\psi(t) = \sqrt{2t\ln(a/t^{n/2})}$ follows a Gamma distribution: $Z \sim \Gamma(n+2,\,1/(n+1))$, which is then transformed to the time increment $c e^{-Z}$. $n$ is the dimension of the Bessel process and $c$ is a constant determined by $n$.}\,\changedtext{resulting first hitting time follows a Gamma distribution after a logarithmic transformation. This distribution can be sampled directly and efficiently using standard routines.}

\citet{deaconu_walk_2017} adapted this framework for the first time to simulate the exit time and exit position of an $n$-dimensional Brownian motion from a domain. By identifying the Euclidean norm of the Brownian motion with a Bessel process, they replaced the fixed spheres of the classical Walk on Spheres with moving spheres\oldtext{ whose radius evolves as $\psi(t) = \sqrt{2t\ln(a/t^{n/2})}$}\changedtext{}. The result is a practical algorithm that jointly samples the hitting time and \oldtext{}\changedtext{the }exit point with \oldtext{$O(|\log\epsilon|)$ expected steps, where $\epsilon$ is the boundary}\,\changedtext{a logarithmic expected number of steps in the} absorption tolerance.

\citet{deaconu2018initial} reinterpreted the method in terms of \emph{heat balls}, whose boundary satisfies a mean value property \oldtext{for caloric functions}\changedtext{}, and extended the WOMS algorithm to the initial-boundary value problem for the heat equation. From the stochastic viewpoint, the solution \oldtext{$u(t,x)$ is expressed via the Feynman-Kac formula as $\mathbb{E}[u(X_\tau, \tau)]$, where $(X_\tau, \tau)$ is the first space-time}\,\changedtext{is expressed by the Feynman--Kac formula as the expected value at the first space--time} exit point of a Brownian path. Their algorithm jointly simulates this exit pair by constructing a Markov chain on heat balls\oldtext{ - level sets of the heat kernel that}\,\changedtext{, which} grow from a point, attain a maximum radius\oldtext{ $\sqrt{2nc/e}$}\changedtext{}, and shrink back to a point. Throughout this series, the logarithmic time coordinate \oldtext{$u = \log(c/s)$ }\changedtext{}and its Gamma distribution\oldtext{ $\Gamma(n/2+1,\,2/n)$}\changedtext{} were known but not treated as a primary geometric parameterization.

Our contribution is to elevate this observation to the primary geometric parameterization\oldtext{: the \emph{space-time radial trace} $(s, r) = (c e^{-u},\; \sqrt{2n c u e^{-u}})$}\,\changedtext{. The resulting \emph{space--time radial trace}} exposes the heat kernel geometry directly and is essential for extending the walk to truncated heat stars, where a ray along the trace may intersect a Neumann boundary before reaching the heat-sphere surface.

In a different direction, \citet{Sabelfeld2017DRD,Sabelfeld2019Global} developed random walk on spheres methods for transient drift-diffusion-reaction equations, sampling the first passage time from a fixed sphere via an alternating series method. The distribution of the first passage time is given by a complicated series expression and must be sampled carefully.
\oldtext{In contrast, our method departs from the stochastic process viewpoint and instead adopts a \emph{boundary integral equation} framework, which naturally decomposes the estimator into double-layer (recursive), single-layer (Neumann), and volume (source) contributions with perfect importance sampling.}\changedtext{Our method uses a boundary integral equation to separate three contributions. The double-layer term propagates the unknown solution through recursive walks. The single-layer term accounts for prescribed Neumann flux. The volume term accounts for the source. We derive importance-sampling rules for these terms below.} 
\oldtext{We use \emph{maximal} heat balls for logarithmic step complexity}\changedtext{We use the largest heat balls allowed by the local geometry and remaining time} and extend to mixed Dirichlet-Neumann conditions via \emph{heat stars}, which are heat balls truncated by the visibility silhouette. \oldtext{This generalization has no analog in prior transient Monte Carlo solvers.}\changedtext{The contribution is the heat-star construction and the associated joint space--time importance sampling.}

\citet{bati2023coupling} propose a Monte Carlo framework for coupled conduction-convection-radiation, approximating transient conduction and convection by finite-difference discretisation and exponential waiting times (a kinetic approximation). This approach enables fast infrared rendering and replay of boundary conditions, but it is not exact. Our work, by contrast, maintains the exact space-time mean value property of heat balls and provides unbiased estimates\oldtext{}\changedtext{, ignoring $\varepsilon$-shell termination error}.

\oldtext{}\changedtext{

Beyond the Walk-on-Spheres family, Walk-on-Boundary (WoB)~\cite{sugimoto2023practical} provides a boundary-integral approach to Monte Carlo PDE solving. For the heat equation, it represents the solution using layer potentials on the space--time boundary~\cite{sabelfeld1991monte,sabelfeld2016stochastic,sugimoto2025toward}. Mixed Dirichlet--Neumann heat problems can also be approached within this framework by extending its boundary integral formulation and sampling scheme. Our contribution is a local heat-star construction with joint space--time importance sampling. WoB admits signed-weight walks on non-convex domains, with variance depending on the sampling design. Section~\ref{sec:wob_comparison} compares WoHSt with a direct pure-Neumann heat WoB estimator.

}

\oldtext{}\changedtext{In concurrent work, \citet{zhou2026grid} extend WoS and WoSt to transient diffusion with mixed Dirichlet--Neumann conditions. Both methods support time-varying data without sequential time stepping, but use different local geometries and sampling constructions. Their method retains spatial balls and stars, using the bounded-ball heat kernel to sample elapsed time conditioned on a spatial boundary point. Our method instead constructs heat balls and truncated heat stars directly in space--time. A uniform direction and a coordinate sampled from the corresponding piecewise Gamma density define a radial trace whose intersection with the heat-star boundary determines both the spatial and temporal step. The distinction is therefore the local geometry and its associated sampling rules.}

\oldtext{}\changedtext{Monte Carlo solvers have already been developed for the Navier--Stokes equations~\cite{rioux2022monte,sugimoto2024velocity}. The naive recursive vorticity formulation of \citet{rioux2022monte} has exponential cost, but their practical implementation avoids this recursive expansion through temporal grid caching. It stores vorticity from the preceding time step and queries that cache during Monte Carlo evaluation of the Biot--Savart integral~\cite[Section 4.3]{rioux2022monte}. The velocity-based solver of \citet{sugimoto2024velocity} uses discrete time steps and operator splitting, with pointwise Monte Carlo estimators for the individual substeps.

The linear scalar heat equation has a more specialized structure than the incompressible Navier--Stokes equations. For the heat equation with prescribed source and boundary data, the explicit heat kernel allows joint sampling of spatial displacement and backward time. Our estimator therefore follows a single recursive continuation without time discretization, operator splitting, or a cache of earlier solution fields. 
}

\subsection{Variance Reduction}

A central challenge in Monte Carlo PDE solvers is variance reduction, since solutions are computed via stochastic sampling and estimates can exhibit substantial noise.
\citet{sawhney2020monte} studied classical techniques including control variates and importance sampling tailored to the WoS setting.
Several subsequent methods reduce variance by reusing information across nearby evaluation points.
Boundary value caching~\cite{miller2023boundary} stores boundary values and normal derivatives to accelerate interior evaluations via boundary integral representations.
Mean value caching~\cite{bakbouk2023mean} exploits volumetric mean-value properties to spatially blend nearby samples inside maximal balls.
Bidirectional WoS~\cite{qi2022bidirectional} draws inspiration from bidirectional path tracing to formulate a bidirectional estimator.
Neural caching approaches~\cite{li2023neural,li2024neural} replace explicit storage with learned predictors that estimate solution values, while harmonic caching~\cite{zhou2025harmonic} reconstructs local solution representations via harmonic expansions.
Off-centered estimators~\cite{czekanski2024walking,bao2025off} reuse samples from overlapping spheres centered at neighboring evaluation points, and statistical weighting~\cite{bao2025off} further suppresses outlier-induced artifacts.
For sparse source terms, path guiding~\cite{huang2025path} adapts techniques from rendering to significantly improve sampling efficiency.\oldtext{}\changedtext{
\citet{huang2026probe} introduce a probe-based estimator to reuse all points visited by WoS walks by leveraging the Markov structure of the process.
\citet{wu2026gradient} adapt the gradient domain reconstruction technique from rendering~\cite{yan2025generalized} to Monte Carlo PDE solvers, enabling gradient-domain reconstruction of the solution.
Walk on Decomposed Subdomains~\cite{jambon2026walk} introduced a hybrid domain decomposition and WoS approach to reduce the number of steps required to reach the target by encoding long-term dependencies into the deterministic solve.}

\oldtext{}\changedtext{WoB also supports variance-reduction strategies from Monte Carlo light transport. \citet{sugimoto2023practical} discuss bidirectional estimators combined through multiple importance sampling (MIS)~\cite{veach1995optimally}, and demonstrate Markov chain Monte Carlo (MCMC) for Dirichlet problems using primary sample space Metropolis light transport~\cite{kelemen2002simple}, a variant of Metropolis light transport~\cite{veach1997metropolis}. These WoB sampling strategies are also discussed in \citet[Chapter~3]{sugimoto2025toward}. These strategies exploit the boundary-integral formulation of WoB and complement the variance-reduction techniques developed for WoS. Their results illustrate the importance of sampling design when assessing WoB performance.}

\section{Background}

We briefly recall the stochastic and boundary-integral viewpoints that underlie our method. Table~\ref{tab:notation} summarizes the notation used throughout the paper.

\begin{table}[!htbp]
\centering
\caption{Notation used throughout the paper.}
\label{tab:notation}
\small
\begin{tabularx}{\linewidth}{l>{\raggedright\arraybackslash}X}
\toprule
\multicolumn{2}{l}{\textbf{Domain \& geometry}} \\
\midrule
$\Omega \subset \mathbb{R}^n$ & \leavevmode\oldtext{Bounded}\,\changedtext{Fixed} spatial domain ($n$: dimension) \\
$\partial\Omega_D$, $\partial\Omega_N$ & Dirichlet and Neumann boundaries \\
$\vec{n}$ & Outward unit normal \\
$\langle\cdot,\cdot\rangle$ & Euclidean inner (dot) product \\
$\mathcal{H}(x, t, c)$ & Heat ball centered at $(x, t)$ with radius $c$ \\
$\mathcal{H}^\star(x, t, c)$ & Heat star: $\mathcal{H} \cap (\Omega \times [t-c, t])$ \\
$\partial\mathcal{H}^\star_{\mathrm{sphere}}$ & Spherical boundary of $\mathcal{H}^\star$ \\
$\partial\mathcal{H}^\star_N$ & Neumann wall of $\mathcal{H}^\star$ \\
\midrule
\multicolumn{2}{l}{\textbf{PDE \& solution}} \\
\midrule
$\Phi(x, t)$ & Fundamental solution of the heat equation \\
$\tau(c)$ & Level-set threshold: $(4\pi c)^{-n/2}$ \\
$u(x, t)$ & Unknown temperature field \\
$g$, $h$, $g_0$ & Dirichlet data, Neumann flux, initial condition \\
$f(x, t)$ & Volumetric source term \\
\midrule
\multicolumn{2}{l}{\textbf{Heat-ball parameterization}} \\
\midrule
$c$ & Heat-ball radius (max.\ value of $\rho$) \\
$s$ & Backward time $t_0 - t$, with $0 < s < c$ \\
$r(s)$ & Cross-section radius: $\sqrt{2n s \log(c/s)}$ \\
\oldtext{$u$}\,\changedtext{$\mu$} & \leavevmode\oldtext{Logarithmic time $\log(c/s)$, $u > 0$}\par\noindent\changedtext{Logarithmic coordinate $\log(\rho/s)=|x-x_0|^2/(2ns)$, with $\rho=c$ on the heat sphere} \\
$\rho$ & Radial coordinate through heat-ball interior \\
$\omega$ & Unit direction on $S^{n-1}$ \\
\midrule
\multicolumn{2}{l}{\textbf{Sampling \& algorithmic}} \\
\midrule
$\Gamma(\alpha, \beta)$ & Gamma distribution (shape $\alpha$, scale $\beta$) \\
\changedtext{$P(a,z)$} & \leavevmode\changedtext{Cumulative probability $\Pr(X\le z)$ for $X\sim\Gamma(a,1)$} \\
$\varepsilon$ & Dirichlet absorption threshold \\
\bottomrule
\end{tabularx}
\end{table}

\subsection{Stochastic Perspective}

For the homogeneous heat equation $\partial_t u = \Delta u$ with pure Dirichlet conditions, the solution admits a stochastic representation via the Feynman--Kac formula:
\begin{equation}
    u(t,x) = \mathbb{E}_{(t,x)}\Bigl[ g(t-\tau, X_{t-\tau})\,\mathbf{1}_{\tau < t} + g_0(X_0)\,\mathbf{1}_{\tau = t} \Bigr],
\end{equation}
where $X_s$ is the position of a Brownian path at physical time $s$, with $X_t = x$, and $\tau = \inf\{s \in (0,t] : X_{t-s} \in \partial\Omega\}$ is the first hitting time of the spatial boundary, with the convention $\tau=t$ if the path reaches the initial time plane first. The path terminates either upon hitting $\partial\Omega_D$ (absorbing at the Dirichlet data $g$) or upon reaching the initial time plane (absorbing at the initial condition $g_0$). This representation reduces the PDE solution to simulating the joint distribution of the first space-time exit pair $(\tau, X_{t-\tau})$ from the cylinder $\Omega \times [0, t]$.

Direct time-stepped simulation of the Brownian path is computationally prohibitive. The Walk on Spheres algorithm~\cite{muller1956some} resolves this for elliptic problems by exploiting the isotropy of Brownian motion to take large spherical steps. For the transient case, the analogous acceleration uses \emph{heat balls}, i.e., space-time regions whose boundary supports a generalized mean value property \oldtext{for caloric functions}\changedtext{}, enabling jointly sampling the backward time step and the spatial displacement in a single large jump. A solution of the homogeneous heat equation, $\partial_t u = \Delta u$, is called a \emph{caloric function}. It is the parabolic counterpart of a harmonic function, and the mean value property just described is what replaces harmonicity in the transient setting. This is the foundation of both the prior random walk on heat balls~\cite{deaconu2018initial} and our Walk on Heat Stars method.

\subsection{Boundary Integral Perspective}

The Walk on Spheres and Walk on Stars algorithms can be understood as methods for solving the \emph{boundary integral equation} (BIE).
For a general Poisson problem $\Delta u = -f$ with mixed boundary conditions, the solution at a point $x$ admits the representation
\begin{align}
    \label{equ:bie_background}
    \alpha(x)\, u(x)
    &= \int_{\partial A} \Bigl[ P^C(x,z)\, u(z) - G^C(x,z)\, \frac{\partial u}{\partial \vec{n}}(z) \Bigr] \mathrm{d}S_z \\
    &+ \int_A G^C(x,y)\, f(y)\,\mathrm{d}y,
\end{align}
which holds for any subdomain $A \subset \Omega$ and any auxiliary domain $C \subset \mathbb{R}^n$ whose Poisson kernel $P^C$ and Green's function $G^C$ are known in closed form, with $\alpha(x)=1$ for $x\in A$ and $\alpha(x)=\frac12$ for $x\in\partial A$.
The BIE reduces the global PDE to integrals over the local subdomain $A$ and its boundary.

\textbf{Walk on Spheres.} For pure Dirichlet problems, choosing $A = C = B(x,R)$ (the largest ball centered at $x$ and contained in $\Omega$) eliminates the normal derivative term since $G^B(x,z)=0$ for $z\in\partial B$.
The BIE simplifies to
\begin{align}
    u(x) = \int_{\partial B(x,R)} P^B(x,z)\, u(z)\,\mathrm{d}S_z + \int_{B(x,R)} G^B(x,y)\, f(y)\,\mathrm{d}y,
\end{align}
yielding a recursive Monte Carlo estimator. The unknown $u(z)$ on the sphere boundary is estimated by further walks, while the source integral is sampled directly.

\oldtext{}\changedtext{Concretely, one samples a point uniformly on the largest inscribed sphere, restarts the walk there, and repeats until the walker is within distance $\varepsilon$ of the Dirichlet boundary. The recursion then terminates with the prescribed Dirichlet value at the closest boundary point. This is \emph{$\varepsilon$-shell absorption}. In the heat-equation analogue (Section~\ref{sec:wohs}), the walk also moves backward in time and may instead terminate near the initial time plane, where it uses the initial condition.}

\textbf{Walk on Stars.} For mixed Dirichlet--Neumann problems, \citet{sawhney2023walk} replace the ball with a star-shaped region $\mathrm{St}(x,R) = \Omega \cap B(x,R)$, where $R$ is chosen as the closest silhouette distance to the Neumann boundary.
Denoting the spherical portion by $\partial\mathrm{St}_B$ and the Neumann wall portion by $\partial\mathrm{St}_N$, the BIE becomes
\begin{align}
    \alpha(x)\, u(x)
    &= \int_{\partial\mathrm{St}} P^B(x,z)\, u(z)\,\mathrm{d}S_z - \int_{\partial\mathrm{St}_N} G^B(x,z)\, h(z)\,\mathrm{d}S_z \\
    &+ \int_{\mathrm{St}} G^B(x,y)\, f(y)\,\mathrm{d}y,
\end{align}
where the normal derivative on $\partial\mathrm{St}_N$ is substituted by the known Neumann data $h$.
The estimator now includes three terms: a recursive double-layer term on $\partial\mathrm{St}$, a Neumann contribution term on $\partial\mathrm{St}_N$, and a source term in $\mathrm{St}$.
Importantly, a uniform random direction cast from $x$ perfectly importance-samples the Poisson kernel $P^B$ on $\partial\mathrm{St}$, with unit weight regardless of whether the ray hits $\partial\mathrm{St}_B$ or $\partial\mathrm{St}_N$.


\section{Methodology}

We develop a Monte Carlo method for estimating the solution of the transient initial-boundary value problem
\begin{equation}
\label{equ:ibvp}
\left\{
\begin{aligned}
    \frac{\partial u}{\partial t} (x, t) &= \Delta u(x, t) \oldtext{-}\changedtext{+} f(x, t), && (x, t) \in \Omega \times [0, T], \\
    u(x, t) &= g(x, t), && (x, t) \in \partial \Omega_D \times [0, T], \\
    \frac{\partial u}{\partial \vec{n}}(x, t) &= h(x, t), && (x, t) \in \partial \Omega_N \times [0, T], \\
    u(x, 0) &= g_0(x), && x \in \Omega,
\end{aligned}
\right.
\end{equation}
posed on a \oldtext{bounded}\,\changedtext{fixed} spatial domain $\Omega \subset \mathbb{R}^n$ with Dirichlet boundary $\partial\Omega_D$ and Neumann boundary $\partial\Omega_N$.\oldtext{}\changedtext{ The local construction also applies to exterior domains, as illustrated in Section~\ref{sec:unbounded}.}\changedtext{ Throughout, the physical domain $\Omega$ is fixed in time and the method does not support moving physical domains.} The source term $f$, boundary data $g$, $h$, and initial condition $g_0$ are assumed sufficiently regular.

Unlike the elliptic equations considered in prior WoS/WoSt works, the transient heat equation couples space and time as a parabolic PDE. For the case $f = 0$, the solution $u$ is called a \emph{caloric function}\oldtext{ (}\,\changedtext{, }the parabolic analogue of a harmonic function. It satisfies the homogeneous heat equation $\partial_t u = \Delta u$ and obeys a mean-value property over heat balls~\cite{watson2012introduction}.

\subsection{Boundary Integral Equation}
\label{sec:problem_setup}

\oldtext{Our method is built on the time-dependent boundary integral formulation of parabolic problems.} \changedtext{We start from describing the time-dependent boundary integral equation of parabolic problems, which is the mathematical foundation of the proposed algorithm.} The fundamental solution of the heat equation in $\mathbb{R}^n$ is
\begin{equation}
    \label{equ:fundamental}
    \Phi(x, t) = (4\pi t)^{-n/2} \exp\!\left(-\frac{|x|^2}{4t}\right),
\end{equation}
and we write $\Phi(y, s)$ for the same kernel evaluated at a spatial offset $y$ and backward time $s$.

For a fixed spatial domain $\Omega$, the standard cylindrical boundary integral representation~\cite{costabel1990boundary} expresses the solution at an interior point $(x, t)$ as the sum of three contributions: propagation of the initial condition, boundary layer potentials in space-time, and the volume source term. Let $G$ denote the heat kernel, and write $G(y,\ell) = \Phi(x-y,\ell)$ for a spatial offset $y$ and lag $\ell$, with $\Phi$ from~\eqref{equ:fundamental}. For the rest of this subsection, $\tau$ denotes physical time, so the kernel inside the layer potentials is evaluated at lag $t-\tau$.
\begin{equation}
    \label{equ:bie_cylinder}
    \begin{aligned}
        \alpha(x) u(x, t) &= \int_{\Omega} G(y,t) \, g_0(y) \, \mathrm{d} y \\
        &+ \int_0^t \int_{\partial\Omega} \Bigl[ G(y,t-\tau) \, \frac{\partial u}{\partial \vec{n}}(y, \tau) - \frac{\partial G}{\partial \vec{n}}(y,t-\tau) \, u(y, \tau) \Bigr] \mathrm{d} S_y \, \mathrm{d} \tau \\
        &+ \int_0^t \int_\Omega G(y,t-\tau) \, f(y, \tau) \, \mathrm{d} y \, \mathrm{d} \tau,
    \end{aligned}
\end{equation}
where $\alpha(x)$ is the solid-angle coefficient ($1$ for interior points, $\frac{1}{2}$ on smooth boundaries). \oldtext{}\changedtext{ In the boundary integral, $G\,\partial u/\partial\vec{n}$ is the single-layer term and $-(\partial G/\partial\vec{n})u$ is the double-layer term. The former weights the solution normal derivative by the Green's function. The latter weights the solution by the normal derivative of the Green's function.}

To accommodate the evolving auxiliary heat balls within the fixed physical domain, we employ a more general \emph{non-cylindrical} boundary integral formulation~\cite{brugger2020boundary}. While the treatment of~\cite{brugger2020boundary} uses Sobolev spaces, we give a self-contained derivation under smooth assumptions in Appendix~\ref{app:noncylindrical_bie}. Let $\{\Omega_\tau\}_{\tau=0}^{t}$ be a family of smoothly evolving spatial domains, defining the space-time region $E = \bigcup_{\tau=0}^t (\Omega_\tau \times \{\tau\})$. For any interior point $(x, t) \in E$, the same kernel $G$ yields the representation
\begin{equation}
    \label{equ:bie_noncylindrical}
    \begin{aligned}
        \alpha(x) u(x, t) &= \int_{\Omega_0} G(y,t) \, g_0(y) \, \mathrm{d} y \\
        &+ \int_0^t \int_{\partial\Omega_\tau} \Bigl[ G(y,t-\tau) \, \frac{\partial u}{\partial \vec{n}}(y, \tau) - \frac{\partial G}{\partial \vec{n}}(y,t-\tau) \, u(y, \tau) \Bigr] \mathrm{d} S_y \, \mathrm{d} \tau \\
        &+ \int_0^t \int_{\partial\Omega_\tau} G(y,t-\tau) \, \langle V, \vec{n} \rangle \, u(y, \tau) \, \mathrm{d} S_y \, \mathrm{d} \tau \\
        &+ \int_0^t \int_{\Omega_\tau} G(y,t-\tau) \, f(y, \tau) \, \mathrm{d} y \, \mathrm{d} \tau,
    \end{aligned}
\end{equation}
where $V = \partial_\tau X(\tau)$ is the velocity of the moving boundary and the third term accounts for convective flux through the deforming surface, which is a consequence of the Reynolds transport theorem. This term vanishes when $\Omega_\tau$ is stationary, recovering the cylindrical formulation. A complete derivation is provided in Appendix~\ref{app:noncylindrical_bie}. In our construction, we will choose $\Omega_\tau$ to be the spatial slice of a heat ball at physical time $\tau$, whose boundary moves with a known velocity, and replace $G$ by the corresponding heat-ball Green's function, which is introduced in Section~\ref{sec:heat_ball}.

\subsection{Heat Ball Geometry and Parameterization}
\label{sec:heat_ball}

\begin{figure}
    \centering
    \includegraphics[width=\linewidth]{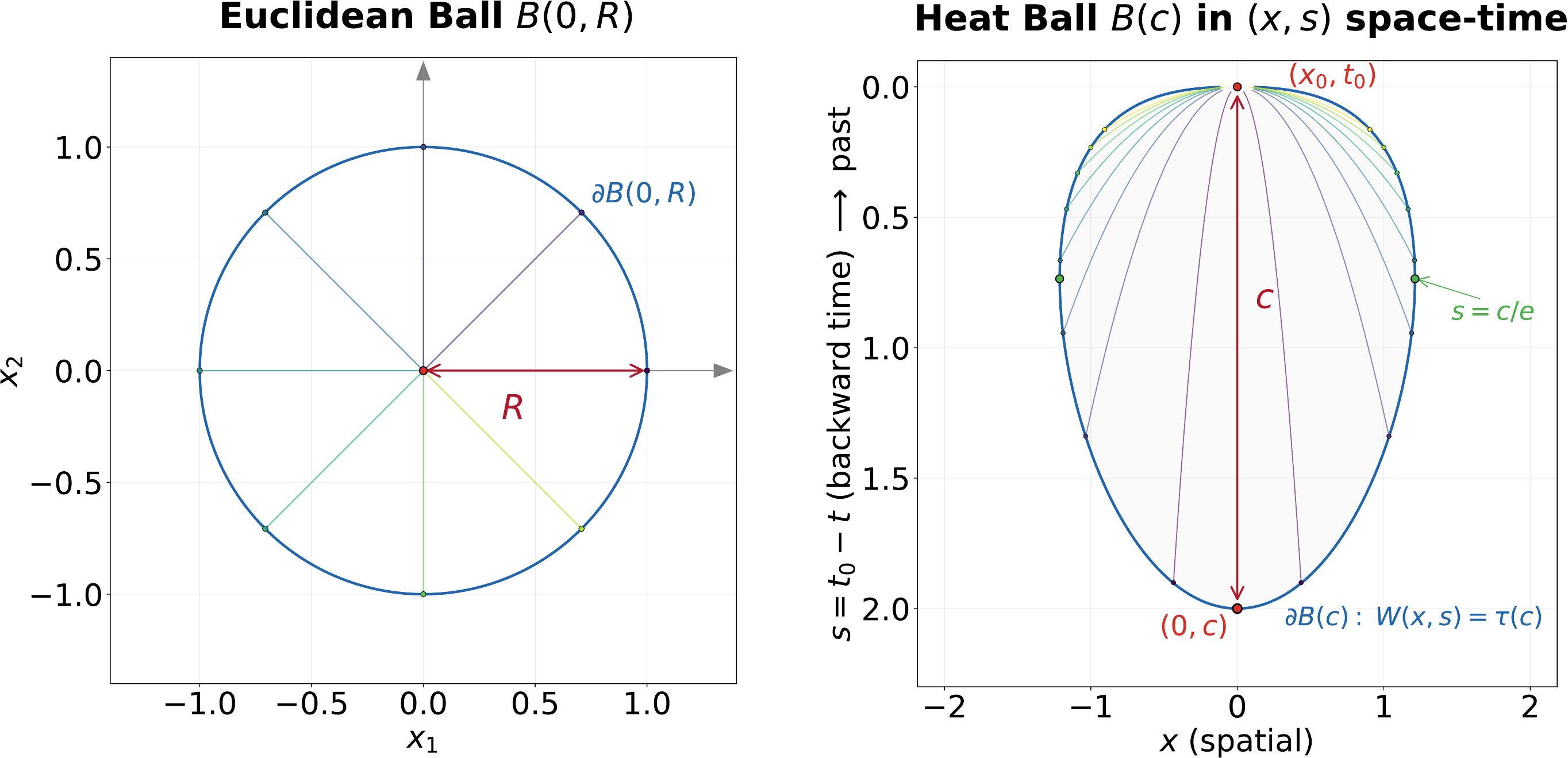}
    \caption{Heat ball geometry. \textbf{Left:} Euclidean ball with straight radial rays $x = x_0 + R\omega$. \textbf{Right:} Heat ball in $(1+1)$-dimensional space-time. The boundary is the heat sphere. Curved radial traces $|x| = \sqrt{2n c \oldtext{u}\changedtext{\mu} e^{-\oldtext{u}\changedtext{\mu}}}$ replace straight rays. The ball expands from a point, reaches maximum radius at $s = c/e$, and contracts back to a point at $s = c$.}
    \label{fig:heat_ball}
\end{figure}

\oldtext{In the elliptic setting, the Walk on Spheres algorithm rests on the mean value property: the solution of Laplace's equation at a point equals its average over any sphere centered there. The sphere's surface is parameterized by a fixed radius $R$ and a direction $\omega \in S^{n-1}$, giving the familiar radial rays $x = x_0 + R\omega$. For the heat equation, the analogous mean value property requires a fundamentally different geometry because space and time scale anisotropically: the fundamental solution depends on the ratio $|x|^2/t$, coupling the two coordinates through the parabolic scaling. The appropriate domain is the \emph{heat ball}~\cite{watson2012introduction}, and its surface is not traversed by straight radial rays but by curved \emph{radial traces} that reflect the parabolic scaling.}

\changedtext{The object that replaces the Euclidean sphere here is the \emph{heat ball}~\cite{watson2012introduction}, a spindle-shaped region of space--time that expands from a point, reaches a maximum radius, and contracts again. This subsection defines it and describes how its volume is parameterized by a radial trace, which is defined by a spatial-temporal direction and a radial coordinate. The sampling strategies below draw both interior and boundary samples from this parameterization.}

Let $\Phi(x, t)$ denote the fundamental solution introduced above. The \emph{heat ball} centered at $(x_0, t_0)$ with radius $c > 0$ is the space-time region
\begin{equation}
    \mathcal{H}(x_0, t_0, c) = \left\{ (y, t) \mid 0 < t_0 - t < c,\; \Phi(x_0 - y, t_0 - t) > \tau(c) \right\},
\end{equation}
where $\tau(c) = (4\pi c)^{-n/2}$ is the level-set threshold. In backward time $s = t_0 - t$, the inequality $\Phi(y, s) > \tau(c)$ reduces to
\begin{equation}
    |x_0 - y| < \sqrt{2n \, s \log\frac{c}{s}}, \qquad 0 < s < c.
\end{equation}
Thus each temporal cross-section is an $(n-1)$-sphere whose radius
\begin{equation}
    r(s) = \sqrt{2n s \log(c/s)}
\end{equation}
grows from $0$ at $s = 0$, attains its maximum $r_{\max} = \sqrt{2nc/e}$ at $s = c/e$ (here and below $e$ denotes the base of the natural logarithm), and contracts back to $0$ at $s = c$. As shown in Figure~\ref{fig:heat_ball}, the heat ball is therefore spindle-shaped in space-time, tapering to a point at both $t = t_0$ and $t = t_0 - c$.

\oldtext{This geometry is best understood through the \emph{logarithmic time} coordinate
\begin{equation}
    u = \log\frac{c}{s}, \qquad u > 0,
\end{equation}
which absorbs the parabolic scaling into a dimensionless variable. Introducing a radial parameter $\rho \in (0, c]$ that sweeps from the center to the boundary of the heat ball, the same substitution $s = \rho e^{-u}$ into $r(s)$ describes the full interior. At the outermost value $\rho = c$, we recover the heat-sphere surface:
\begin{equation}
    \label{equ:radial_trace}
    s = \rho\,e^{-u}, \qquad |x - x_0| = \sqrt{2n\,u\,s}.
\end{equation}
Varying $\rho$ over $(0, c]$ at fixed $(u, \omega)$ sweeps a \emph{space-time radial ray} through the heat-ball interior, a path along which the ratio $|x|^2 / s = 2n u$ is constant. These radial rays are the parabolic analogue of Euclidean radial lines: the ray starts at the center $(\rho = 0)$, follows a curved trajectory governed by the parabolic scaling, and terminates on the heat-sphere surface at $\rho = c$. They form the backbone of our sampling strategy (Section~\ref{sec:directional_sampling}).}

\changedtext{On the heat-sphere boundary, define the \emph{logarithmic time}
\begin{equation}
    \mu = \log\frac{c}{s}, \qquad s = t_0 - t.
\end{equation}
It ranges over $\mu>0$, tending to $+\infty$ as $s\to0$ and vanishing at the far tip $s=c$. Together with a spatial direction $\omega\in S^{n-1}$, it determines a \emph{radial trace}: a curve obtained by varying $\rho\in(0,c]$ at fixed $(\mu,\omega)$,
\begin{equation}
    \label{equ:radial_trace}
    x(\rho)=x_0+\sqrt{2n\rho\mu e^{-\mu}}\;\omega, \qquad
    t(\rho)=t_0-\rho e^{-\mu}.
\end{equation}
The trace starts at $(x_0,t_0)$ as $\rho\to0$ and meets the heat sphere at $\rho=c$. In the interior, the coordinate relation is $\mu=\log(\rho/s)$, rather than $\log(c/s)$. Conversely, for $s=t_0-t>0$ and $x\ne x_0$, the inverse coordinates are $\mu=|x-x_0|^2/(2ns)$, $\omega=(x-x_0)/|x-x_0|$, and $\rho=s e^\mu$. These coordinates cover the heat-ball interior apart from the central time axis, a set of zero volume where $\mu=0$ and the direction is undefined. This parameterization is the backbone of our sampling strategy (Section~\ref{sec:directional_sampling}).}

The \emph{heat sphere} $\partial\mathcal{H}(x_0, t_0, c)$ is the boundary of the heat ball, satisfying
\begin{equation}
    |x - x_0|^2 = 2n (t_0 - t) \log\frac{c}{t_0 - t}.
\end{equation}
When the heat ball is chosen as the domain in the non-cylindrical boundary integral equation~\eqref{equ:bie_noncylindrical}, the initial term $\int_{\Omega_0} G \, g_0 \, \mathrm{d}y$ vanishes because $\Omega_0$ degenerates to a single point.

To exploit this geometry for Monte Carlo estimation, we construct a Green's function that vanishes on the heat-sphere boundary by subtracting the level-set constant:
\begin{equation}
    G(x, t; x_0, t_0) = \Phi(x_0 - x, t_0 - t) - \tau(c),
\end{equation}
so that $G = 0$ identically on $\partial\mathcal{H}(x_0, t_0, c)$. Substituting $G$ into~\eqref{equ:bie_noncylindrical} and noting that the vanishing boundary condition eliminates the single-layer term, we obtain the \emph{non-uniform mean value property} for the heat ball:
\begin{align}
    \label{equ:mean_value}
    u(x_0, t_0) = &-\int_0^c \int_{\partial \Omega_s} \frac{\partial G}{\partial \vec{n}} \, u(y, t_0 - s) \; \mathrm{d} S_y \, \mathrm{d}s  \\
    &\;+\; \int_0^c \int_{\Omega_s} G(y, s) \, f(y, t_0 - s) \; \mathrm{d}y \, \mathrm{d}s,
\end{align}
where $\Omega_s = \{y : |x_0 - y| < r(s)\}$ is the spatial ball of radius $r(s)$ at backward time $s$.

\subsection{Walk on Heat Spheres}
\label{sec:wos}

\oldtext{}\changedtext{As a direct application of the boundary integral equation, we recover the Walk on Heat Spheres algorithm. \citet{deaconu2018initial} obtained this algorithm from the stochastic side, as the joint first exit of a Brownian path. Here it follows directly from~\eqref{equ:bie_noncylindrical}, by inserting the heat-ball Green's function $G = \Phi - \tau(c)$ and factorizing the double-layer kernel. It is the pure-Dirichlet building block of the heat-star walk developed in Section~\ref{sec:wohs}. We use $\Gamma(\alpha,\beta)$ for the Gamma distribution with shape $\alpha>0$ and scale $\beta>0$, and $\Gamma(z;\alpha,\beta)$ for its density. Standard library routines sample this distribution directly and efficiently.}

\begin{algorithm}[t]
\caption{Walk on Heat Spheres}
\label{alg:wos}
\begin{algorithmic}[1]
\Require Current state $(x, t)$, termination threshold $\varepsilon$
\Ensure Estimate $\widehat{u}(x, t)$
\State $R \gets \Call{distance}{x, \partial\Omega}$
\State $c \gets \min\!\bigl(t,\; e R^2 / (2n)\bigr)$
\State $\oldtext{u}\changedtext{\mu} \sim \Gamma\!\bigl(\frac{n}{2}+1,\; \frac{2}{n}\bigr)$, \quad $\omega \sim \Call{Uniform}{S^{n-1}}$
\State $s \gets c\,e^{-\oldtext{u}\changedtext{\mu}}$, \quad $r \gets \sqrt{2n\,c\,\oldtext{u}\changedtext{\mu}\,e^{-\oldtext{u}\changedtext{\mu}}}$
\State $x' \gets x + r\,\omega$, \quad $t' \gets t - s$
\State $\widehat{S} \gets \Call{SampleSource}{x, t, c, \omega\changedtext{,\infty}}$ \Comment{Volume source term}
\If{$\Call{distance}{x', \partial\Omega} < \varepsilon$ \textbf{or} $t' < \varepsilon$}
    \State \Return $\Call{boundaryOrInitialValue}{x', t'} + \widehat{S}$
\Else
    \State \Return $\Call{WalkOnHeatSpheres}{x', t', \varepsilon} + \widehat{S}$
\EndIf
\end{algorithmic}
\end{algorithm}

\begin{figure}
    \centering
    \includegraphics[width=\linewidth]{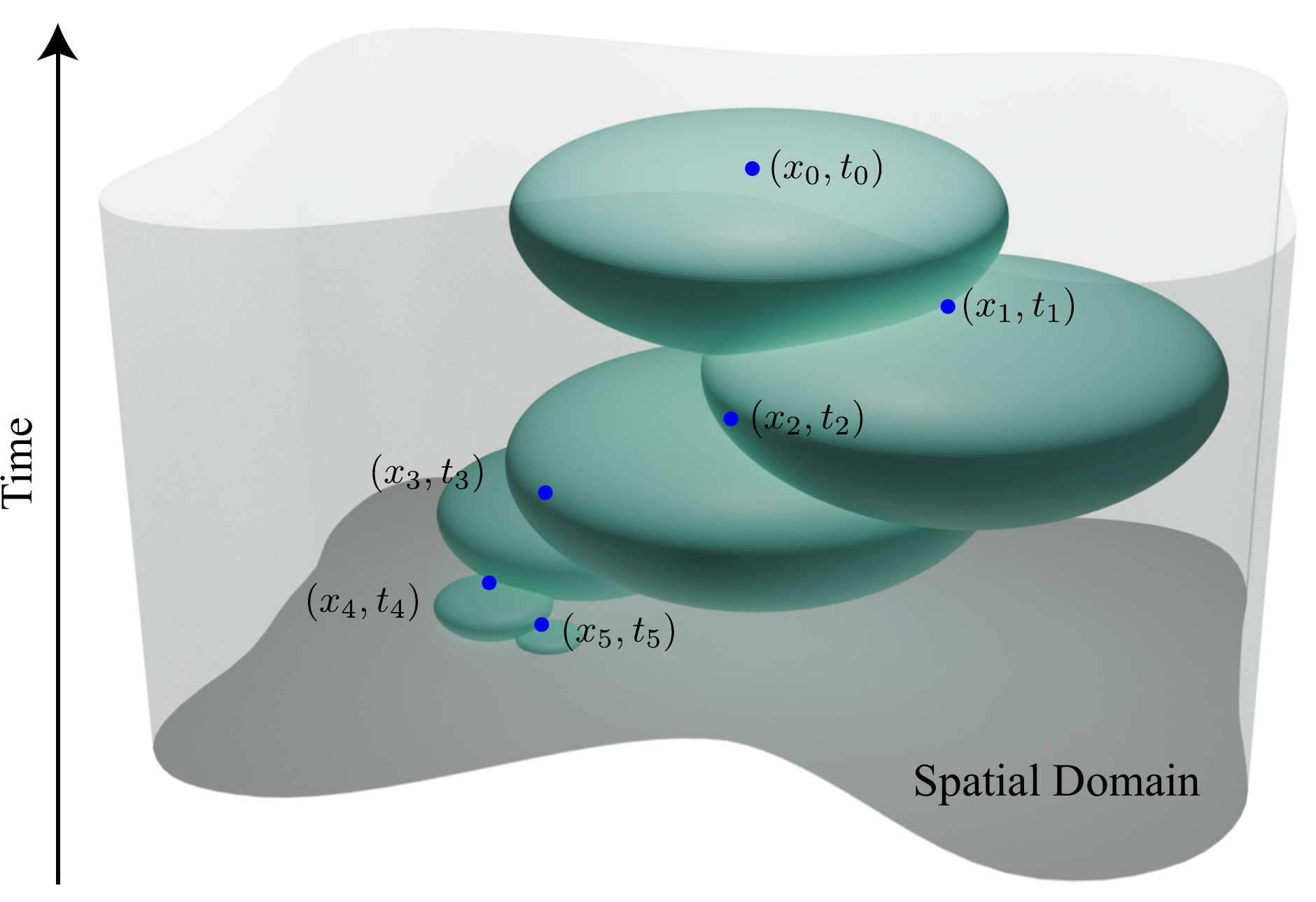}
    \caption{Walk on Heat Spheres (WoHS). The next state $(x_{k+1}, t_{k+1})$ is sampled from the heat-sphere boundary, and the estimator recursively evaluates $\widehat{u}(x_{k+1}, t_{k+1})$ with unit weight. The source term is sampled from the interior of the heat ball.}
    \label{fig:wohs}
\end{figure}

The normal-derivative kernel $-\partial G / \partial \vec{n}$ on the heat sphere admits a remarkable simplification. On $\partial\Omega_s$, the outward normal is radial ($x \cdot \vec{n} = r(s)$) and $\Phi(r(s), s) = \tau(c)$ is constant. Hence
\begin{equation}
    -\frac{\partial G}{\partial \vec{n}} = \frac{r(s)}{2s} \, \tau(c).
\end{equation}
Substituting $r(s) = \sqrt{2n s \log(c/s)}$ and changing variable to $\oldtext{u}\changedtext{\mu} = \log(c/s)$ \oldtext{with $\mathrm{d}s = c e^{-u} \mathrm{d}u$}\changedtext{with Jacobian $|\mathrm{d}s|=c e^{-\mu}\,\mathrm{d}\mu$} transforms the joint space-time measure into
\begin{equation}
    -\frac{\partial G}{\partial \vec{n}} \; \mathrm{d}\sigma \; \mathrm{d}s
    = \frac{1}{|S^{n-1}|} \; \Gamma\!\left(\oldtext{u}\changedtext{\mu};\, \frac{n}{2}+1,\; \frac{2}{n}\right) \mathrm{d}\oldtext{u}\changedtext{\mu} \; \mathrm{d}\omega,
\end{equation}
where $\mathrm{d}\sigma$ is the surface measure on $\partial\Omega_s$ and \oldtext{$\mathrm{d}\omega$ is the uniform probability measure on $S^{n-1}$. (For a complete derivation, see Appendix~\ref{app:sampling_directional}.)}\changedtext{$\mathrm{d}\omega$ is surface measure on $S^{n-1}$, so $\mathrm{d}\omega/|S^{n-1}|$ is the uniform probability measure. For a complete derivation, see Appendix~\ref{app:sampling_directional}.}

\oldtext{}\changedtext{The shape $n/2+1$ and scale $2/n$ are determined by the heat-ball geometry. After substituting $\mu=\log(c/s)$, the transformed double-layer measure contains the factor $\mu^{n/2}e^{-n\mu/2}$, which gives these parameters. They depend only on the spatial dimension and are not tuned empirically.}

Thus the sampling procedure for the next state $(x_{k+1}, t_{k+1})$ reduces to two independent draws:
\begin{enumerate}
    \item Sample $\oldtext{u}\changedtext{\mu} \sim \Gamma(\frac{n}{2}+1, \frac{2}{n})$ and set $s = c e^{-\oldtext{u}\changedtext{\mu}}$.
    \item Sample $\omega \sim \mathrm{Uniform}(S^{n-1})$ and set $x_{k+1} = x_k + r(s)\,\omega$, $t_{k+1} = t_k - s$.
\end{enumerate}

It remains to determine the heat-ball radius $c$ at each step. The spatial radius $r(s)$ attains its maximum $r_{\max} = \sqrt{2nc/e}$. Let $R = \mathrm{dist}(x_0, \partial\Omega)$ be the Euclidean distance to the spatial boundary. To keep the entire heat ball inside the domain we require $c < t_0$ (the ball must not extend beyond the initial time) and $r_{\max} < R$. Thus we set
\begin{equation}
    c = \min\!\left(t_0,\; \frac{e R^2}{2n}\right).
\end{equation}
When $R$ is small relative to $t_0$, the heat ball shrinks accordingly. The walk terminates upon absorption at the Dirichlet boundary or the initial time plane.

Figure~\ref{fig:wohs} illustrates the procedure of the WoHS walk. At the current state $(x_k, t_k)$, the largest heat ball contained in the domain is constructed (center). The next state $(x_{k+1}, t_{k+1})$ is sampled from the heat-sphere boundary by drawing a logarithmic time $\oldtext{u}\changedtext{\mu} \sim \Gamma$ and a direction $\omega \sim \mathrm{Uniform}(S^{n-1})$ independently, then tracing the radial curve to the boundary. The walk recurses from $(x_{k+1}, t_{k+1})$ until absorption at the Dirichlet boundary or the initial time plane. When a source term $f$ is present, an auxiliary point is sampled inside the heat ball \oldtext{(dashed arrow) and contributes a deterministic correction to the estimator}\,\changedtext{to estimate the source contribution.} The source sampling strategy is detailed in Section~\ref{sec:source_sampling}.

When $f \equiv 0$, the source term vanishes and the WoHS estimator~\eqref{eq:wohs_wos_estimator} reduces to a pure recursive walk on heat-sphere boundaries, recovering exactly the Walk on Heat Balls algorithm of \citet{deaconu2018initial}.

\estimatorbox{Walk on Heat Spheres Estimator}{%
A recursive single-sample estimator for \eqref{equ:mean_value} at a point $(x_k, t_k)$ in the domain is given by
\begin{align}
    \label{eq:wohs_wos_estimator}
    \widehat{u}(x_k, t_k) \; \coloneqq \;
    &\frac{ \bigl(-\frac{\partial G}{\partial \vec{n}}\bigr)(x_k, t_k;\, x_{k+1}, t_{k+1}) \; \widehat{u}(x_{k+1}, t_{k+1}) }{ p^{\partial\mathcal{H}(x_k,c)}\!(x_{k+1}, t_{k+1}) } \nonumber\\
    &\;+\; \frac{ G_{x_k,c}(y_{k+1}, s_{k+1}) \; f(y_{k+1}, t_k - s_{k+1}) }{ p^{\mathcal{H}(x_k,c)}\!(y_{k+1}, s_{k+1}) },
\end{align}
where
\begin{itemize}
    \item $(x_{k+1}, t_{k+1})$ lies on the heat-sphere boundary $\partial\mathcal{H}(x_k, c)$ and is sampled from $p^{\partial\mathcal{H}(x_k,c)}$, while $(y_{k+1}, s_{k+1})$ lies inside the heat ball $\mathcal{H}(x_k, c)$ and is sampled from $p^{\mathcal{H}(x_k,c)}$,
    \item the heat-ball radius $c$ is chosen so that $\mathcal{H}(x_k, c)$ is contained entirely in the domain (pure Dirichlet case),
    \item $G_{x_k,c}(y, s) \coloneqq \Phi(x_k - y, s) - \tau(c)$ is the heat-ball Green's function, and the double-layer kernel $-\partial G/\partial \vec{n}$ is importance-sampled exactly, yielding unit weight in the recursive term.
\end{itemize}
}

\subsection{Walk on Heat Stars}
\label{sec:wohs}

\oldtext{We now introduce the transient analogue of Walk on Stars~\cite{sawhney2023walk}: \emph{Walk on Heat Stars} (WoHSt).}
\changedtext{We now introduce \emph{Walk on Heat Stars} (WoHSt), the transient analogue of Walk on Stars~\cite{sawhney2023walk}.} The \emph{heat-star} region is the intersection of the heat ball with the spatial domain:
\begin{equation}
    \mathcal{H}^\star(x_0, t_0, c) \;=\; \mathcal{H}(x_0, t_0, c) \cap (\Omega \times [t_0 - c, t_0]).
\end{equation}
Its space-time boundary $\partial\mathcal{H}^\star$ comprises two disjoint components (Figure~\ref{fig:heatstar}):
\begin{align}
    \partial\mathcal{H}^\star_{\mathrm{sphere}} &= \partial\mathcal{H}(x_0, t_0, c) \cap (\Omega \times [t_0 - c, t_0]), \label{equ:boundary_sphere} \\
    \partial\mathcal{H}^\star_{N} &= \mathcal{H}(x_0, t_0, c) \cap (\partial\Omega_N \times [t_0 - c, t_0]). \label{equ:boundary_neumann}
\end{align}
As in WoSt, the Dirichlet boundary is never included in the star. Dirichlet conditions are handled by $\varepsilon$-shell absorption.

\subsubsection{Boundary Integral Decomposition}

Substituting the heat-star region into the non-cylindrical BIE~\eqref{equ:bie_noncylindrical}, the initial term vanishes because $\mathcal{H}_0^\star = \emptyset$. The Green's function $G = \Phi - \tau(c)$ vanishes by construction on $\partial\mathcal{H}^\star_{\mathrm{sphere}}$, eliminating the single-layer potential there. On the stationary Neumann wall $\partial\mathcal{H}^\star_N$, we have $\langle V, \vec{n} \rangle = 0$ and the boundary condition prescribes $\partial u / \partial \vec{n} = h$. The representation therefore decomposes into three terms:
\begin{align}
    \label{equ:bie_decomposed}
    u(x_0, t_0) &= \underbrace{\int_{\partial\mathcal{H}^\star} \!\!\left(-\frac{\partial G}{\partial \vec{n}}\right) u \; \mathrm{d}S \, \mathrm{d}s}_{\text{double-layer on } \partial\mathcal{H}^\star}
    \;+\; \underbrace{\int_{\partial\mathcal{H}^\star_N} \!\! G \, h \; \mathrm{d}S \, \mathrm{d}s}_{\text{single-layer on } \partial\mathcal{H}^\star_N} \\
    &\;+\; \underbrace{\int_{\mathcal{H}^\star} G \, f \; \mathrm{d}y \, \mathrm{d}s}_{\text{volume source}}.
\end{align}

\begin{figure}
    \centering
    \oldtext{}\changedtext{\includegraphics[width=\linewidth]{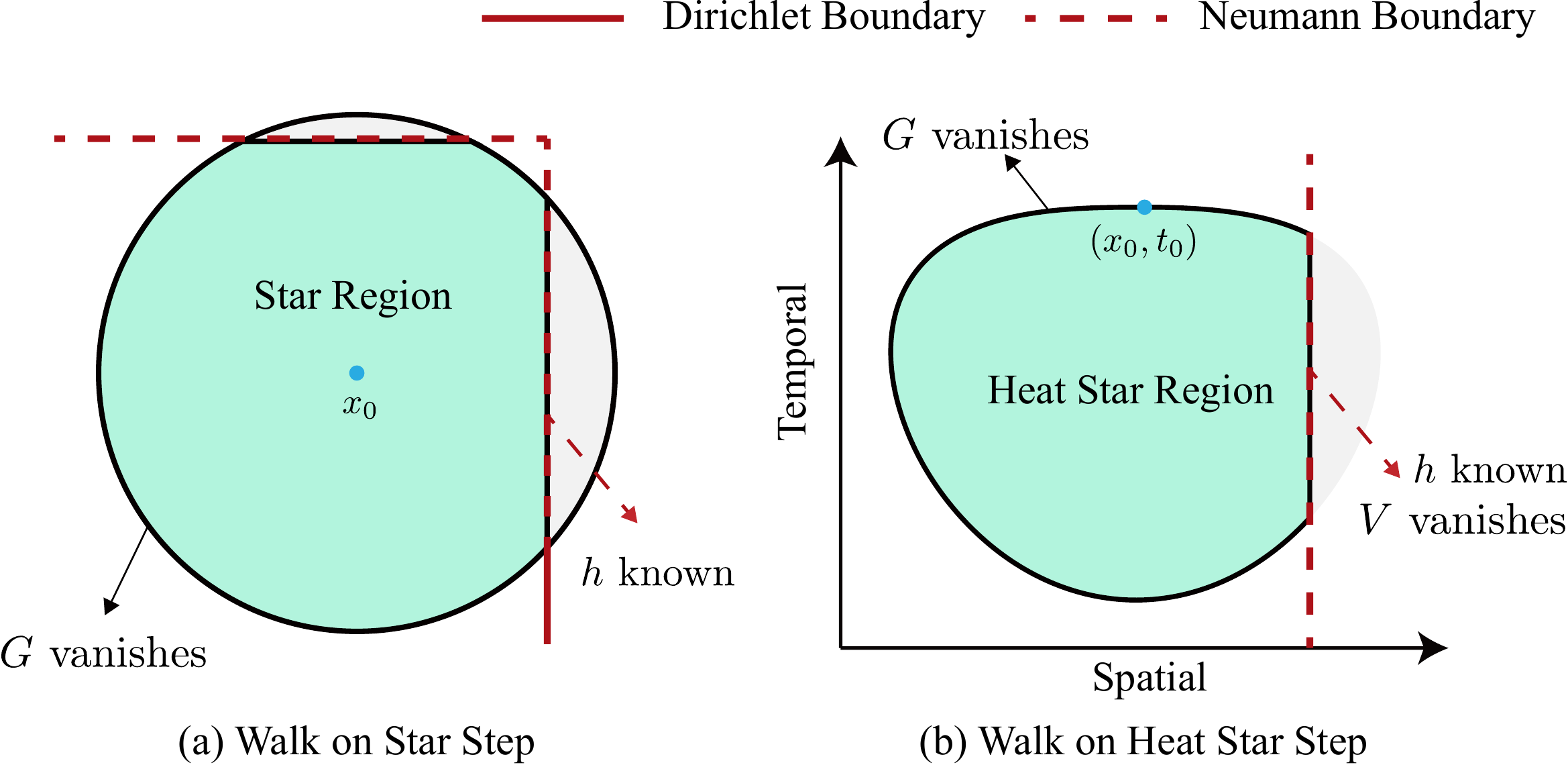}}
    \caption{From Walk on Stars to Walk on Heat Stars. \textbf{Left:} One step of the standard (elliptic) Walk on Stars~\cite{sawhney2023walk}. A ball is expanded from the current point $x_k$. Its intersection with $\Omega$ forms the star-shaped region. The walker samples the next state on the star boundary $\partial\mathrm{St}$, which may lie on the spherical part $\partial\mathrm{St}_B$ or on the wall $\partial\mathrm{St}_N$. \textbf{Right:} The transient analogue. The ball is replaced by a heat ball, the star by the heat-star region $\mathcal{H}^\star = \mathcal{H}(x_k, t_k, c) \cap (\Omega \times [t_k-c, t_k])$, and the boundary decomposes into the spherical component $\partial\mathcal{H}^\star_{\mathrm{sphere}}$ and the Neumann wall $\partial\mathcal{H}^\star_N$.}
    \label{fig:heatstar}
\end{figure}

Figure~\ref{fig:heatstar} compares the elliptic Walk on Stars (left) with our parabolic Walk on Heat Stars (right). Structurally, both estimators share the same three-term form: a recursive double-layer integral over the \oldtext{spherical}\changedtext{heat-star} boundary,\oldtext{ a deterministic single-layer correction}\,\changedtext{  a single-layer contribution} over the Neumann wall, and a volumetric source term. The difference is that the Euclidean ball and star-shaped region are replaced by their space-time counterparts (the heat ball and heat star), and the sampling distributions are lifted from the spatial domain to the joint space-time measure governed by the heat kernel.

\subsubsection{The WoHSt Estimator}
\label{sec:wohst_estimator}

Interpreted probabilistically, the double-layer integral governs the recursive random walk. \oldtext{the kernel $-\partial G / \partial \vec{n}$ serves as the transition density from the current state to the next point on $\partial\mathcal{H}^\star$. The single-layer integral contributes a deterministic correction whenever the walker lands on $\partial\mathcal{H}^\star_N$, proportional to the Neumann data $h$ weighted by the reflected Green's function.}\changedtext{We choose a probability density $p^{\partial\mathcal{H}^\star}$ proportional to the nonnegative double-layer kernel on the local heat star, accounting for its normalization through $\alpha(x_k)$ in the estimator below. The single-layer integral is estimated separately using auxiliary samples on $\partial\mathcal{H}^\star_N$, weighted by the Green's function and the prescribed Neumann data $h$.} The volume term accounts for the source $f$ and can be estimated by sampling auxiliary points inside $\mathcal{H}^\star$. The exact importance sampling of $-\partial G / \partial \vec{n}$ and the associated source and flux sampling strategies are detailed in Section~\ref{sec:sampling}.

\oldtext{Since the boundary integral equation~\eqref{equ:bie_decomposed} is an exact representation of the solution, and each sampling strategy draws from the corresponding integral measure (with unit importance weight), the recursive estimator is unbiased: $\mathbb{E}[\widehat{u}(x_k, t_k)] = u(x_k, t_k)$.} \changedtext{The densities below are importance-sampling choices for the three integrals in~\eqref{equ:bie_decomposed}. Each sampled contribution is divided by its own density. Our directional rule gives unit weight for the recursive term, whereas the flux and source contributions retain their corresponding weights. The representation yields an unbiased estimator with exact recursion and boundary termination. The implemented $\varepsilon$-shell stopping rule introduces a termination approximation.}

\estimatorbox{Walk on Heat Stars Estimator}{%
A recursive single-sample estimator for \eqref{equ:bie_decomposed} at a point $(x_k, t_k) \in \Omega \times \mathbb{R}^+$ is given by
\begin{align}
    \label{eq:wohst_estimator}
    \widehat{u}(x_k, t_k) \; \coloneqq \;
    &\frac{ \bigl(-\frac{\partial G}{\partial \vec{n}}\bigr)(x_k, t_k;\, x_{k+1}, t_{k+1}) \; \widehat{u}(x_{k+1}, t_{k+1}) }{ \alpha(x_k) \; p^{\partial\mathcal{H}^\star}\!(x_{k+1}, t_{k+1}) } \nonumber\\
    &\;+\; \frac{ G_{x_k,c}(z_{k+1}, s_{k+1}) \; h(z_{k+1}, t_k - s_{k+1}) }{ \alpha(x_k) \; p^{\partial\mathcal{H}^\star_N}\!(z_{k+1}, s_{k+1}) } \nonumber\\
    &\;+\; \frac{ G_{x_k,c}(y_{k+1}, s_{k+1}) \; f(y_{k+1}, t_k - s_{k+1}) }{ \alpha(x_k) \; p^{\mathcal{H}^\star}\!(y_{k+1}, s_{k+1}) },
\end{align}
where
\begin{itemize}
    \item the space-time points $(x_{k+1}, t_{k+1}) \in \partial\mathcal{H}^\star$, $(z_{k+1}, s_{k+1}) \in \partial\mathcal{H}^\star_N$, and $(y_{k+1}, s_{k+1}) \in \mathcal{H}^\star$ are sampled from the probability densities $p^{\partial\mathcal{H}^\star}$, $p^{\partial\mathcal{H}^\star_N}$, and $p^{\mathcal{H}^\star}$, resp.
    \item $c$ is chosen so that $\mathcal{H}^\star(x_k, c)$ is star-shaped with respect to $(x_k, t_k)$,
    \item $\alpha(x_k) = 1/2$ if $x_k \in \partial\Omega_N$ and $1$ otherwise,
    \item $G_{x_k,c}(y, s) \coloneqq \Phi(x_k - y, s) - \tau(c)$ is the heat-ball Green's function.
\end{itemize}
}

\subsubsection{Algorithm}

Algorithm~\ref{alg:wohs} implements WoHSt using the subroutines defined in Section~\ref{sec:sampling}. At each step, the heat-ball radius $c$ is determined from the silhouette distance to $\partial\Omega_N$. The Neumann correction is accumulated via \textsc{SampleFlux}. The recursive walk step is performed by \textsc{SampleDirectional}, which handles the $(\oldtext{u}\changedtext{\mu}, \omega)$ sampling, ray casting, and branch decision. Finally, the source term is sampled via \textsc{SampleSource}, which reuses the direction $\omega$ and the \oldtext{clamped step distance $r$ returned by \textsc{SampleDirectional} for the containment check. (The source integral runs over the heat-star interior, so its spatial extent in direction $\omega$ is exactly $r$.)}\changedtext{wall distance $d_{\mathrm{wall}}$ returned by \textsc{SampleDirectional} for the containment check.} The returned spatial displacement $r$ is clamped by the Dirichlet distance $d_D$ as a safety measure before the recursive call. Far from boundaries, SampleDirectional reduces to a standard Walk on Heat Spheres step (Algorithm~\ref{alg:wos}) with unit weight.

\begin{algorithm}[t]
\caption{Walk on Heat Stars}
\label{alg:wohs}
\begin{algorithmic}[1]
\Require Current state $(x, t)$, termination threshold $\varepsilon$, flag $\mathit{on\_neumann}$, normal $n_N$
\Ensure Estimate $\widehat{u}(x, t)$
\State $d_D \gets \Call{distance}{x, \partial\Omega_D}$
\If{$d_D < \varepsilon$}
    \State \Return $g(\Call{closestPoint}{x, \partial\Omega_D},\, t)$
\EndIf
\If{$t < \varepsilon$}
    \State \Return $g_0(x)$ \Comment{Initial condition}
\EndIf
\State $d_{\mathrm{star}} \gets \min\!\bigl(\Call{silhouetteDistance}{x, \partial\Omega_N},\; d_D\bigr)$
\State $c \gets \min\!\bigl(t,\; e \cdot d_{\mathrm{star}}^2 \,/\, (2n)\bigr)$
\State $\widehat{u} \gets 0$
\State \changedtext{$\alpha \gets 1$}
\State \changedtext{\textbf{if} $\mathit{on\_neumann}$ \textbf{then} $\alpha \gets 1/2$}
\State $\widehat{u} \mathrel{+}= \Call{SampleFlux}{x, t, c}\changedtext{\,/\,\alpha}$ \Comment{Neumann correction}
\State $(x', t', \mathit{on\_neumann}, n_N, r, \omega\changedtext{, d_{\mathrm{wall}}}) \gets \Call{SampleDirectional}{x, t, c, \mathit{on\_neumann}, n_N}$
\State $\widehat{u} \mathrel{+}= \Call{SampleSource}{x, t, c, \omega, \oldtext{r}\changedtext{d_{\mathrm{wall}}}}$ \Comment{Volume source term\oldtext{; heat-star extent in direction $\omega$ is $r$}}
\State $r \gets \min(r,\; d_D)$ \Comment{Clamp spatial step by Dirichlet distance}
\State $x' \gets x + r\,\omega$ \Comment{Recompute position with clamped step}
\State \Return $\widehat{u} + \Call{WalkOnHeatStars}{x', t', \varepsilon, \mathit{on\_neumann}, n_N}$
\end{algorithmic}
\end{algorithm}

\subsection{Sampling Strategies}
\label{sec:sampling}

\oldtext{The WoHSt estimator~\eqref{eq:wohst_estimator} involves three sampling distributions: the double-layer kernel $-\partial G / \partial \vec{n}$ on $\partial\mathcal{H}^\star$ (driving the recursive walk), the single-layer correction on $\partial\mathcal{H}^\star_N$, and the volumetric source term in $\mathcal{H}^\star$. Each admits an exact importance-sampling strategy derived directly from the heat ball geometry of Section~\ref{sec:heat_ball}.}

\changedtext{The WoHSt estimator~\eqref{eq:wohst_estimator} contains three integral contributions: the recursive double-layer term, the Neumann flux term, and the volume source term. We design an importance-sampling algorithm to estimate each contribution, using the heat-ball geometry of Section~\ref{sec:heat_ball}.} We present each as a self-contained subroutine.

\subsubsection{Directional Sampling under Heat Ball Geometry}
\label{sec:directional_sampling}

The double-layer kernel $-\partial G / \partial \vec{n}$ governs the random walk. At each step, the walker transitions from $(x, t)$ to a new state $(x', t')$ on $\partial\mathcal{H}^\star$, after which the process recurs. The central challenge is to sample $(x', t')$ from a density proportional to $-\partial G / \partial \vec{n}$ on the heat-star boundary.

In the elliptic Walk on Stars, the Poisson kernel coincides with the solid angle and is exactly importance-sampled by casting a uniform random direction from the current point~\cite{sawhney2023walk}. This elegant construction, known as directional sampling, yields unit importance weight on both the spherical and Neumann components of the star-shaped boundary. Our key insight is that the same principle extends to the parabolic setting when the geometry is properly understood.

Although the heat kernel's space-time anisotropy makes the double-layer kernel more complex than its elliptic counterpart, the radial trace parameterization~\eqref{equ:radial_trace} reveals a factorization that restores the simplicity of directional sampling. Specifically, the kernel separates into \oldtext{independent temporal and directional factors}\changedtext{a uniform directional factor and a temporal factor conditional on the direction}, enabling a two-stage sampling procedure that is the parabolic analogue of the elliptic ray cast.

\begin{proposition}
\label{prop:directional_sampling}
\oldtext{The double-layer kernel $-\partial G / \partial \vec{n}$ factorizes into a Gamma distribution over the logarithmic time $u$ and a uniform distribution over the spatial direction $\omega$}\changedtext{The double-layer kernel $-\partial G / \partial \vec{n}$ factorizes into a uniform distribution over the spatial direction $\omega$ and a piecewise Gamma density $\Gamma_\star(\mu;c,d)$ in $\mu=|x|^2/(2ns)$ conditional on $\omega$}:
\oldtext{\begin{equation}
    -\frac{\partial G}{\partial \vec{n}} (x, s) \; \mathrm{d} \sigma \; \mathrm{d} s
    = \frac{1}{|S^{n-1}|} \cdot \Gamma\!\left(u;\, \frac{n}{2}+1,\; \frac{2}{n}\right) \mathrm{d} \omega \; \mathrm{d} u,
\end{equation}}\changedtext{\begin{equation}
\label{eq:directional_piecewise_gamma}
 -\frac{\partial G}{\partial\vec n}\,\mathrm d\sigma\,\mathrm ds
 =\frac{1}{|S^{n-1}|}\,\Gamma_\star(\mu;c,d)\,\mathrm d\omega\,\mathrm d\mu ,
\end{equation}}
where $\mathrm{d}\sigma$ is the surface measure \changedtext{on the spatial boundary at backward time $s$} and \changedtext{$d=d_{\mathrm{wall}}(\omega)$ is the visibility distance to the Neumann wall ($d=\infty$ if there is no hit). Writing $r_{\mathrm{sphere}}(\mu)=\sqrt{2nc\mu e^{-\mu}}$, we define
\begin{equation}
\label{eq:heat_star_gamma}
 \Gamma_\star(\mu;c,d)=
 \begin{cases}
 \Gamma\!\left(\mu;\frac n2+1,\frac 2n\right), & r_{\mathrm{sphere}}(\mu)\le d,\\[2pt]
 \Gamma\!\left(\mu;\frac n2,\frac 2n\right), & r_{\mathrm{sphere}}(\mu)>d,
 \end{cases}
\end{equation}
where }$\Gamma(\cdot; \alpha, \beta)$ denotes the Gamma density with shape $\alpha$ and scale $\beta$.
\end{proposition}

\oldtext{This factorization is not restricted to the full heat-sphere boundary; it extends to the truncated heat-star boundary as well, yielding unit importance weight on \emph{both} the sphere and Neumann branches of WoHSt.} \changedtext{This factorization reveals the double-layer kernel on the truncated heat-star boundary can be perfectly importance sampled.} Consequently, the double-layer component of the estimator is unbiased. \oldtext{A complete derivation is provided in}\changedtext{A complete derivation of the decomposition and a concrete sampler for $\Gamma_\star$ are given in} Appendix~\ref{app:sampling_directional}.

\begin{figure}
    \centering
    \oldtext{}\changedtext{\includegraphics[width=\linewidth]{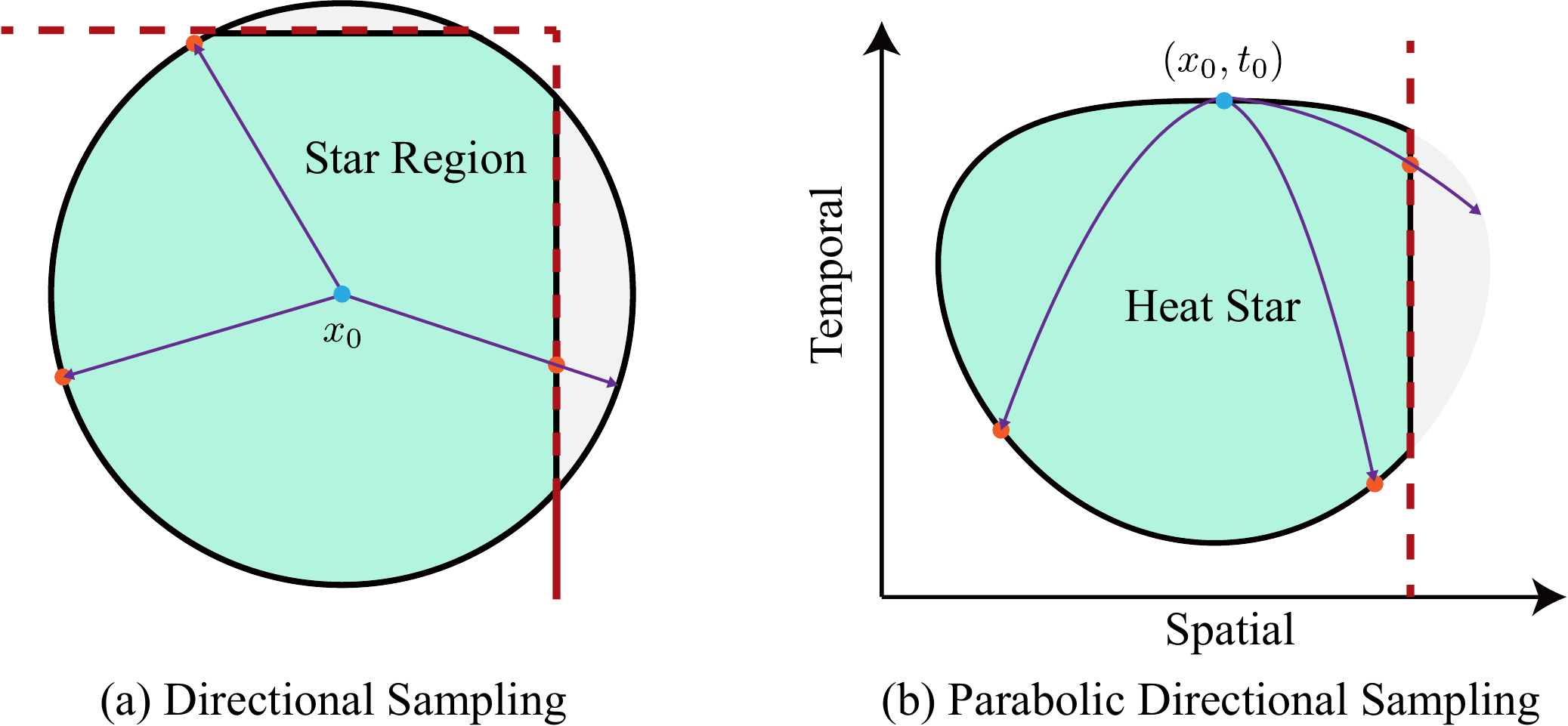}}
    \caption{Directional sampling: Euclidean versus heat-ball geometry. \textbf{Left:} In the elliptic Walk on Stars, a uniform direction $\omega \sim \mathrm{Uniform}(S^{n-1})$ is drawn from the current point $x_k$, and a ray is cast to the sphere boundary at distance $R$. The Poisson kernel is importance-sampled exactly, giving unit weight regardless of whether the ray hits the spherical or Neumann component. \textbf{Right:} In Walk on Heat Stars, the sampling is lifted to the joint space-time domain. \oldtext{A logarithmic time $u \sim \Gamma(\frac{n}{2}+1, \frac{2}{n})$ and a direction $\omega \sim \mathrm{Uniform}(S^{n-1})$ are drawn independently; the radial trace $(s, r) = (\rho e^{-u}, \sqrt{2n \rho u e^{-u}})$ is then followed to the heat-sphere surface. The double-layer kernel factorizes into independent temporal and directional components, and the same unit-weight property holds on the truncated heat-star boundary.}\changedtext{A uniform direction $\omega$ fixes the wall distance $d$. A draw $\mu\sim\Gamma_\star(c,d)$ determines where the radial trace meets the heat sphere or the Neumann wall.}}
    \label{fig:sampling_directional}
\end{figure}

Geometrically, the factorization realizes a parabolic counterpart to elliptic directional sampling. As illustrated in Figure~\ref{fig:sampling_directional}, a \emph{radial trace} (a parabolic curve in space-time) is launched from the current point $(x_0, t_0)$. Its spatial direction is determined by a uniform sample $\omega \sim \mathrm{Uniform}(S^{n-1})$\oldtext{, while its temporal extent is governed by a logarithmic time $u \sim \Gamma(\frac{n}{2}+1, \frac{2}{n})$}. The trace is then intersected with the heat-star boundary to obtain the next state $(x', t')$. \oldtext{ In the elliptic limit, the radial trace collapses to a straight ray of fixed radius $R$; in the parabolic case, the trace follows a parabolic curve $|x| = \sqrt{2n c u e^{-u}}$ that reflects the heat-kernel scaling $|x| \sim \sqrt{s}$.}

Algorithmically, this geometric picture requires no new primitives beyond standard ray casting. Given a direction $\omega$, \changedtext{we first flip it if the current point lies on the Neumann wall and $\omega\cdot n_N>0$. Then }we compute the visibility distance $d_{\mathrm{wall}}$ to the Neumann boundary. Given the \oldtext{sampled time $u$}\changedtext{draw $\mu\sim\Gamma_\star(c,d_{\mathrm{wall}})$}, we compute the heat-sphere distance $r_{\mathrm{sphere}} = \oldtext{\sqrt{2n c u e^{-u}}}\changedtext{\sqrt{2n c \mu e^{-\mu}}}$.

\begin{itemize}
    \item \textbf{Neumann branch} ($r_{\mathrm{sphere}} > d_{\mathrm{wall}}$). The radial trace strikes the Neumann wall before reaching the heat-sphere surface. The spatial step is clamped to the wall: $r = d_{\mathrm{wall}}$. The time step $s$ is determined by inverting the radial trace relation $\oldtext{u}\changedtext{\mu} = r^2/(2ns)$, giving $s = d_{\mathrm{wall}}^2 / (2n \oldtext{u}\changedtext{\mu})$. The next state is $x' = x + d_{\mathrm{wall}}\,\omega$, $t' = t - d_{\mathrm{wall}}^2/(2n \oldtext{u}\changedtext{\mu})$. The walker lands on $\partial\Omega_N$, where the Neumann correction is accumulated and the walk continues from the wall.
    \item \textbf{Sphere branch} ($r_{\mathrm{sphere}} \le d_{\mathrm{wall}}$). The radial trace reaches the heat-sphere surface unobstructed. The spatial and temporal steps follow directly from the $(\oldtext{u}\changedtext{\mu}, \omega)$ sample: $r = r_{\mathrm{sphere}}$, $s = c e^{-\oldtext{u}\changedtext{\mu}}$. The next state is $x' = x + r_{\mathrm{sphere}}\,\omega$, $t' = t - c e^{-\oldtext{u}\changedtext{\mu}}$, and the recursive walk continues from the heat-sphere interior.
\end{itemize}

In either case, the construction guarantees unit importance weight with no additional correction. In the algorithm below, \textsc{hitNeumann} is true exactly when the ray along $\omega$ intersects $\partial\Omega_N$ (i.e., \textsc{visibilityDistance} returns a finite value), and \textsc{hitNormal} returns the outward normal at the intersection point.  Algorithm~\ref{alg:sampling_directional} summarizes the procedure.

\begin{algorithm}
\caption{SampleDirectional$(x, t, c, \textit{on\_neumann}, n_N)$}
\label{alg:sampling_directional}
\begin{algorithmic}[1]
\Require Current state $(x, t)$, heat-ball radius $c$, boundary flags $(\textit{on\_neumann}, n_N)$
\Ensure Next state $(x', t')$ and updated flags, all with importance weight $w = 1$
\State \oldtext{$u \sim \Gamma\!\bigl(\frac{n}{2}+1,\; \frac{2}{n}\bigr)$ \Comment{Logarithmic diffusion time}\par\noindent}\changedtext{$\omega \sim \Call{Uniform}{S^{n-1}}$}
\State \changedtext{\textbf{if} $\textit{on\_neumann}$ \textbf{and} $\omega\cdot n_N>0$ \textbf{then} $\omega\gets-\omega$}
\State \oldtext{$\omega \sim \Call{Uniform}{S^{n-1}}$\par\noindent}\changedtext{$d_{\mathrm{wall}} \gets \Call{visibilityDistance}{x, \omega, \partial\Omega_N}$}
\State \oldtext{$r_{\mathrm{sphere}} \gets \sqrt{2n\,c\,u\,e^{-u}}$\par\noindent}\changedtext{$\mu\sim\Gamma_\star(c,d_{\mathrm{wall}})$}
\State \oldtext{$d_{\mathrm{wall}} \gets \Call{visibilityDistance}{x, \omega, \partial\Omega_N}$\par\noindent}\changedtext{$r_{\mathrm{sphere}}\gets\sqrt{2nc\mu e^{-\mu}}$}
\If{$\Call{hitNeumann}{}$ \textbf{and} $r_{\mathrm{sphere}} > d_{\mathrm{wall}}$}
    \State $s \gets d_{\mathrm{wall}}^2 \,/\, (2n \oldtext{u}\changedtext{\mu})$ \Comment{Neumann branch}
    \State $r \gets d_{\mathrm{wall}}$
    \State $\textit{on\_neumann} \gets \text{true}$, \quad $n_N \gets \Call{hitNormal}{}$
\Else
    \State $s \gets c\,e^{-\oldtext{u}\changedtext{\mu}}$ \Comment{Heat-sphere branch}
    \State $r \gets r_{\mathrm{sphere}}$
    \State $\textit{on\_neumann} \gets \text{false}$
\EndIf
\State \Return $(x + r\,\omega,\; t - s,\; \textit{on\_neumann},\; n_N,\; r,\; \omega\changedtext{,\; d_{\mathrm{wall}}})$
\end{algorithmic}
\end{algorithm}

\subsubsection{Neumann Contribution Sampling}
\label{sec:flux_sampling}

The single-layer term accounts for the prescribed Neumann flux $h = \partial u / \partial \vec{n}$ on $\partial\Omega_N$. Geometrically, the heat star $\mathcal{H}^\star$ intersects $\partial\Omega_N$ in a curved space-time patch. At each backward time $s \in (0, c)$, the intersection is the portion of $\partial\Omega_N$ lying within the spatial ball $B(x, r(s))$. The integrand weights each point by the Green's function $G(x-y, s)$ and the Neumann datum $h(y, t-s)$.

We sample this contribution in two stages. First, we draw a time $s$ from \oldtext{the same Gamma--Uniform distribution used for the double-layer kernel (Section~\ref{sec:directional_sampling})}\changedtext{the heat-sphere time distribution (Section~\ref{sec:heat_ball})}, which yields the radius $r(s) = \sqrt{2n s \log(c/s)}$ defining the spatial support. Second, we sample a point $y$ uniformly on $\partial\Omega_N \cap B(x, r(s))$, i.e., the visible Neumann patch within the heat ball at that instant. The geometric primitive for sampling the ball--Neumann intersection is taken directly from the elliptic WoSt method of~\citet{sawhney2023walk}, requiring no modification for the parabolic setting.

\begin{algorithm}[t]
\caption{SampleFlux$(x, t, c)$}
\label{alg:sampling_flux}
\begin{algorithmic}[1]
\Require Current state $(x, t)$, heat-ball radius $c$
\Ensure Neumann contribution $\widehat{I}_h$
\State $\oldtext{u}\changedtext{\mu} \sim \Gamma\!\bigl(\frac{n}{2}+1,\; \frac{2}{n}\bigr)$, \quad $s \gets c\,e^{-\oldtext{u}\changedtext{\mu}}$
\State $r \gets \sqrt{2n\,s \oldtext{u}\changedtext{\mu}}$
\State $y, p_y \sim \Call{SampleNeumannBoundary}{x, r}$
\If{$\vert y - x \vert > r$}
    \State \Return $0$ \Comment{Outside heat star}
\EndIf
\State \Return $G(x-y, s) \cdot h(y, t-s) \;/\; \bigl(p_S(s)\,p_y\bigr)$
\end{algorithmic}
\end{algorithm}

\subsubsection{Source Term Sampling}
\label{sec:source_sampling}

The volume source integral $I_f = \int_{\mathcal{H}^\star} G(y,s) \, f(y, t-s) \;\mathrm{d}y\,\mathrm{d}s$ runs over the heat-star interior. Direct sampling in $(y, s)$-space is complicated by the weight $G(y,s) = \Phi(x-y, s) - \tau(c)$, which couples the spatial and temporal coordinates. We resolve this by applying the same radial trace transformation used for the boundary (Section~\ref{sec:directional_sampling}), now extended to the heat-ball interior:

\begin{proposition}
\label{prop:source_sampling}
The source integral admits a factorized importance-sampling scheme under the radial trace parameterization:
\begin{equation}
    G(y, s) \; \mathrm{d} y \, \mathrm{d}s
    = Z \, p_\rho(\rho) \, \oldtext{p_u}\changedtext{p_\mu}(\oldtext{u}\changedtext{\mu}) \, p_\omega(\omega) \; \mathrm{d} \rho \, \mathrm{d} \oldtext{u}\changedtext{\mu} \, \mathrm{d} \omega,
\end{equation}
where $p_\omega$ is uniform on $S^{n-1}$, $\oldtext{p_u}\changedtext{p_\mu}$ \oldtext{is the Gamma distribution}\changedtext{is the Gamma density} $\Gamma\bigl(\frac{n}{2},\, \frac{1}{n/2+1}\bigr)$, $p_\rho$ is the power-law density
\begin{equation}
    p_\rho(\rho) = \frac{n + 2}{c n}\left[1 - \left(\frac{\rho}{c}\right)^{\!\frac{n}{2}}\right], \qquad 0 < \rho < c,
\end{equation}
and $Z = c\bigl(\frac{n}{n+2}\bigr)^{n/2+1}$ is the analytically computed constant.
\end{proposition}

Here $\rho$ is a heat-spherical radial coordinate (distinct from the heat-sphere radius $r(s)$), taking values in $(0, c)$. Under this transformation, the Green's function simplifies to $G = \tau(\rho) - \tau(c)$, depending \emph{only on $\rho$}, and the volume element factorizes into independent $\rho$, $\oldtext{u}\changedtext{\mu}$, and $\omega$ components. The resulting sampling procedure is simple. Draw $(\rho, \oldtext{u}\changedtext{\mu}, \omega)$ from the factorized distribution, map to $(y, s)$ via $s = \rho e^{-\oldtext{u}\changedtext{\mu}}$ and $|y| = \sqrt{2n \rho \oldtext{u}\changedtext{\mu} e^{-\oldtext{u}\changedtext{\mu}}}$, and evaluate $f(y, t-s)$. The direction $\omega$ is reused from the directional sampling step, and the containment check $|y - x| \le \oldtext{r}\changedtext{d_{\mathrm{wall}}}$ uses the \oldtext{heat-star extent}\changedtext{wall distance} in direction $\omega$ returned by \textsc{SampleDirectional}. If $y$ falls outside, the contribution is zero. The constant $Z$ absorbs the integral of $G$ over the heat ball, so no explicit Green's function evaluation is needed. Algorithm~\ref{alg:sampling_source} summarizes the subroutine.

\begin{algorithm}[t]
\caption{SampleSource$(x, t, c, \omega, \oldtext{r}\changedtext{d_{\mathrm{wall}}})$}
\label{alg:sampling_source}
\begin{algorithmic}[1]
\Require Current state $(x, t)$, heat-ball radius $c$, direction $\omega$ (reused from \textsc{SampleDirectional}), \oldtext{heat-star extent $r$}\changedtext{wall distance $d_{\mathrm{wall}}$} in direction $\omega$
\Ensure Source estimate $\widehat{I}_f$
\State $\oldtext{u}\changedtext{\mu} \sim \Gamma\!\bigl(\frac{n}{2},\; \frac{1}{n/2+1}\bigr)$
\State $\rho \sim 1 - (\rho/c)^{\,n/2}$ on $(0, c)$ \Comment{Rejection: draw $\rho \sim U(0,c)$, accept with probability $1 - (\rho/c)^{n/2}$}
\State $s \gets \rho\,e^{-\oldtext{u}\changedtext{\mu}}$, \quad $y \gets x + \sqrt{2n\,\rho\,\oldtext{u}\changedtext{\mu}\,e^{-\oldtext{u}\changedtext{\mu}}}\;\omega$
\If{$\vert y - x \vert > \oldtext{r}\changedtext{d_{\mathrm{wall}}}$}
\State \Return $0$ \Comment{Outside heat star}
\EndIf
\State \Return $Z \cdot f(y, t-s)$ \quad where $Z = c\bigl(\frac{n}{n+2}\bigr)^{n/2+1}$
\end{algorithmic}
\end{algorithm}

\subsection{Gradient Estimation}
\label{sec:gradient}

Many applications, such as heat flux computation, shape optimization, and thermal stress analysis, require not only the solution $u(x,t)$ but also its spatial gradient $\nabla_x u(x,t)$. We derive a gradient estimator that does not require recursion on $\nabla u$. The recursive variable is $u$ itself, estimated by the standard WoHSt walk, while the gradient is expressed as a weighted \oldtext{boundary}\,\changedtext{surface and volume} integral of $u$. Throughout this section the source term is assumed to vanish, $f \equiv 0$.

\subsubsection{Gradient Formula}

The derivation follows the same volumetric mean-value approach used for the Laplace equation\oldtext{ (see Appendix}\,\changedtext{~\cite{sawhney2020monte} (see Appendix}~\ref{app:gradient_derivation} for the full derivation and the Laplace analogy). Starting from the volumetric heat mean value property~\eqref{equ:volume_mvp}, differentiating with respect to $x_0$, and applying the divergence theorem on each spatial ball $B_r = B(x_0, r(s))$ yields the following representation.

\begin{proposition}
\label{prop:gradient_formula}
Let $u$ be a source-free caloric function, $\partial_t u = \Delta u$ in $\Omega \times \mathbb{R}^+$. For any heat ball $\mathcal{H}(x_0, t_0, c)$ contained in the domain, the spatial gradient admits the representation
\begin{equation}
    \label{equ:grad_final}
    \begin{aligned}
    \nabla_{x_0} u(x_0, t_0) = \tau(c) \int_0^c
    \Bigg[
    &\frac{r^2}{4 s^2} \oint_{\partial B_r} \!\! u \; \omega \; \mathrm{d}S \\
    &\;-\; \frac{1}{2 s^2} \int_{B_r} \! u \; (y-x_0) \; \mathrm{d}y
    \Bigg]
    \mathrm{d}s,
    \end{aligned}
\end{equation}
where $r(s) = \sqrt{2n s \log(c/s)}$, $B_r = B(x_0, r(s))$, and $\omega = (y-x_0)/r$. The integrand contains only $u$, not $\nabla u$.
\end{proposition}

The formula separates naturally into a surface term $S_i$ (the $\oint$ integral) and a volume term $V_i$ (the $\int_{B_r}$ integral), with $\nabla_i u = S_i + V_i$. The volume term is absent in the Laplace case because the heat mean value property weight $w(r) = r^2/(4s^2)$ is non-constant, so its gradient $\nabla_y w = (y-x_0)/(2s^2)$ contributes an additional integral via the divergence theorem. In the Laplace case $w \equiv 1$ and this term vanishes, leaving only the surface integral (see Appendix~\ref{app:gradient_derivation}).

For implementation, the two terms expand to
\begin{align}
    S_i &= \tau(c) \int_0^c \frac{r(s)^{n+1}}{4 s^2} \int_{S^{n-1}} u(x_0+r(s)\,\omega, t_0-s) \; \omega_i \; \mathrm{d}\omega \, \mathrm{d}s, \label{equ:surf_term} \\
    V_i &= -\tau(c) \int_0^c \frac{1}{2 s^2} \int_0^{r(s)} r^n \int_{S^{n-1}} u(x_0+r\,\omega, t_0-s) \; \omega_i \; \mathrm{d}\omega \, \mathrm{d}r \, \mathrm{d}s. \label{equ:vol_term}
\end{align}

\estimatorbox{WoHSt Gradient Estimator}{%
At a point $(x_k, t_k)$, one step draws $\oldtext{u_g}\changedtext{\mu_g} \sim \Gamma\bigl(\frac{n+3}{2}, \frac{2}{n-1}\bigr)$, sets $s = c e^{-\oldtext{u_g}\changedtext{\mu_g}}$, $R = \sqrt{2n s\log(c/s)}$, and draws $\omega_S, \omega_V \sim \text{Uniform}(S^{n-1})$ independently.  Then:
\begin{align}
    \label{eq:wohs_grad_vol}
    \widehat{\nabla u}(x_k, t_k) \; \coloneqq \;
    &\frac{Z_S}{N} \sum_{j=1}^N \omega_S^{(j)} \; \widehat{u}\bigl(x_k + R\,\omega_S^{(j)},\; t_k - s^{(j)}\bigr) \\
    &\;-\; \frac{Z_V}{N} \sum_{j=1}^N \omega_V^{(j)} \; \widehat{u}\bigl(x_k + r^{(j)}\omega_V^{(j)},\; t_k - s^{(j)}\bigr), \nonumber
\end{align}
where $r^{(j)} = R \cdot (t^{(j)})^{1/(n+1)}$ with $t^{(j)} \sim U(0,1)$, and $\widehat{u}(\cdot)$ is the \emph{value} estimate from the standard WoHSt recursive walk (Algorithm~\ref{alg:wohs}). The two $\widehat{u}(\cdot)$ evaluations spawn independent recursive walks that share the same $s^{(j)}$ but use different spatial locations. The current formulation is derived for the source-free case. Incorporating the source term $f$\oldtext{ and Neumann boundary contributions}\changedtext{} via the same volumetric mean value property machinery is left to future work.
}

\subsubsection{Sampling Strategy}

Both integrands share the same $s$-marginal density $p_S(s) \propto r(s)^{n+1}/s^2$.  Using the substitution $\oldtext{u_g}\changedtext{\mu_g} = \log(c/s)$, this becomes a \textbf{Gamma distribution}:
\begin{equation}
    \;
    \begin{aligned}
    \oldtext{u_g}\changedtext{\mu_g} &\sim \Gamma\!\left(\frac{n+3}{2},\; \frac{2}{n-1}\right), \\[2pt]
    s &= c\,e^{-\oldtext{u_g}\changedtext{\mu_g}}, \\[2pt]
    R &= \sqrt{2n s \log(c/s)}
    \end{aligned}
    \end{equation}
which differs from the value walk's $\Gamma(\frac{n}{2}+1, \frac{2}{n})$. For the volume term, the radial coordinate inside the ball is sampled from the power law $p(r \mid s) \propto r^n$ on $[0, R]$, i.e.\ $r = R \cdot t^{1/(n+1)}$ with $t \sim U(0,1)$. The direction $\omega$ is sampled uniformly on $S^{n-1}$ for both terms (independently).

The normalization constants are computed analytically by evaluating the Gamma integrals:
\begin{align}
    Z_S &= \tau(c)\,\frac{|S^{n-1}|}{4} \int_0^c \frac{r(s)^{n+1}}{s^2}\,\mathrm{d}s \nonumber\\
        &= \frac{\tau(c)\,|S^{n-1}|}{4}\,(2n)^{\frac{n+1}{2}} c^{\frac{n-1}{2}}\,
           \frac{\Gamma\!(\frac{n+3}{2})}{(\frac{n-1}{2})^{\frac{n+3}{2}}}, \\[4pt]
    Z_V &= \tau(c)\,\frac{|S^{n-1}|}{2(n+1)} \int_0^c \frac{r(s)^{n+1}}{s^2}\,\mathrm{d}s
        = \frac{2}{n+1}\,Z_S.
\end{align}
\oldtext{Both are dimension-dependent constants, precomputed once and reused for all queries. } 

\oldtext{
For pure Dirichlet problems without source, the gradient estimate telescopes as a weighted sum of $u$-values along the walk; the recursive variable is $u$ only, with no recursion on $\nabla u$.
}

\changedtext{

\subsubsection{Antithetic Sampling and Neumann Boundary}

We employ antithetic sampling~\cite{rioux2022monte} to reduce variance in the gradient estimator. For each sampled direction $\omega$, we also evaluate the integrand at $-\omega$ and average the two contributions. This exploits the symmetry of the heat kernel and reduces the raw gradient error at the tested budgets.

For problems with Neumann boundaries, the first step instead takes the largest heat ball that is entirely contained in $\Omega$, so that this step is a pure Walk on Heat Spheres step, and the walk returns to WoHSt from the second step onward. Ignoring $\varepsilon$-shell termination error, the estimator remains unbiased because the boundary integral representation holds for any subdomain of $\Omega$, and the change removes the Neumann correction from the largest step, where a wall hit would produce the largest single contribution.

}

\subsection{Variance Reduction}
\label{sec:variance}

\oldtext{To reduce the variance of WoHSt estimates, we adopt the heteroscedastic regression-based denoising technique of~\citet{bao2026monte}, extended to the space-time domain.}\changedtext{As optional post-processing of the raw WoHSt estimates, we adopt the heteroscedastic regression-based denoising technique of~\citet{bao2026monte}, extended to the space-time domain. The denoiser is trained from scratch on the raw estimates.} For a set of query points $\{(x_j, t_j)\}$ with raw WoHSt estimates $\widehat{u}_j$, we jointly regress the mean $u(x,t)$ and log-variance $\log\sigma^2(x,t)$ using a SIREN network~\cite{sitzmann2020implicit} trained with the $\beta$-NLL loss~\cite{seitzer2022pitfalls}:
\begin{equation}
    \mathcal{L}_{\beta}
    = \frac{1}{N} \sum_{i=1}^N
    \bigl\lfloor \sigma^{2\beta}_i \bigr\rfloor
    \left[
        \frac{(\widehat{u}_\theta(x_i, t_i) - \widehat{u}_i)^2}{2\sigma^2_\theta(x_i, t_i)}
        + \frac{1}{2}\log \sigma^2_\theta(x_i, t_i)
    \right],
\end{equation}
where $\lfloor\cdot\rfloor$ denotes a stop-gradient and $\beta = 0.5$. The only modification from~\citet{bao2026monte} is the inclusion of the time coordinate $t$ in the network input $(x, t)$, allowing the denoiser to exploit temporal correlations in the transient estimates.

\section{Experiments}
\label{sec:experiments}

We evaluate the accuracy, efficiency, and flexibility of WoHSt on \oldtext{a suite of transient heat conduction problems. All experiments use a fixed per-run time budget to ensure fair comparison, and we report mean squared error (MSE) against the ground truth.}\,\changedtext{transient heat conduction problems, reporting mean squared error (MSE) against analytical or numerical references. For an unbiased average of $N$ independent samples with finite variance, the expected MSE scales as $O(N^{-1})$. The corresponding root mean squared error (RMSE) scales as $O(N^{-1/2})$. Our convergence plots report MSE. For time-budget plots, the analogous inverse-time rate assumes approximately constant average cost per sample.

Unless explicitly labeled as denoised, the reported Monte Carlo estimates and errors are raw results, computed without learned post-processing. Compute time in the raw convergence plots measures Monte Carlo estimation, without denoiser training or inference.}

\subsection{Implementation Details}

WoHSt is implemented in C++ with geometric queries accelerated by FCPW~\cite{FCPW}. \oldtext{The heteroscedastic regression denoiser is implemented in PyTorch and trained with the same hyperparameters as~\citet{bao2026monte}. All experiments run on an Intel Core i7-10700K CPU with thread-level parallelization.}\,\changedtext{The Monte Carlo experiments run on an Intel Core i7-10700K CPU with thread-level parallelization. The optional heteroscedastic regression denoiser is implemented in PyTorch with the same hyperparameters as~\citet{bao2026monte}. For the 45 mixed-boundary spatial cases, we fit a separate network to samples on a $256\times256$ training grid. Fitting takes $27.79$\,s per field on average on an NVIDIA GeForce RTX~4060 Laptop GPU, in addition to the Monte Carlo budget. Denoising is applied only to results explicitly labeled as denoised. Monte Carlo-only timings exclude this additional cost. For the archived FEM comparisons in Section~\ref{sec:fem_comparison}, the captions instead report the recorded cost of each displayed raw or post-processed field.}

\subsection{Validation on Analytical Solutions}

\begin{figure}[ht]
    \centering
    \oldtext{\includegraphics[width=\linewidth]{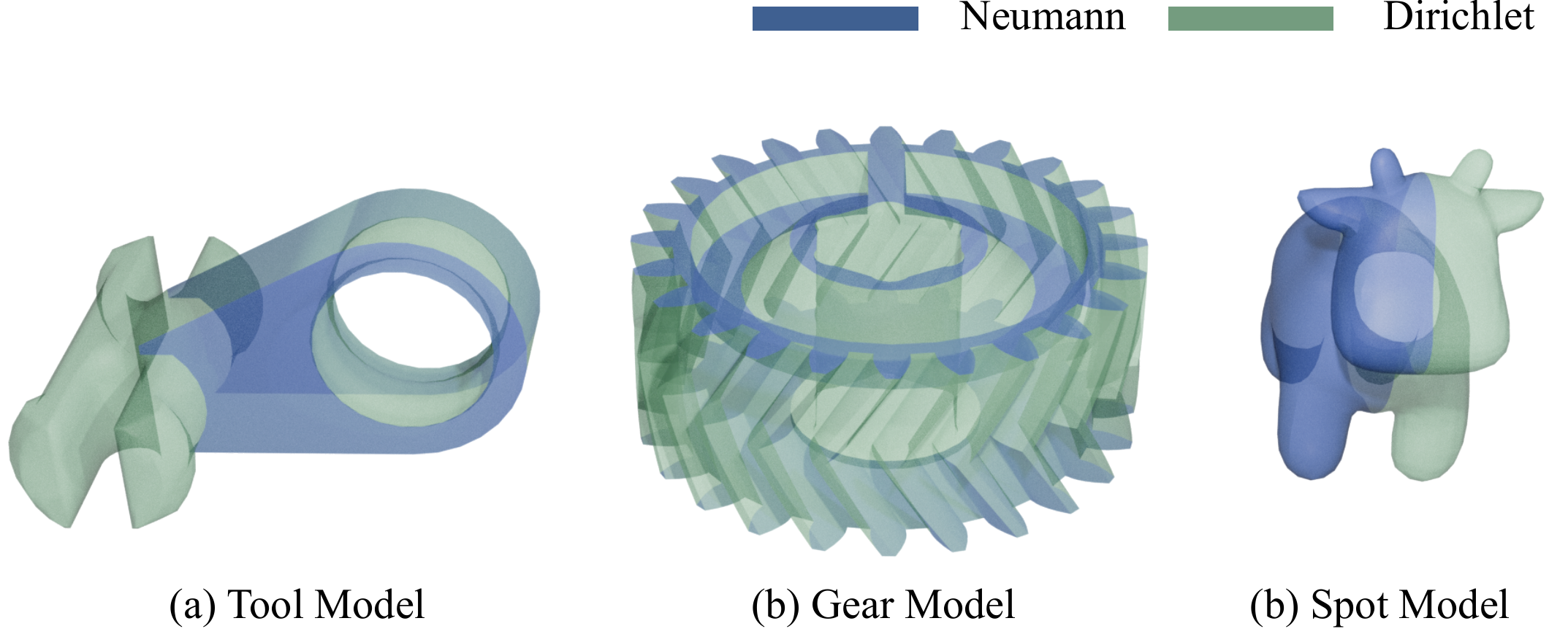}\par\smallskip}
    \changedtext{\includegraphics[width=\linewidth]{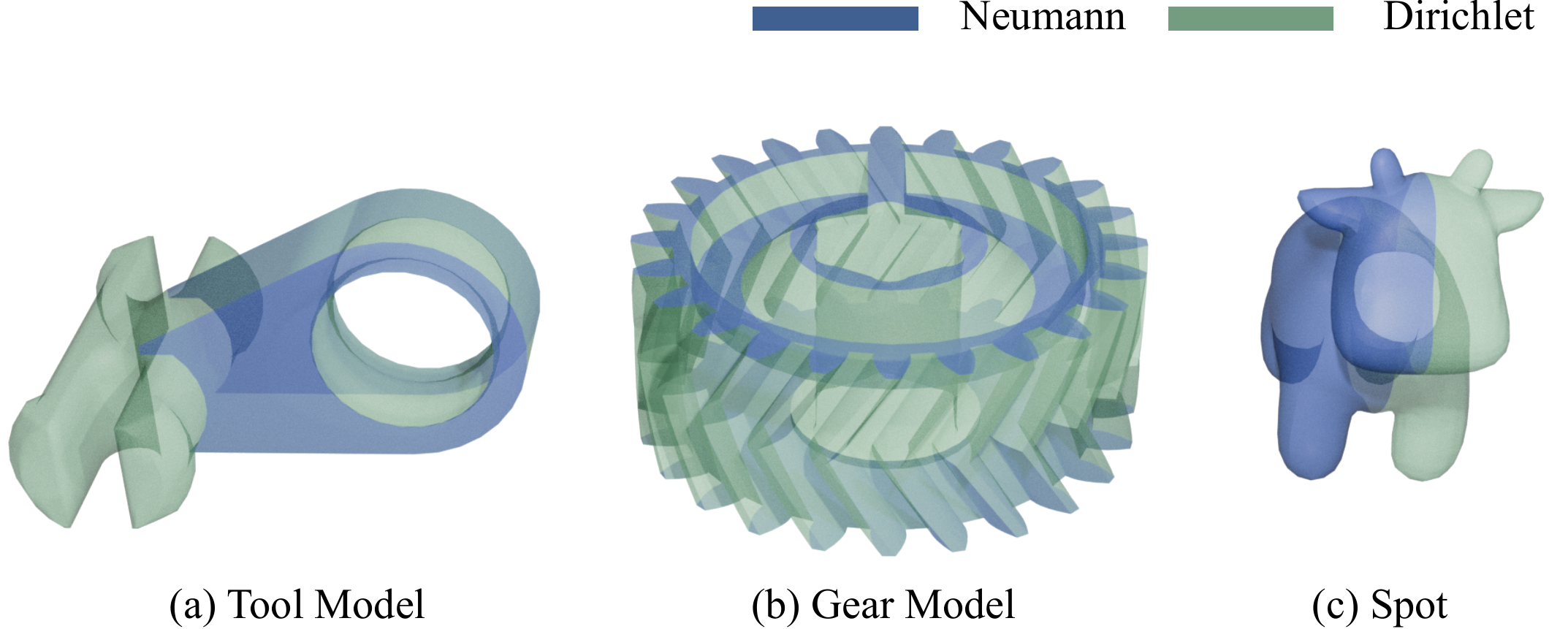}}
    \caption{Test geometries and boundary conditions. We evaluate WoHSt on three domains (a tool, a gear-shaped domain, \oldtext{and a spot)}\changedtext{and Spot)} with varying boundary complexity. Each domain is tested under three spatial frequencies $\omega \in \{\pi, 2\pi, 4\pi\}$, which control the solution's oscillatory complexity. The color coding indicates the Dirichlet (green) and Neumann (blue) partitions of the boundary.}
    \label{fig:models}
\end{figure}

We validate WoHSt against the family of separable solutions
\begin{equation}
    u(x, y, z, t) = \cos(2\pi t) \cos(\omega x) \cos(\omega y) \cos(\omega z),
\end{equation}
where the spatial frequency $\omega \in \{\pi, 2\pi, 4\pi\}$ controls the solution's oscillatory complexity. This ansatz satisfies the heat equation with a source term $f = \partial_t u - \Delta u$ and can be adapted to Dirichlet, Neumann, or mixed boundary conditions by appropriate choice of $\omega$ and domain size. We test on three geometries (a tool, a gear-shaped domain, \oldtext{and a spot)}\changedtext{and Spot)} spanning a range of boundary complexity. Models and Dirichlet-Neumann partitions are shown in Figure~\ref{fig:models}.

\subsubsection{Spatial Accuracy at Fixed Time Instants}

We first measure spatial accuracy by evaluating WoHSt on a dense slice of each domain at five fixed time instants $t \in \{0.5, 1.0, 1.5, 2.0, 2.5\}$. At each time stamp, we evaluate the Monte Carlo estimators over a $512 \times 512$ dense grid. For each configuration we allocate a 500-second compute budget and record the final MSE averaged over all query points on the slice.

\begin{table*}[ht]
\centering
\caption{\changedtext{Raw spatial MSE after a 500\,s Monte Carlo budget. The WoHS columns use pure-Dirichlet boundaries, and the WoHSt columns use mixed Dirichlet--Neumann boundaries as shown in Figure~\ref{fig:models}.}}
\label{tab:mse_all}
\begin{tabular}{llcccccccccc}
\toprule
Model & $\omega$ & \multicolumn{5}{c}{WoHS (Dirichlet)} & \multicolumn{5}{c}{\changedtext{WoHSt (mixed)}} \\
 &  & t=0.5 & t=1.0 & t=1.5 & t=2.0 & t=2.5 & t=0.5 & t=1.0 & t=1.5 & t=2.0 & t=2.5 \\
\midrule
tool & $\pi$ & 1.52e-05 & 1.52e-05 & 1.50e-05 & 1.52e-05 & 1.54e-05 & \changedtext{1.55e-04} & \changedtext{1.55e-04} & \changedtext{1.56e-04} & \changedtext{1.55e-04} & \changedtext{1.55e-04} \\
 & $2\pi$ & 6.24e-05 & 6.26e-05 & 6.26e-05 & 6.20e-05 & 6.23e-05 & \changedtext{6.37e-04} & \changedtext{6.37e-04} & \changedtext{6.38e-04} & \changedtext{6.40e-04} & \changedtext{6.37e-04} \\
 & $4\pi$ & 2.98e-04 & 2.95e-04 & 2.97e-04 & 2.95e-04 & 3.01e-04 & \changedtext{1.83e-03} & \changedtext{1.85e-03} & \changedtext{1.85e-03} & \changedtext{1.87e-03} & \changedtext{1.86e-03} \\
\midrule
gear & $\pi$ & 2.03e-04 & 2.07e-04 & 2.07e-04 & 2.05e-04 & 2.16e-04 & \changedtext{5.09e-04} & \changedtext{5.08e-04} & \changedtext{5.10e-04} & \changedtext{5.09e-04} & \changedtext{5.10e-04} \\
 & $2\pi$ & 7.77e-04 & 7.75e-04 & 7.75e-04 & 7.76e-04 & 7.68e-04 & \changedtext{2.17e-03} & \changedtext{2.17e-03} & \changedtext{2.18e-03} & \changedtext{2.16e-03} & \changedtext{2.18e-03} \\
 & $4\pi$ & 3.83e-03 & 3.79e-03 & 3.78e-03 & 3.86e-03 & 3.85e-03 & \changedtext{1.29e-02} & \changedtext{1.31e-02} & \changedtext{1.29e-02} & \changedtext{1.28e-02} & \changedtext{1.31e-02} \\
\midrule
\changedtext{Spot} & $\pi$ & 3.00e-05 & 3.07e-05 & 3.04e-05 & 3.07e-05 & 3.08e-05 & \changedtext{1.94e-02} & \changedtext{2.01e-02} & \changedtext{2.01e-02} & \changedtext{2.01e-02} & \changedtext{2.01e-02} \\
 & $2\pi$ & 1.41e-04 & 1.40e-04 & 1.40e-04 & 1.40e-04 & 1.39e-04 & \changedtext{5.70e-02} & \changedtext{5.73e-02} & \changedtext{5.75e-02} & \changedtext{5.75e-02} & \changedtext{5.75e-02} \\
 & $4\pi$ & 6.91e-04 & 6.94e-04 & 6.95e-04 & 6.92e-04 & 6.99e-04 & \changedtext{4.03e-01} & \changedtext{4.07e-01} & \changedtext{4.08e-01} & \changedtext{4.08e-01} & \changedtext{4.08e-01} \\
\midrule
\bottomrule
\end{tabular}
\end{table*}

\begin{figure*}
    \centering
    \changedtext{\includegraphics[width=\linewidth]{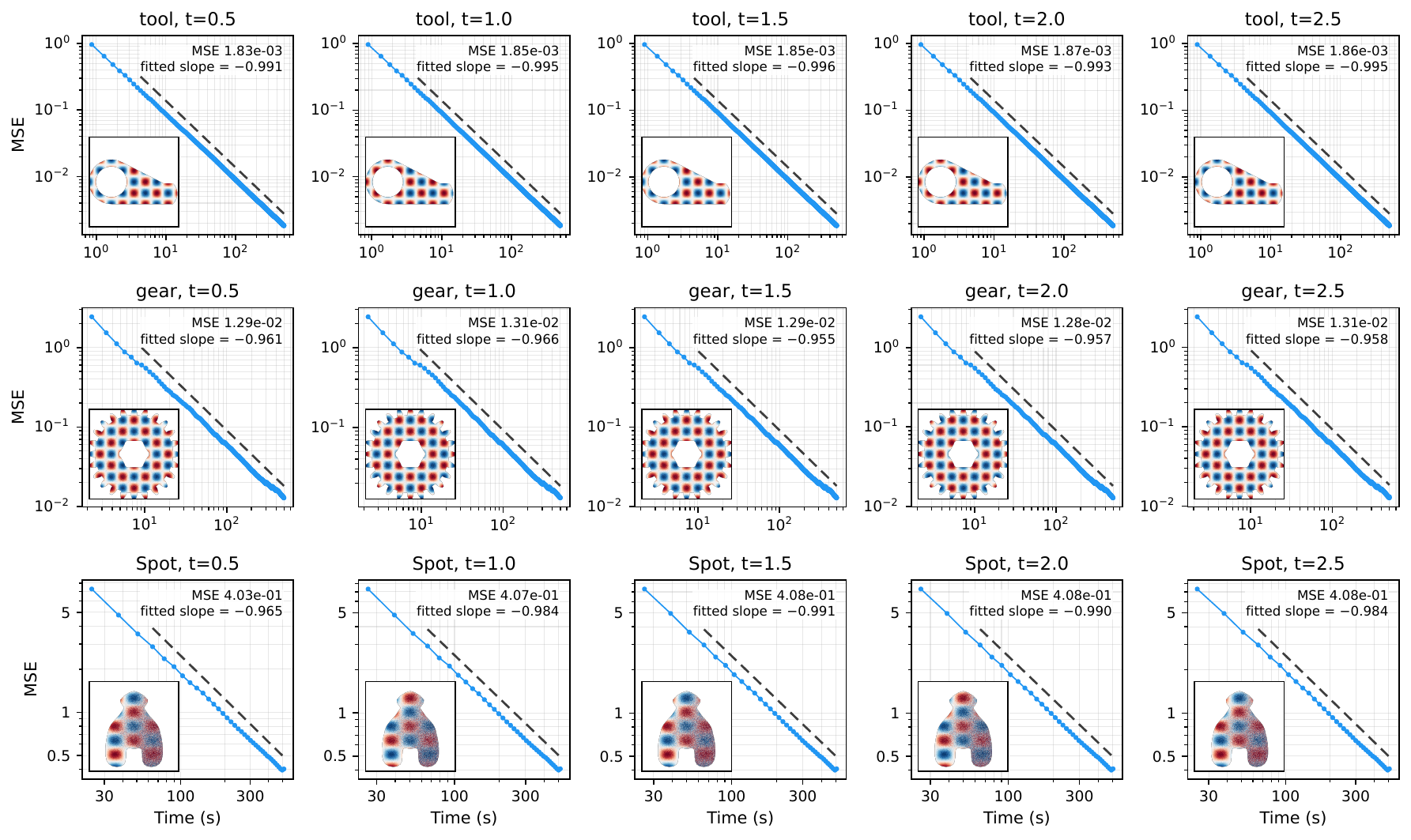}}
    \caption{\oldtext{Convergence of Walk on Heat Stars at the spatial frequency $\omega = 4\pi$. Each panel shows MSE as a function of compute time for one geometry--time-instant pair, with the final estimated solution field and ground truth inset in the lower-left corner. The decay confirms the expected Monte Carlo convergence rate.}\changedtext{Convergence of Walk on Heat Stars at the spatial frequency $\omega=4\pi$. Each panel shows the full MSE of the raw estimates against Monte Carlo computation time, with the final estimated field inset on a common $[-1,1]$ scale. The annotations give the final MSE and the fitted log--log slope. Dashed lines indicate a reference slope of $-1$.}}
    \label{fig:convergence_wohs}
\end{figure*}

\oldtext{Table~\ref{tab:mse_all} reports the spatial MSE for all configurations. Across all geometries and both algorithms, error increases with spatial frequency $\omega$, reflecting the greater difficulty of resolving fine oscillatory detail under a fixed compute budget.

For pure Dirichlet problems, WoHS applies when the entire heat ball fits inside the domain. Its MSE ranges from $10^{-5}$ to $10^{-3}$ on the tool and spot geometries, and $10^{-4}$ to $4\times10^{-3}$ on the gear.

For mixed Dirichlet--Neumann problems, WoHSt combines a recursive double-layer walk, a single-layer Neumann correction, and a volumetric source term. On the tool geometry its MSE ranges from $2\times10^{-4}$ to $3\times10^{-3}$; on the gear, from $10^{-3}$ to $2\times10^{-2}$. On the spot geometry, where half the boundary is Neumann, the extensive Neumann boundary conditions yield MSE up to $4\times10^{-1}$ at the highest frequency. WoHSt handles domains with Neumann boundaries via heat-star truncation, a capability that WoHS does not provide.}\changedtext{Table~\ref{tab:mse_all} reports raw spatial MSE for WoHS with Dirichlet boundaries and WoHSt with mixed boundaries. For both sets of boundary conditions, the error increases with spatial frequency on each geometry.}\oldtext{Figure~\ref{fig:convergence_wohs} shows the convergence curves of Walk on Heat Stars at $\omega = 4\pi$, the most challenging frequency. Each panel plots MSE against compute time, with the final estimated solution field and ground truth inset in the lower-left corner. Across all three geometries and five time instants, the solution fields closely match the ground truth and the MSE decays at the expected rate, confirming that the importance-sampling construction remains statistically well-behaved even on complex, non-convex domains.} \changedtext{Figure~\ref{fig:convergence_wohs} shows the convergence curves at $\omega=4\pi$. We fit the slope of log MSE against log computation time using all recorded checkpoints.}

\oldtext{}\changedtext{

\begin{figure*}
    \centering
    \includegraphics[width=\linewidth]{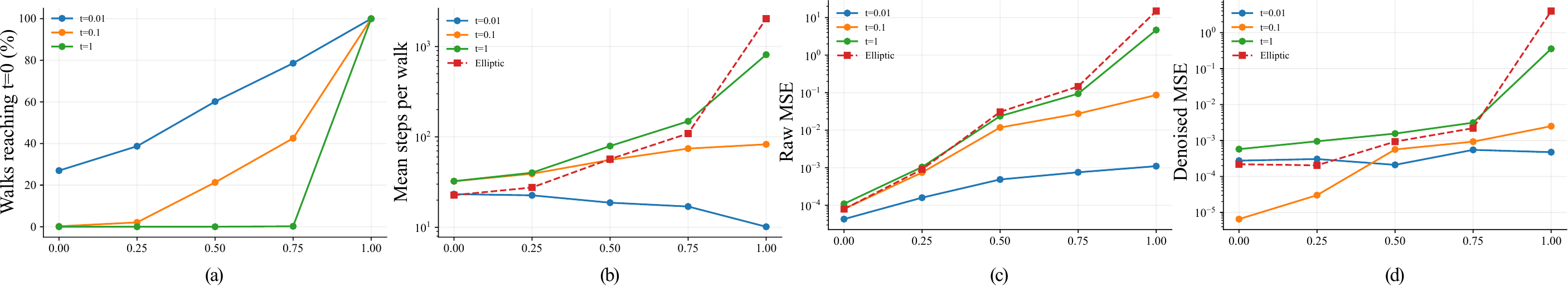}
    \caption{Effect of Neumann boundary fraction on WoHSt accuracy on Spot. All x-axes give the Neumann area fraction. (a) Percentage of transient walks reaching the initial time plane. (b) Mean walk length. (c) Raw MSE. (d) Denoised MSE. The elliptic comparison uses unregularized Poisson solves at fractions below one and a Tikhonov-regularized solve with $\lambda=1$ at the pure-Neumann endpoint.}
    \label{fig:neumann_fraction}
\end{figure*}

\begin{figure*}
    \centering
    \includegraphics[width=\linewidth]{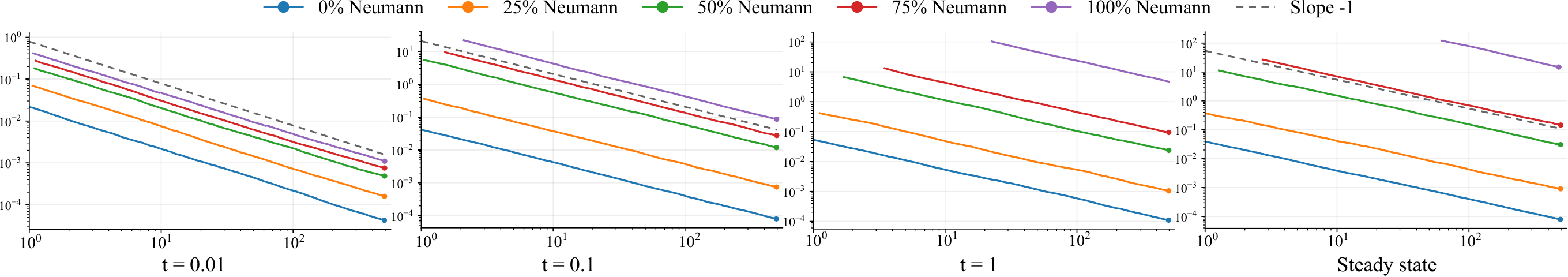}
    \caption{Raw MSE versus Monte Carlo computation time at different Neumann fractions, for three transient query times and the elliptic comparison. The pure-Neumann elliptic endpoint uses Tikhonov regularization with $\lambda=1$.}
    \label{fig:neumann_fraction_mse}
\end{figure*}

We evaluate Neumann area fractions in $\{0,0.25,0.5,0.75,1\}$ on Spot at $t\in\{0.01,0.1,1\}$. Figure~\ref{fig:neumann_fraction} shows that the fraction of walks reaching the initial time plane increases with Neumann fraction and reaches one for pure-Neumann problems. Mean walk length increases at $t=0.1$ and $t=1$, but decreases at $t=0.01$, where the nearby initial plane terminates walks quickly. Raw MSE increases with Neumann fraction. Denoising reduces error, but substantial error remains at the pure-Neumann endpoint, especially at $t=1$.

For comparison, we include elliptic WoSt solves~\cite{sawhney2023walk} on the same geometry. Figure~\ref{fig:neumann_fraction_mse} reports the raw MSE-time curves. This study varies boundary fraction and query time within one family of partitions.

\begin{figure*}
    \centering
    \includegraphics[width=\linewidth]{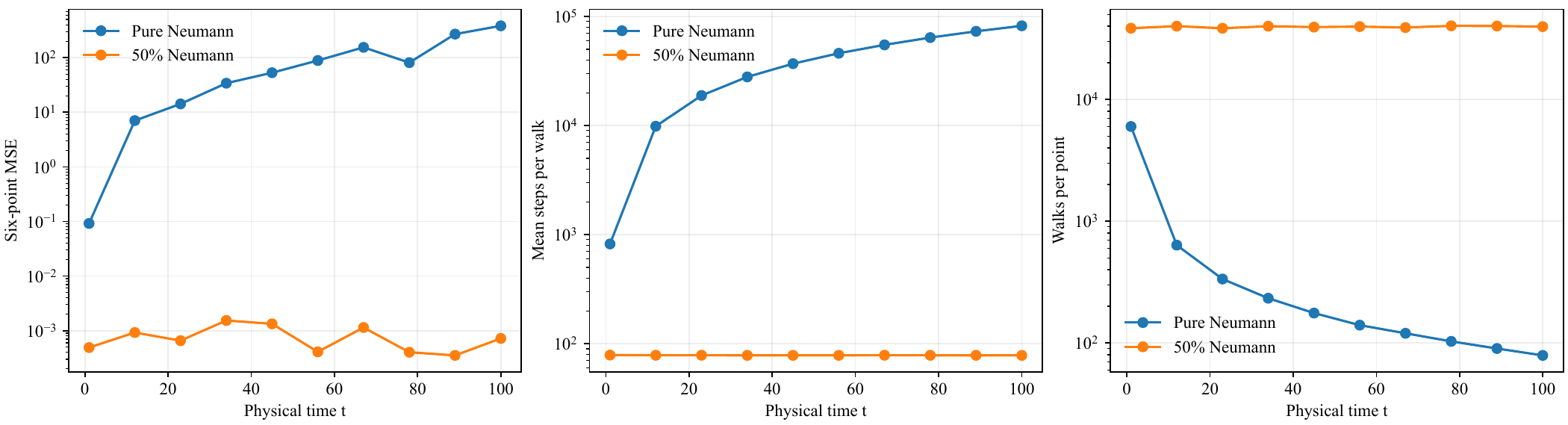}
    \caption{Long-time behavior at the same six interior sensors on Spot, comparing pure-Neumann and 50\%-Neumann boundaries with a 100-second budget per timestamp. (a) Raw six-point MSE. (b) Mean steps per walk. (c) Completed walks per point under the fixed budget.}
    \label{fig:large_time}
\end{figure*}

Figure~\ref{fig:large_time} evaluates query times from $t=1$ to $t=100$ at six fixed interior sensors on Spot, using 100 seconds per timestamp. For the tested 50\%-Neumann partition, MSE, mean walk length, and completed walks per point remain approximately stable because almost all walks terminate at the Dirichlet boundary. For pure-Neumann boundaries, all walks must reach the initial time plane. Mean walk length increases with query time, fewer walks are completed within the fixed budget, and MSE increases as query time increases.
}

\subsubsection{Evaluation on Temporal Domain}

\begin{table}[ht]
\centering
\caption{\changedtext{RelMSE of raw WoHSt estimates with Dirichlet conditions on the entire boundary ($\omega=2\pi$, 4096 walks per point and time, 100 equally spaced times on $[1,5]$).}}
\label{tab:temporal_relmse}
\changedtext{%
\begin{tabular}{lccccc}
\toprule
Model & Point 1 & Point 2 & Point 3 & Point 4 & Point 5 \\
\midrule
tool & 3.82e-03 & 7.54e-06 & 3.18e-04 & 3.94e-04 & 5.22e-03 \\
gear & 1.20e-01 & 4.22e-03 & 2.56e-02 & 2.78e-02 & 9.74e-05 \\
Spot & 1.60e-04 & 1.20e-03 & 3.21e-03 & 2.64e-03 & 9.76e-04 \\
\bottomrule
\end{tabular}}
\end{table}

\begin{figure*}
    \centering
    \changedtext{\includegraphics[width=\linewidth]{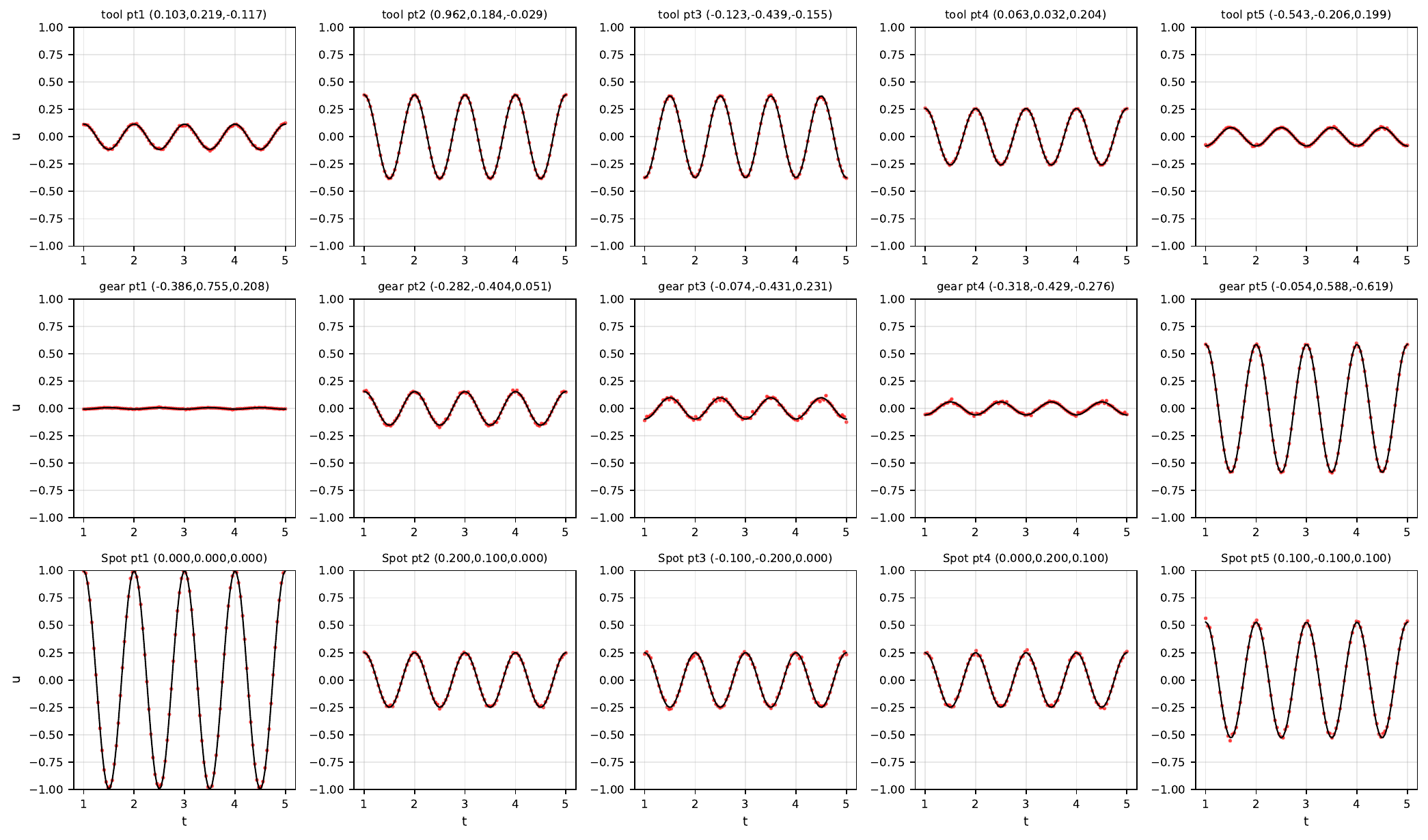}}
    \caption{\oldtext{Temporal convergence of Walk on Heat Stars at $\omega = 4\pi$. Each plot shows the solution at a single spatial point as a function of time: the black curve is the analytical ground truth, and the red points are WoHSt estimates evaluated at individual time instants. The relative MSE (RelMSE) is reported in Table~\ref{tab:temporal_relmse}.}\changedtext{Temporal accuracy of Walk on Heat Stars with all-Dirichlet boundaries at $\omega=2\pi$. Each plot shows the solution at one spatial point. The black curve is the analytical solution, and the red points are raw estimates from the Monte Carlo estimator. All panels use the range $[-1,1]$. Table~\ref{tab:temporal_relmse} uses all estimates, including those outside the plotted range.}}
    \label{fig:temporal_convergence}
\end{figure*}

We also evaluate WoHSt on the temporal domain by fixing five spatial points in each geometry and recording the solution as a function of time. Figure~\ref{fig:temporal_convergence} plots the temporal curves, and Table~\ref{tab:temporal_relmse} reports the relative MSE (RelMSE) for each point and geometry. \changedtext{For this experiment, we prescribe the analytical temperature on the entire boundary of each model. We compute RelMSE by dividing the mean squared error over the 100 query times by the temporal variance of the analytical solution at that point. The largest RelMSE is $0.120$ at point~1 on the gear, where the analytical temperature has the smallest temporal variance among the five gear sensors.}

\subsubsection{\changedtext{Solution-Field Denoising}}

\begin{figure*}
    \centering
    \changedtext{\includegraphics[width=\linewidth]{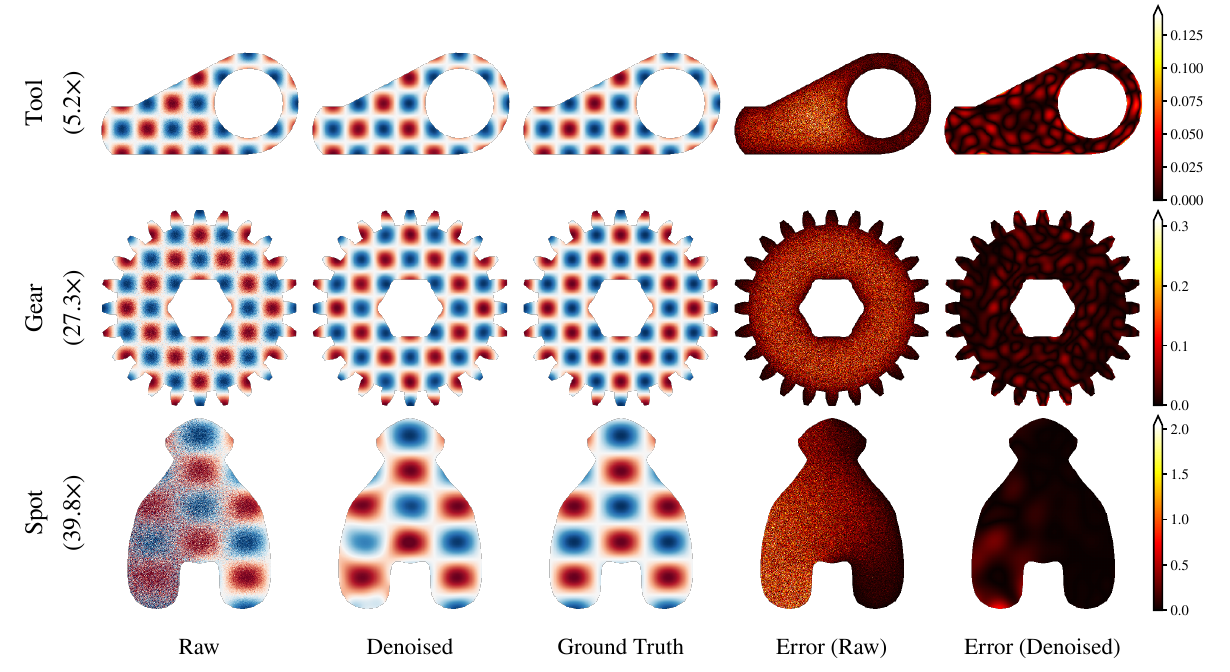}}
    \caption{\oldtext{Effect of heteroscedastic regression denoising on Walk on Heat Stars estimates. The raw Monte Carlo estimates (left) exhibit high variance, while the denoised estimates (right) closely match the ground truth.}\changedtext{Heteroscedastic regression denoising on the tool, gear, and Spot at $t=1.5$ and $\omega=4\pi$. The first three columns show the raw estimate, denoised estimate, and ground truth on a common $[-1,1]$ scale. The last two show absolute errors.}}
    \label{fig:solution_denoising}
\end{figure*}

\oldtext{For the solution estimation, we also present heteroscedastic regression denoising results. Figure~\ref{fig:solution_denoising} compares the raw WoHSt estimates with the denoised estimates for a representative slice of the gear geometry at $t=2.5$ and $\omega=4\pi$. The raw estimates show significant variance, while the denoised version closely matches the ground truth, demonstrating the effectiveness of the regression-based variance reduction technique.}\changedtext{We train for 2000 epochs on the $256\times256$ subsampling of each raw field and evaluate both estimates at all interior points of the original $512\times512$ slice. Figure~\ref{fig:solution_denoising} shows the middle of the five benchmark times, $t=1.5$, at $\omega=4\pi$. Denoising reduces the full MSE in all 45 mixed-boundary cases. Across the 15 cases per geometry, the median ratio of raw to denoised MSE is $30.2$ for the tool, $17.8$ for the gear, and $36.6$ for Spot. Table~\ref{tab:denoiser_resolution} compares training grids of $128\times128$, $256\times256$, and $512\times512$ on the same 45 cases. The $256\times256$ resolution gives the lowest average MSE at about a third of the time of the $512\times512$ grid.}

\begin{table}[ht]
\centering
\caption{Effect of the training-grid resolution of the denoiser over the 45 mixed-boundary spatial cases. All cases use the same raw Monte Carlo fields, computed with a fixed 500\,s budget. The time column is the mean per-case denoising time. ``Best'' counts the cases in which the denoised estimate at that resolution has the lowest MSE of the three grids.}
\label{tab:denoiser_resolution}
\begin{tabular}{lccc}
\toprule
Training grid & Best (of 45) & Mean MSE & Time (s) \\
\midrule
$128\times128$ & 7 & $3.614\times10^{-3}$ & 13.48 \\
$256\times256$ & \textbf{27} & $\mathbf{1.836\times10^{-3}}$ & 27.79 \\
$512\times512$ & 11 & $2.083\times10^{-3}$ & 96.63 \\
\bottomrule
\end{tabular}
\end{table}

\subsubsection{Gradient Estimation}

To validate the gradient estimator, we compute $\nabla u$ at a dense grid of points in the ball geometry, using a caloric function without a source term
\begin{equation}
    u(x,t) = \exp\left(\langle p, x \rangle + \vert p \vert^2 t\right),
\end{equation}
where $p$ is a fixed vector. This function satisfies the heat equation without source and has a known analytical gradient $\nabla u = u \, p$. \oldtext{We run the WoHSt gradient estimator with 1024 samples for each integral and compare the estimated gradients to the ground truth.}\changedtext{}

\oldtext{}

\changedtext{\begin{figure*}
    \centering
    \includegraphics[width=\linewidth]{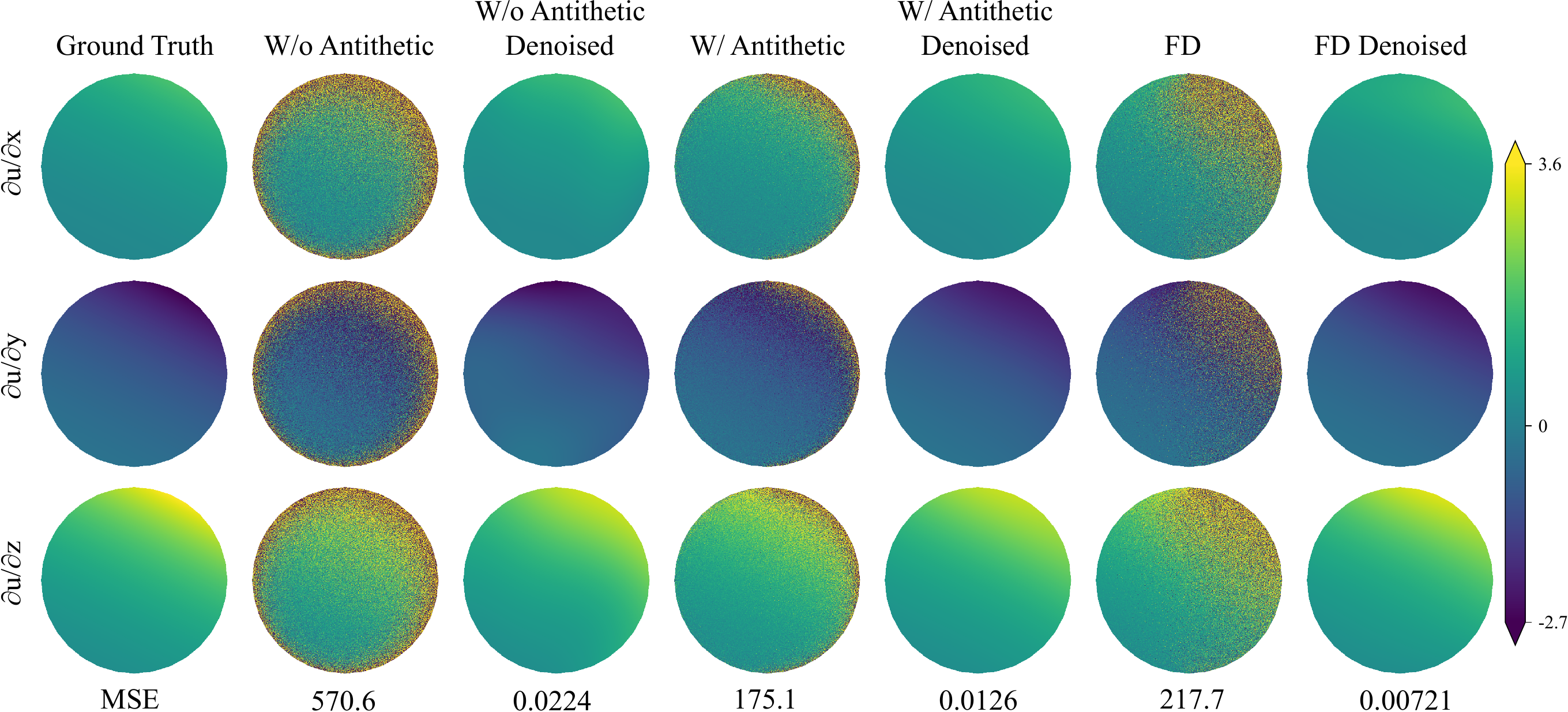}
    \caption{Gradient estimation on a ball geometry in which half of the boundary carries Neumann data. We compare the proposed estimator with and without antithetic sampling, and with and without heteroscedastic denoising, together with a finite-difference estimator that shares its random number sequence between the two neighbouring query points. The columns without the label ``Denoised'' show raw estimates. Each MSE value is computed from the estimate in its own column. Errors before and after learned post-processing are reported separately.}
    \label{fig:gradient_comparison_sphere}
    \label{fig:quiver}
\end{figure*}}

\oldtext{}

\changedtext{\begin{figure*}
    \centering
    \includegraphics[width=\linewidth]{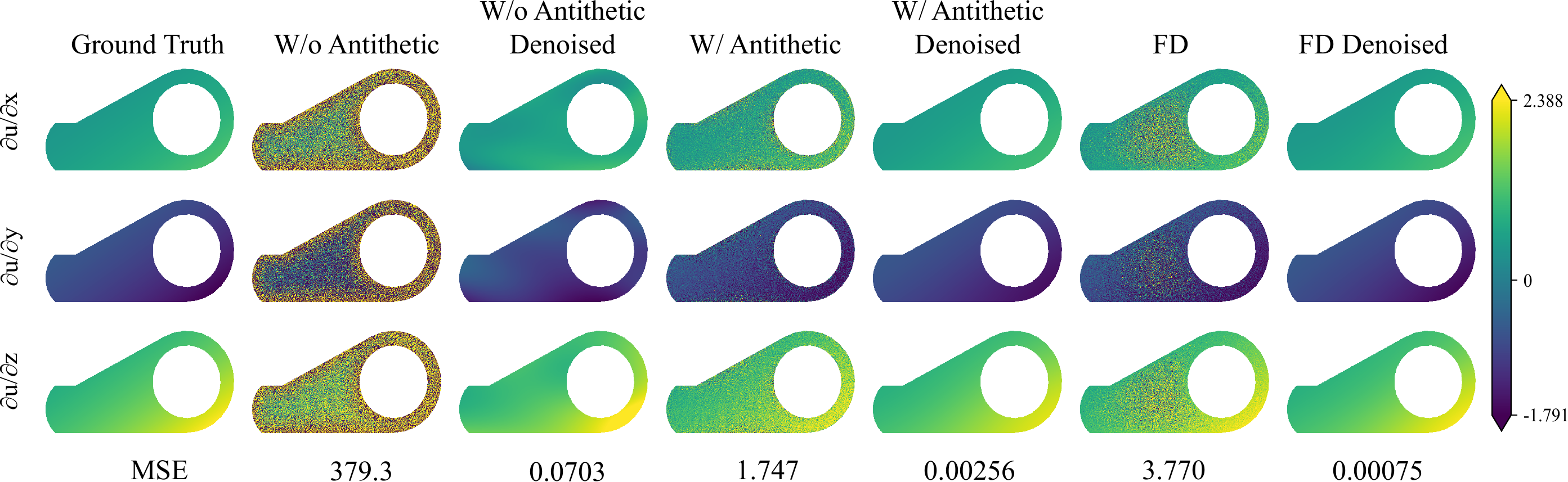}
    \caption{Gradient estimation on the tool geometry with mixed boundary conditions. We compare the proposed estimator with and without antithetic sampling, and with and without heteroscedastic denoising, together with a finite-difference estimator that shares its random number sequence between the two neighbouring query points. Before denoising the proposed estimator is the more accurate of the two, while after denoising the finite-difference estimator is more accurate. The columns without the label ``Denoised'' show raw estimates. Each MSE value is computed from the estimate in its own column. Errors before and after learned post-processing are reported separately.}
    \label{fig:gradient_comparison_tool}
    \label{fig:gradient_estimation}
\end{figure*}}

\oldtext{Although the MSE of the gradient estimates decreases with more samples as expected, the variance is substantially higher than the value estimates. To produce readable results, we apply the same heteroscedastic regression denoiser to the gradient estimates. Figure~\ref{fig:quiver} visualizes the ground truth and denoised gradient fields, as well as the error map and MSE convergence curve. The denoised gradient field matches the ground truth reasonably well, and the MSE decreases with more samples, confirming the correctness of the gradient formula. However, we note that the raw (pre-denoising) estimates are too noisy for direct practical use, and we regard the gradient estimator in its current form as a proof-of-concept whose robustness needs improvement before deployment in applications such as shape optimization.}\changedtext{}

\oldtext{Figure~\ref{fig:gradient_estimation} provides a detailed view of the denoised gradient field components alongside the ground truth. The alignment confirms that the gradient formula, combined with the heteroscedastic denoiser, can recover spatial derivatives with reasonable accuracy. We also implemented a version of the gradient estimator based on the standard Walk on Heat Spheres algorithm and finite differences; however, this approach yielded significantly higher variance even with a relatively large step length $\epsilon = 10^{-2}$, confirming that the volumetric mean value property approach is preferable despite its current variance limitations.}

\changedtext{We evaluate the source-free gradient estimator on a ball with half Neumann boundary and on the tool geometry with the mixed boundary configuration in Figure~\ref{fig:models}. Figures~\ref{fig:gradient_comparison_sphere} and~\ref{fig:gradient_comparison_tool} separate three estimators: the proposed method without antithetic sampling, the proposed method with antithetic sampling, and finite differences with step size $0.01$ and shared random numbers. Each estimator is shown before and after learned denoising, with a separate MSE for each result. Comparing the raw columns measures the effect of the estimator and antithetic sampling. Comparing each raw column with its denoised counterpart measures the additional effect of learned post-processing. In these tests, antithetic sampling reduces raw gradient error, and the proposed antithetic estimator is more accurate than raw finite differences. After denoising, finite differences have the lower error.}

\oldtext{}\changedtext{

\subsection{Unbounded Domains}
\label{sec:unbounded}

\begin{figure*}
    \centering
    \includegraphics[width=\linewidth]{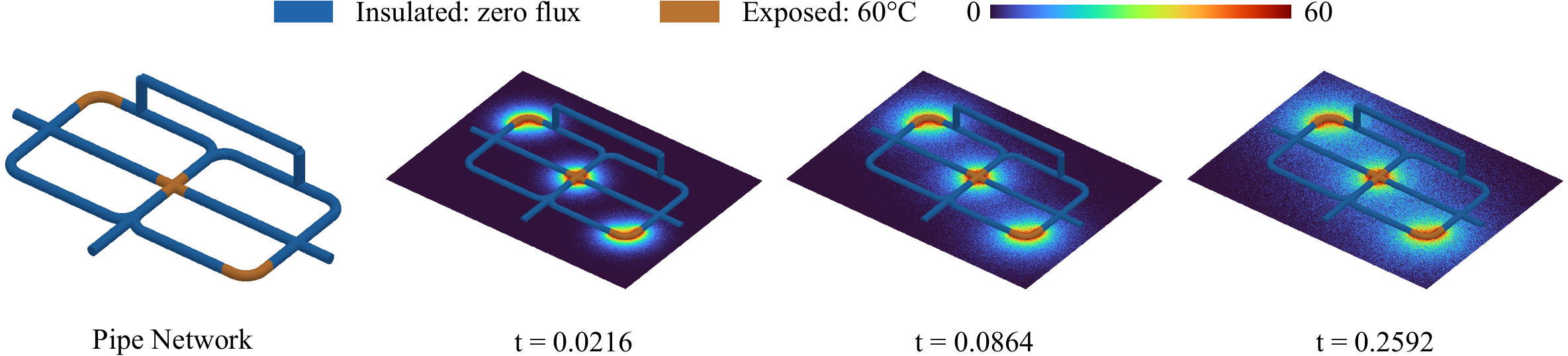}
    \caption{Pipe network embedded in an unbounded medium. \textbf{Left:} Geometry of the network. Blue segments are insulated and carry zero Neumann flux, while brown segments are held at a fixed temperature of $60^\circ$C and act as Dirichlet sources. \textbf{Right:} WoHSt estimates of the temperature field on a fixed slice at physical times of 6 hours, 1 day and 3 days, converted to solver time by Eq.~\eqref{eq:pipe_time_scale}. The surrounding medium is treated as an unbounded domain.}
    \label{fig:unbounded_pipe}
\end{figure*}

The experiments above use bounded spatial domains. For elliptic problems on unbounded domains, a walk may escape to infinity without reaching a boundary. Existing methods address this through a Kelvin transformation~\cite{nabizadeh2021kelvin} or analytical escape and return sampling~\cite{zhou1994brownian}. For finite-time heat problems, the initial time plane provides an additional termination boundary. A WoHS or WoHSt walk moves backward in time and terminates either at a Dirichlet boundary or, through the temporal stopping tolerance, at the initial time plane. The remaining time also bounds the heat-ball size, even far from the physical boundary. The initial condition therefore supplies the terminal value for walks that do not reach a Dirichlet boundary, without an additional rule for escape to infinity.

Figure~\ref{fig:unbounded_pipe} applies WoHSt to a pipe network embedded in such a medium. Almost the entire pipe surface is insulated, so it carries zero Neumann flux, while a few short segments are held at a fixed temperature of $60^\circ$C and act as Dirichlet sources. One spatial unit in the pipe geometry represents $L=1\,\mathrm{m}$, and we assume a thermal diffusivity of $\kappa=10^{-6}\,\mathrm{m}^2/\mathrm{s}$ for the surrounding medium. With positions scaled by $L$, the physical heat equation $\partial_{t_{\mathrm{phys}}}u=\kappa\Delta_{x_{\mathrm{phys}}}u$ takes the unit-diffusivity form used by WoHSt when solver time is
\begin{equation}
    t=\frac{\kappa\,t_{\mathrm{phys}}}{L^2}
     =\frac{t_{\mathrm{phys}}}{10^6\,\mathrm{s}}.
    \label{eq:pipe_time_scale}
\end{equation}
For example, 6 hours is $21{,}600\,\mathrm{s}$, giving $t=0.0216$. The same conversion gives $t=0.0864$ for 1 day and $t=0.2592$ for 3 days. At the earliest instant the field consists of separate plumes around the exposed segments. As time grows the plumes spread along the pipes and merge into a single smooth field. The panels illustrate heat spreading through the surrounding unbounded medium over these query times.
}

\oldtext{}\changedtext{

\subsection{Comparison with Walk-on-Boundary Estimator}
\label{sec:wob_comparison}

\begin{figure*}
    \centering
    \includegraphics[width=\linewidth]{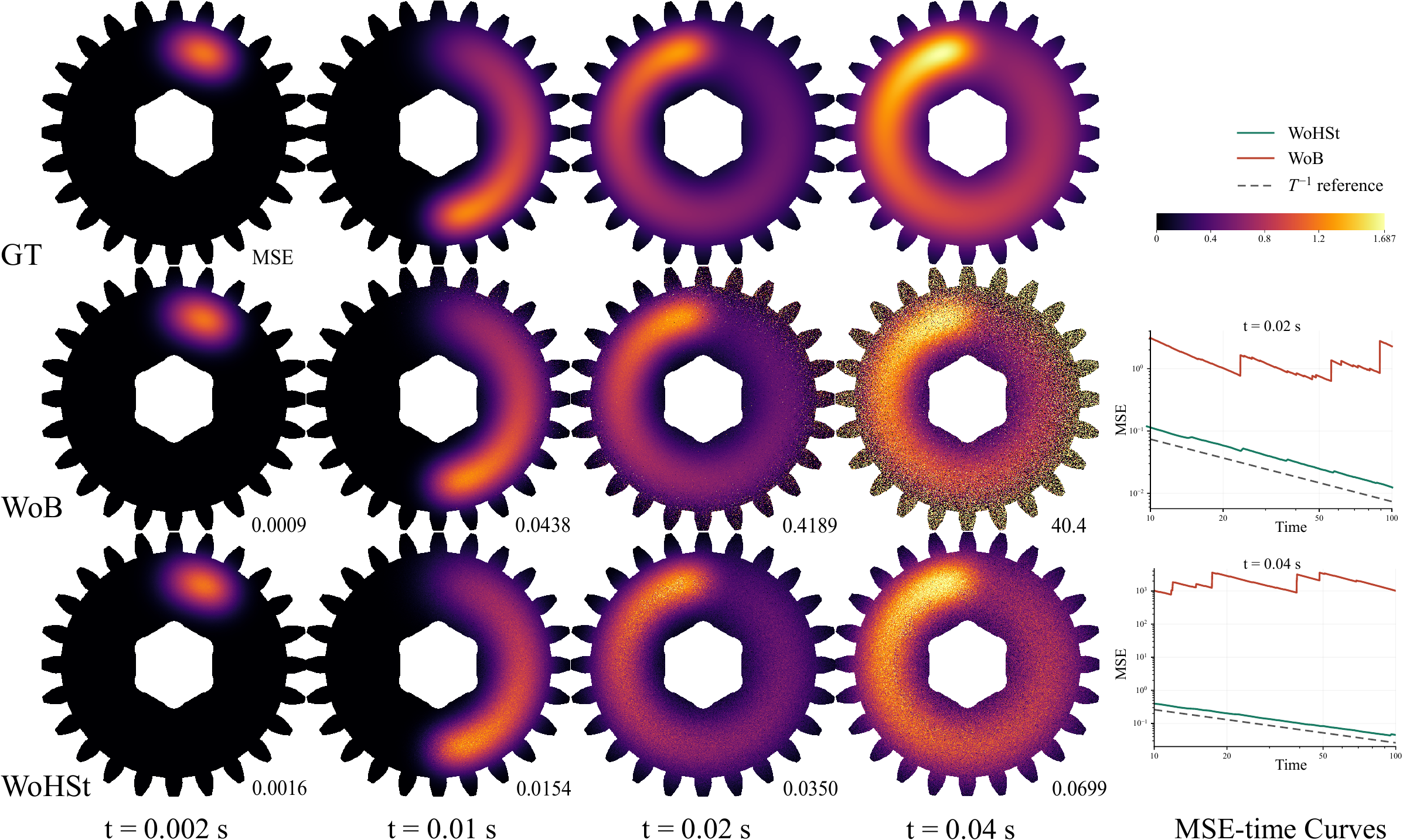}
    \caption{Comparison of WoHSt and Walk-on-Boundary (WoB) estimators on a gear geometry with pure Neumann boundary conditions. The first row shows the moving-source reference from~\eqref{eq:moving_source_reference}, the second row shows WoB estimates, and the third row shows WoHSt estimates. In this comparison, WoB has slightly lower MSE at the shortest query time, while WoHSt has lower MSE at the later times.}
    \label{fig:wob_comparison}
\end{figure*}

For simplicity, we compare against a pure-Neumann heat WoB baseline within the established framework~\cite{sabelfeld1991monte,sabelfeld2016stochastic,sugimoto2025toward}. A mixed-boundary extension is possible, but is not needed for this comparison of Neumann handling. Both methods use the same geometry, query locations, time instants, spatial acceleration structures, threading, and compute budgets.

The comparison uses a moving Gaussian source with zero initial temperature. We use the unit-diffusivity equation $u_t=\Delta u+f$ and solver time $t$ throughout these comparisons. The source is $f(x,t)=A\exp(-\|x-c(t)\|^2/(2\sigma^2))$ with $A=1000$. Convolving it with the whole-space heat kernel gives
\begin{equation}
\label{eq:moving_source_reference}
\begin{aligned}
u_{\mathrm{ref}}(x,t)
&=\int_0^t A\left(\frac{\sigma^2}{v(s,t)}\right)^{3/2}
\exp\!\left(-\frac{\|x-c(s)\|^2}{2v(s,t)}\right)\,\mathrm{d}s, \\
v(s,t)&=\sigma^2+2(t-s).
\end{aligned}
\end{equation}
This expression includes heat deposited at all previous source positions. We restrict it to the test geometry and prescribe $h=\nabla u_{\mathrm{ref}}\cdot n$ on its boundary, obtaining a pure-Neumann problem with a known solution. The normal derivative is computed by differentiating the integrand. The remaining time integral is evaluated by composite Gauss--Legendre quadrature and checked independently with a higher-order rule and smaller panels. 

For an interior point $x\in\Omega$, the direct pure-Neumann formulation expresses $u$ in terms of its unknown boundary trace $\phi$:
\begin{equation}
\label{eq:wob_direct_representation}
\begin{aligned}
u(x,t) &= -\int_0^t\!\int_{\partial\Omega}
\frac{\partial Z}{\partial n_y}(x,t;y,\tau)\,\phi(y,\tau)\,\mathrm{d}S_y\,\mathrm{d}\tau \\
&\quad+\int_0^t\!\int_{\partial\Omega}
Z(x,t;y,\tau)\,h(y,\tau)\,\mathrm{d}S_y\,\mathrm{d}\tau \\
&\quad+\int_\Omega Z(x,t;y,0)\,g_0(y)\,\mathrm{d}y \\
&\quad+\int_0^t\!\int_\Omega
Z(x,t;y,\tau)\,f(y,\tau)\,\mathrm{d}y\,\mathrm{d}\tau.
\end{aligned}
\end{equation}
Here $\phi=u|_{\partial\Omega}$ is the boundary temperature, $n_y=n(y)$ points outward, and $h$, $g_0$, and $f$ are the prescribed Neumann flux, initial condition, and source. Following the heat-kernel notation of \citet[Chapter~4]{sugimoto2025toward}, $Z(x,t;y,\tau)=\Phi(x-y,t-\tau)$. Taking the interior limit at a smooth boundary point gives the recursive equation for $\phi$:
\begin{equation}
\label{eq:wob_direct_bie}
\begin{aligned}
\phi(x,t) &= -2\int_0^t\!\int_{\partial\Omega}
\frac{\partial Z}{\partial n_y}(x,t;y,\tau)\,\phi(y,\tau)\,\mathrm{d}S_y\,\mathrm{d}\tau \\
&\quad+2\int_0^t\!\int_{\partial\Omega}
Z(x,t;y,\tau)\,h(y,\tau)\,\mathrm{d}S_y\,\mathrm{d}\tau \\
&\quad+2\int_\Omega Z(x,t;y,0)\,g_0(y)\,\mathrm{d}y \\
&\quad+2\int_0^t\!\int_\Omega
Z(x,t;y,\tau)\,f(y,\tau)\,\mathrm{d}y\,\mathrm{d}\tau.
\end{aligned}
\end{equation}
The first term continues the walk on the boundary, while the remaining terms evaluate prescribed data.

In 3D, each step samples a uniform direction $\omega$ and $a\sim\Gamma(3/2,1)$. We trace one ray from an interior point and both directions of a line from a boundary point, excluding the origin face. Each intersection $y$ is assigned time $\tau=t-\|y-x\|^2/(4a)$ and retained if $\tau>0$. Each traced direction also contributes an initial-time endpoint $x+2\sqrt{ta}\,\omega$ if it lies inside $\Omega$. We select one of these $\kappa_k$ candidates uniformly. Starting from $W_0=1$, a boundary selection updates the weight to $W_{k+1}=W_k\kappa_k\operatorname{sgn}(n(y)\cdot(y-x))$ and continues at $(y,\tau)$. An initial-time selection has sign $+1$ and terminates with the weighted value of $g_0$. Thus $\kappa_k$ includes initial-time endpoints as well as boundary intersections. Source and flux estimates are accumulated along the walk. This comparison uses neither bidirectional MIS nor Metropolis path sampling.

On non-convex geometry, multiple boundary intersections can produce factors $\kappa_k>1$ that amplify path contributions. The normal-derivative kernel also changes sign, requiring cancellation between positive and negative contributions. Both effects can increase variance. Related signed intersection-count weights appear in the forward heat-Neumann estimator of~\citet[Eq.~(4.46)]{sabelfeld2016stochastic}, and \citet[Section~7.2 and Figure~12]{miller2024walkin} discuss this variance mechanism for elliptic WoB. Small geometric features can further produce short spatial steps and small time increments, allowing more weight factors to accumulate before termination~\cite[Chapter~4]{sugimoto2025toward}. WoHSt uses a local star-shaped region with one intersection along each radial trace and unit recursive weight, avoiding this amplification mechanism. Its source and flux estimates can still have substantial variance.

Figure~\ref{fig:wob_comparison} shows slightly lower MSE for WoB at the shortest query time. At later times, WoB exhibits larger errors and noisy fields near the boundary, consistent with these variance mechanisms. The experiment does not isolate their individual contributions.

\subsection{Comparison with Finite Element Methods}
\label{sec:fem_comparison}

\begin{figure*}
    \centering
    \includegraphics[width=\linewidth]{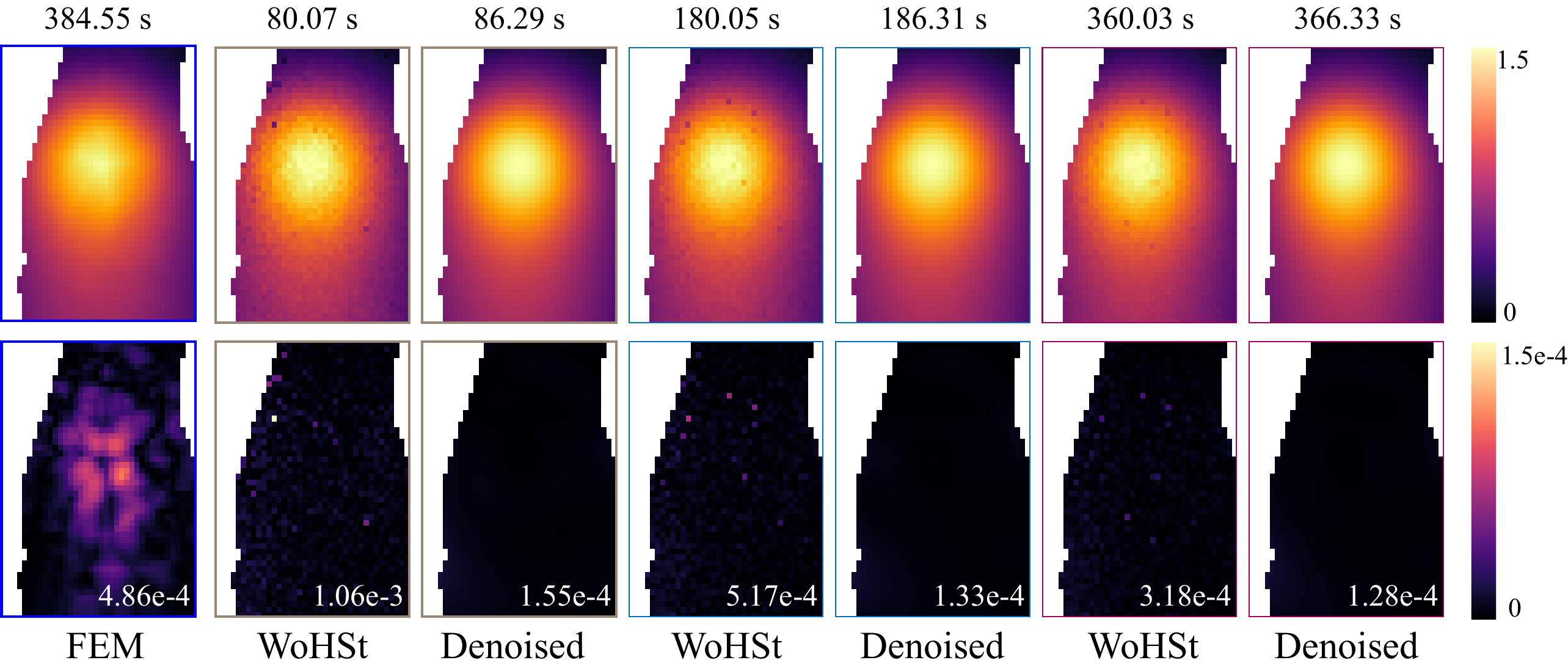}
    \caption{Local solution estimates on SimJEB 223 with mixed boundary conditions at $t=0.02$. The raw WoHSt columns show the saved sampling prefixes used for the adjacent denoised fields. Each column reports its measured cost, including post-processing where used. FEM cost includes meshing and the solver process.}
    \label{fig:fem_comparison}
\end{figure*}

\begin{figure*}
    \centering
    \includegraphics[width=\linewidth]{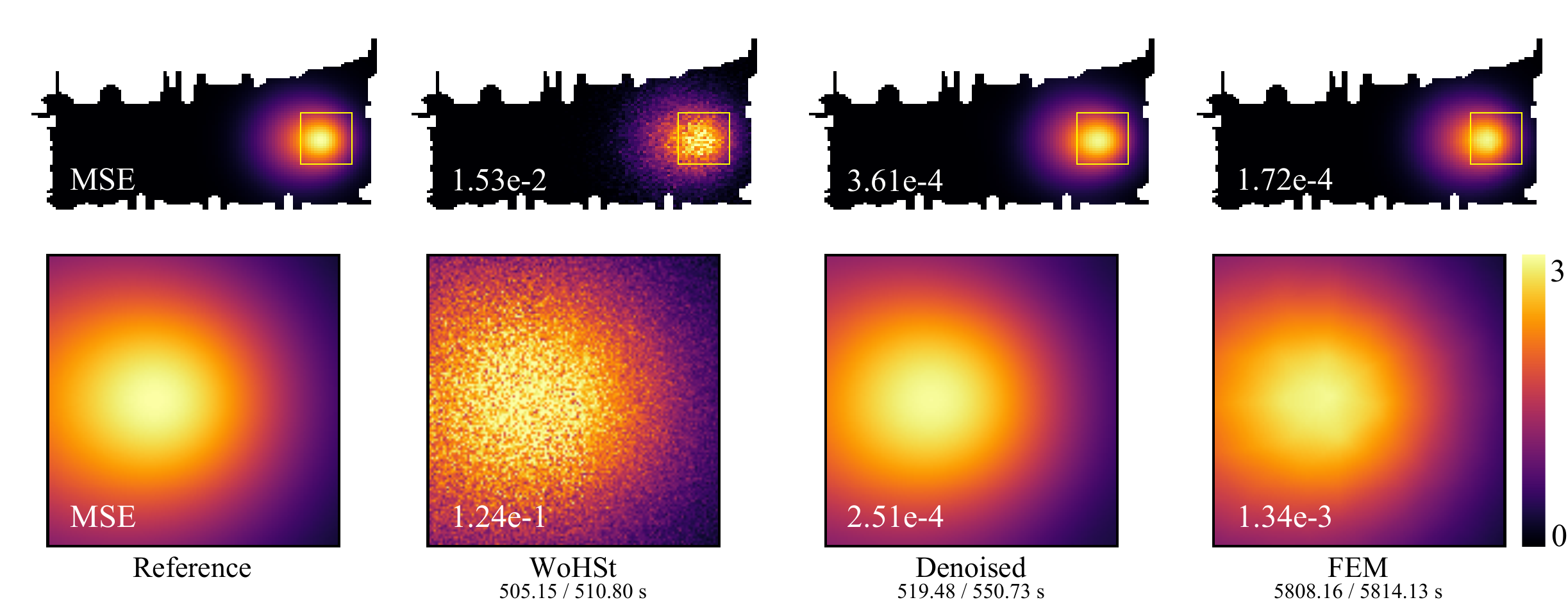}
    \caption{Engine model with mixed boundary conditions at $t=0.02$. The columns show the reference solution, raw WoHSt estimates, denoised estimates, and FEM estimates, respectively. The top row evaluates a broader slice, and the bottom row evaluates a separately sampled local region. Timing labels give the measured top-row / bottom-row costs, including post-processing where used and local FEM field evaluation. The nominal WoHSt sampling allowance excludes post-processing.}
    \label{fig:fem_comparison_engine}
\end{figure*}

We compare WoHSt with the finite element solver deal.II~\cite{2025:arndt.bangerth.ea:deal}, using volumetric meshes generated by fTetWild (FastTetWild)~\cite{hu2020fast}. The experiments examine local dense solution evaluation, where WoHSt can allocate computation directly to the requested region, progressively improve its estimates with a flexible sampling budget, and avoid volumetric mesh generation.

The two models are SimJEB model 223~\cite{whalen2021simjeb} (Figure~\ref{fig:fem_comparison}) and a four-cylinder engine (Figure~\ref{fig:fem_comparison_engine}). Both surfaces are meshed with fTetWild~\cite{hu2020fast} at a target edge length of $0.05$ and a surface envelope of $3\times10^{-4}$ of the bounding-box diagonal, with at most 80 optimization iterations.

The FEM baseline uses continuous linear tetrahedral elements, a consistent mass matrix and backward Euler over 1440 steps up to $t=0.02$ (time step $\Delta t = 0.02/1440 \approx 1.39\times10^{-5}$). Every step is solved by conjugate gradients with Jacobi preconditioning. Both geometries use the moving-source reference construction in Section~\ref{sec:wob_comparison}. 

\begin{table}[ht]
\centering
\caption{Configuration and recorded costs of the FEM comparisons ($\sigma=0.06$ for SimJEB and $0.08$ for the engine). FEM process time includes setup, assembly, time stepping, and output. Engine timing excludes computer sleep. Its stage timers omit small retry-startup and repair-serialization overhead. Totals are computed from unrounded records. Local engine field evaluation adds 5.97\,s to the listed base cost.}
\label{tab:fem_config}
\begin{tabular}{lcc}
\toprule
& SimJEB 223 & Engine \\
\midrule
Input surface & 43,564 tri. & 9,994,566 tri. \\
Surface after simplification & 21,782 tri. & none \\
Tetrahedra & 59,860 & 1,335,594 \\
Vertices (scalar DOFs) & 15,196 & 313,861 \\
Boundary faces (D\,/\,N) & 1,682\,/\,16,672 & 40,997\,/\,286,681 \\
Shared input preparation & -- & 11.19\,s \\
Meshing process & 301.08\,s & 4,130.13\,s \\
Repair and validation & -- & 50.87\,s \\
FEM process & 83.47\,s & 1,615.97\,s \\
Recorded total & \textbf{384.55\,s} & \textbf{5,808.16\,s} \\
\bottomrule
\end{tabular}
\end{table} 

\begin{figure}
    \centering
    \includegraphics[width=\linewidth]{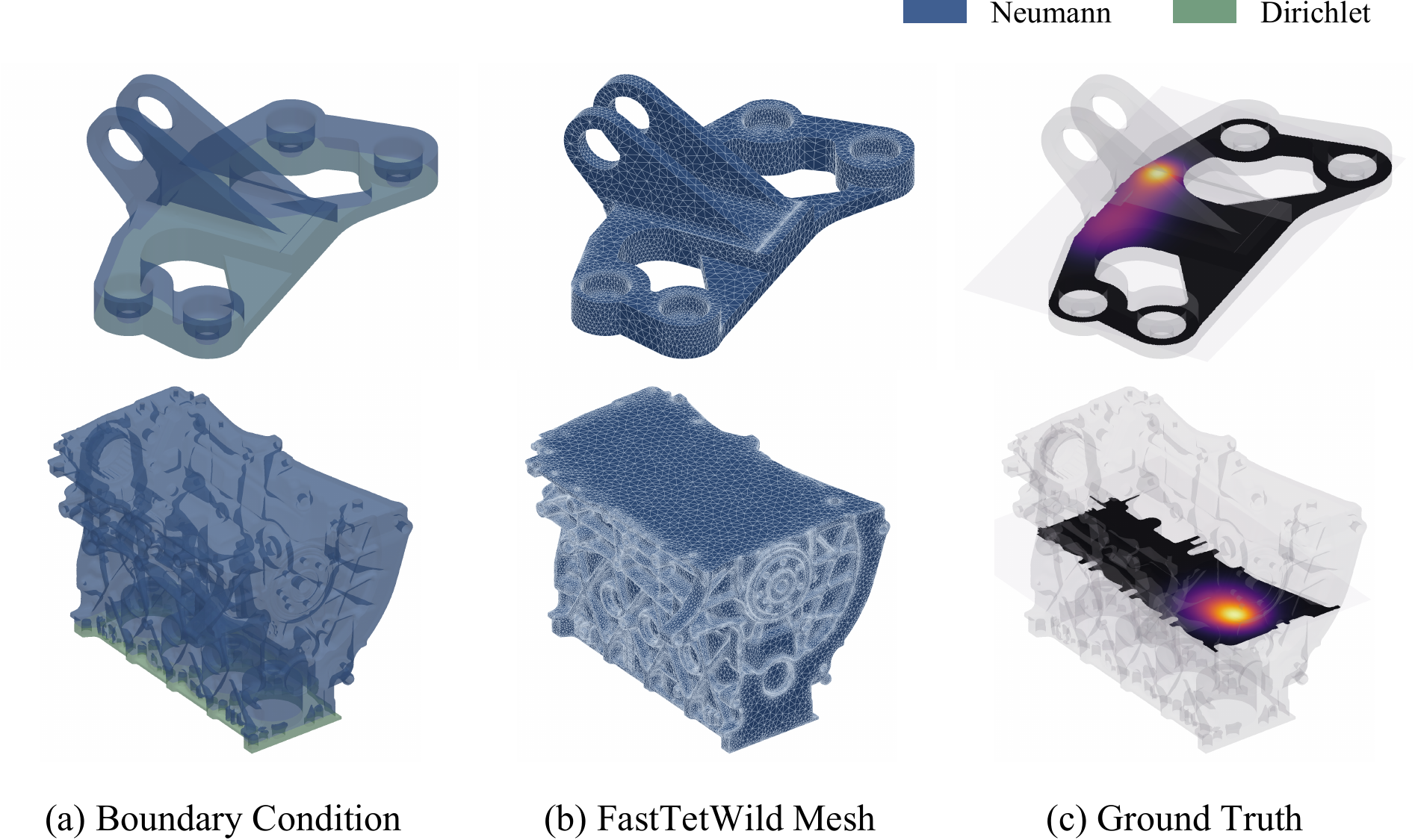}
    \caption{Geometry, boundary conditions, and reference solution for the two FEM test models: SimJEB 223 (top row) and the engine (bottom row). \textbf{(a)} Boundary conditions (Dirichlet green, Neumann blue). \textbf{(b)} Tetrahedral mesh generated by fTetWild. \textbf{(c)} Reference solution used to define the manufactured problem, shown at $t=1$ in solver units. The quantitative comparisons in Figures~\ref{fig:fem_comparison} and~\ref{fig:fem_comparison_engine} are evaluated at $t=0.02$.}
    \label{fig:fem_jet_geometry}
\end{figure}

Figure~\ref{fig:fem_comparison} compares local estimates on SimJEB 223 against the analytical moving-source reference at increasing Monte Carlo budgets. Figure~\ref{fig:fem_comparison_engine} extends the comparison to broader-slice and local estimates on the full-resolution engine. Table~\ref{tab:fem_config} reports the FEM cost breakdown, while the figure labels give the measured costs of the displayed raw and post-processed fields.

These examples illustrate the practical benefit of grid-free Monte Carlo for local dense evaluation. FEM requires volumetric meshing and a global solve before the requested solution values can be evaluated. WoHSt works directly with the boundary geometry and progressively accumulates estimates at the requested points. The SimJEB comparison shows how the sampling budget can be varied, while the engine example demonstrates local evaluation on a complex surface whose tetrahedral mesh is expensive to construct and requires repair. With optional denoising, the local fields achieve competitive accuracy at the reported costs. The figures retain both raw and denoised errors and include post-processing in the corresponding timings.

Taken together, these comparisons indicate that WoHSt is the better choice when the requested output is a local dense field on a geometrically complex surface whose volumetric meshing is expensive or fragile, and when a flexible, progressively refined sampling budget is acceptable. For full-domain solutions on meshing-friendly geometries, where a global solve amortizes the meshing cost over many queries, FEM remains the more direct choice.
}

\subsection{Application: Heat Simulation}

In this section, we compare the proposed Walk on Heat Stars method with a deterministic finite element solver on a heat sink temperature problem. \changedtext{We then use WoHSt to evaluate surface cooling on the gearbox in Figure~\ref{fig:gearbox_teaser}.}

\begin{figure*}
    \centering
    \changedtext{\includegraphics[width=\linewidth]{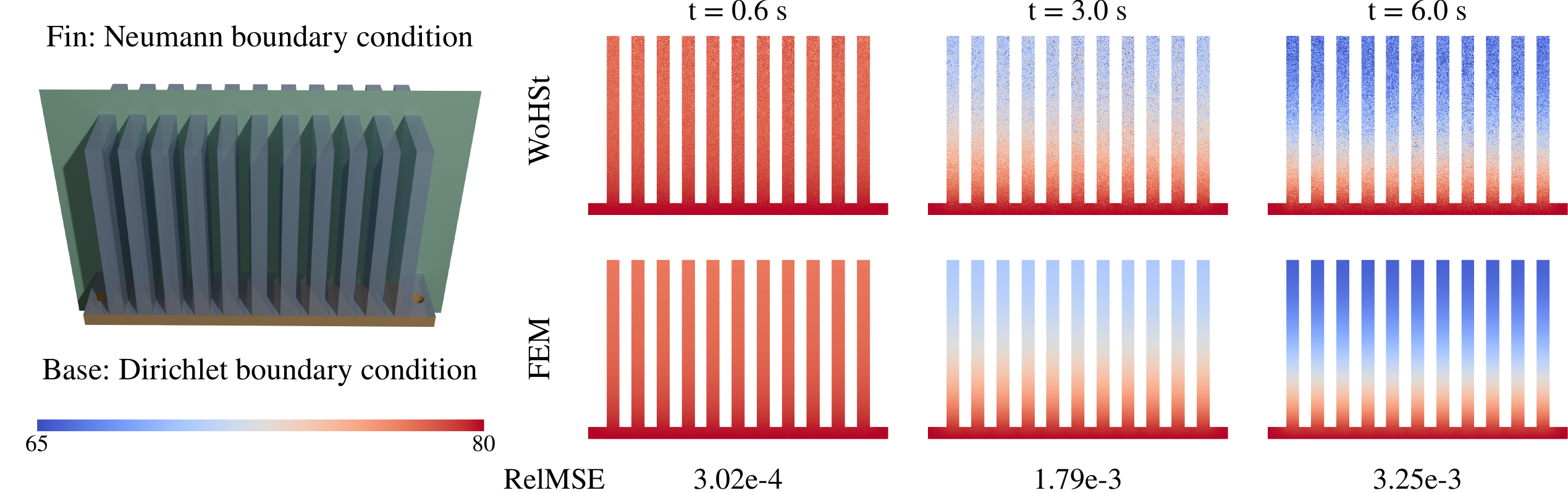}}
    \caption{\oldtext{WoHSt estimates of transient heat conduction in a heat sink geometry with mixed boundary conditions. \textbf{Left:} Geometry and boundary conditions. The domain is a 3D heat sink with Dirichlet (fixed temperature) boundaries on the outer faces and Neumann boundaries on the inner faces. \textbf{Right:} Walk on Heat Stars estimation and FEM results at different time instants.}\changedtext{Transient heat conduction in a heat sink with mixed boundary conditions. \textbf{Left:} The base is held at the initial temperature of $80\,^{\circ}\mathrm{C}$, and a uniform Neumann flux cools the fins. \textbf{Right:} WoHSt estimates (top, 128 walks per pixel) and the FEM reference (bottom) on the same slice. All six fields use the color scale at left, in $^{\circ}\mathrm{C}$. RelMSE is the squared error summed over the slice divided by the summed squared FEM temperature, both in solver units.}}
    \label{fig:heat_sink}
\end{figure*}

As shown in Figure~\ref{fig:heat_sink}, we consider a 3D heat sink geometry with mixed Dirichlet and Neumann boundary conditions. \oldtext{Initially, the heat sink is at a uniform temperature.}\changedtext{Initially, the heat sink is at a uniform temperature of $80\,^{\circ}\mathrm{C}$.} \oldtext{At the time $t = 0$, the base of the heat sink is subjected to a new temperature profile.}\changedtext{The base is held at this temperature, i.e., a Dirichlet condition, throughout the simulation.} The fins in the heat sink have Neumann boundary conditions, transferring heat to the surrounding air. We run both the Walk on Heat Stars method and a finite element solver to compute the transient temperature distribution in the heat sink over time. The results show that WoHSt agrees with the FEM reference computed by fTetWild~\cite{hu2020fast} and deal.II~\cite{2025:arndt.bangerth.ea:deal}, accurately capturing the transient heat conduction behavior in the complex geometry. This demonstrates the practical applicability of WoHSt for simulating heat transfer in engineering contexts where complex geometries and mixed boundary conditions are common.

\oldtext{To show the geometric flexibility of WoHSt, we also apply it to a more complex turbine housing geometry with a pure Neumann setup, simulating the transient heat conduction and cooling process. The initial condition is a uniform high temperature, and the boundary condition is a uniform heat flux on the outer surface. Unlike the elliptic case, the pure Neumann setup is well-posed under the constraint of initial temperature and hence Tikhonov regularization is not needed. We run the WoHSt method at three different time instants with a fixed spatial slice; each estimate uses 16 random walks. We also present the estimation at three fixed spatial points as a function of time, simulating the sensor readings during the cooling process.
The results shown in Figure~\ref{fig:housing} demonstrate that WoHSt can handle complex geometry and accurately capture the transient cooling process, confirming its potential for real-world engineering applications.}
\changedtext{Figure~\ref{fig:gearbox_teaser} shows a gearbox represented by nearly 10 million surface triangles, which cannot be meshed with fTetWild under 10 hours time budget. Starting from a uniform temperature, we impose a constant outward normal derivative on the entire surface, with zero volumetric source. We report temperature changes from the initial state in solver units. For each pixel in a $128\times128$ view, we cast a camera ray and evaluate the solution at its first intersection with the gearbox. At each point we use 16 walks at $t=0.1$ and another 16 at $t=0.2$. Computing the two fields takes 27 min 24 s at $t=0.1$ and 53 min 36 s at $t=0.2$ of wall-clock time.} \changedtext{At the three marked surface sensors, we use 128 walks per sensor at each of 100 equally spaced times from $t=0.001$ to $0.5$, which takes 15 min 6 s of wall-clock time. Although the prescribed flux is uniform, the sensors cool at different rates. The results demonstrate that WoHSt can handle extremely complex geometries and provide transient heat transfer estimates without requiring volumetric meshing.}

\section{Limitations and Future Work}



\paragraph{Robin Boundary Conditions}

In the elliptic case, \citet{miller2024walkin} extended the Walk-on-Stars method to handle Robin boundary conditions by employing the Brakhage-Werner trick together with an adaptive radius selection strategy. The extension to the parabolic case is valuable since Robin conditions are common in heat transfer problems, modeling convective heat exchange with the environment. Developing a robust WoHSt variant that can handle Robin conditions is an important direction for future work.

\paragraph{Derivative Estimation}

\oldtext{Although we derived a gradient estimator under the parabolic setting, it currently applies only to source-free problems; the source term $f$ and Neumann boundary flux contributions are not handled by the present formulation. Moreover, the variance is still high compared to the value estimator due to the $\tau(c)$ prefactor in the gradient formula (Eq.~\ref{equ:grad_final}). In practice, the raw gradient estimates are too noisy for direct use without the heteroscedastic regression denoiser; we therefore view the current gradient estimator as a proof-of-concept rather than a production-ready tool. Under the elliptic setting, \citet{yu2025robust,yu2024differential} developed robust derivative estimators that significantly reduce variance near Neumann boundaries via a Robin boundary reformulation. Extending these techniques to the parabolic setting is an important direction for future work, as accurate and efficient gradient estimation is crucial}\,\changedtext{The gradient estimator is derived for source-free problems. Extending it to nonzero sources requires an additional derivation. Mixed Dirichlet--Neumann problems are handled by taking the first gradient step in an interior heat ball and evaluating the resulting solution values with WoHSt, which includes prescribed Neumann flux. The new antithetic-sampling experiments reduce raw gradient error, while learned denoising provides a separate post-processing gain. Under the elliptic setting, \citet{yu2025robust,yu2024differential} developed robust derivative estimators that reduce variance near Neumann boundaries via a Robin boundary reformulation. Extending these techniques to the parabolic setting remains an important direction for applications such as shape optimization and thermal stress analysis.}

\paragraph{Variance Reduction}

\changedtext{The variance-reduction toolbox for Monte Carlo PDE solvers was developed mainly for elliptic problems, where control variates, boundary-value caching, and the reuse of samples across overlapping spheres or neighbouring queries are available~\cite{sawhney2020monte,miller2023boundary,bakbouk2023mean,czekanski2024walking,bao2025off}. These constructions do not carry over directly to heat-star walks. The transition measure is a space--time measure, so boundary data cannot be cached at a fixed spatial location, and estimates at different query times do not share the same local geometry. To the best of our knowledge, no non-learning variance-reduction scheme has been established for value estimation with walks on heat balls or heat stars. For this reason, the present implementation relies on a learned denoiser as optional post-processing.}

\section{Conclusion}

\oldtext{We presented Walk on Heat Stars, a  novel Monte Carlo method for solving transient heat conduction problems on complex geometries. By deriving a non-cylindrical boundary integral representation and an associated importance sampling strategy, WoHSt enables efficient estimation of the solution at arbitrary space-time points without requiring the entire heat ball to fit inside the domain. We also derived a gradient estimator for source-free problems that expresses spatial derivatives as weighted boundary integrals of the solution, avoiding recursion on $\nabla u$, though it does not yet handle source terms or Neumann flux contributions and its current variance necessitates denoising for practical use. Our experiments on analytical solutions demonstrate that WoHSt achieves accurate estimates across a range of geometries and spatial frequencies, with convergence at the expected Monte Carlo rate. Future work includes extending WoHSt to handle Robin boundary conditions, developing robust derivative estimators with reduced variance, and exploring hybrid strategies that combine WoHSt with classical time-stepping for full-domain problems.}\,\changedtext{We presented Walk on Heat Stars, a Monte Carlo method that extends the local star-shaped construction of WoSt to the parabolic heat equation on fixed physical domains. Its heat-ball parameterization and joint space--time sampling rules support mixed Dirichlet--Neumann conditions and prescribed source terms without requiring the entire heat ball to fit inside the domain. We also derived a source-free gradient estimator that recurses on solution values rather than gradients. An interior first step permits its use with mixed boundaries. Antithetic sampling improves the raw gradient estimates in our tests, and learned denoising remains optional post-processing. The fitted raw-MSE slopes indicate approximately inverse-time decay over the measured budgets, while the Neumann-fraction and long-time studies expose substantial variance in pure-Neumann regimes. Future work includes nonzero-source gradient estimation, Robin boundary conditions, more robust variance reduction, and hybrid methods for full-domain problems.}


\bibliographystyle{ACM-Reference-Format}
\bibliography{ref}

\appendix

\section{Derivation of the Non-Cylindrical Boundary Integral Equation}
\label{app:noncylindrical_bie}

We derive the non-cylindrical boundary integral representation~\eqref{equ:bie_noncylindrical} from first principles. \oldtext{}\changedtext{The proof combines three classical ingredients, namely Green's second identity (spatial), the Reynolds transport theorem (moving domain), and the adjoint heat kernel identity.} A general treatment in Sobolev spaces can be found in~\cite{brugger2020boundary}. Here we give a simpler proof under smoothness assumptions.

Let $\{\Omega_\tau\}_{0<\tau<T}$ be a family of smoothly evolving spatial domains in $\mathbb{R}^n$, with boundaries $\Gamma_\tau = \partial\Omega_\tau$ of class $C^2$, outward unit normal $\vec{n}$, and boundary velocity $V(\tau, y)$. Define the moving space-time domain $Q_T = \bigcup_{0<\tau<T} \{\tau\} \times \Omega_\tau$ with lateral boundary $\Sigma_T = \bigcup_{0<\tau<T} \{\tau\} \times \Gamma_\tau$. Assume $u \in C^{2,1}(Q_T)$ satisfies \oldtext{$(\partial_\tau - \Delta_y) u = 0$}\,\changedtext{$(\partial_\tau - \Delta_y) u = f$} in $Q_T$ with \oldtext{$u(0,\cdot)=0$}\,\changedtext{$u(0,\cdot)=g_0$} on $\Omega_0$. Fix an observation point $(t,x)$ with $x \in \Omega_t$, and define the forward heat kernel
\begin{equation}
    G(\tau, y) \equiv G(t,\tau,x,y)
    = \begin{cases}
        \dfrac{1}{(4\pi(t-\tau))^{n/2}}
        \exp\!\left(-\dfrac{|x-y|^2}{4(t-\tau)}\right), & \tau < t, \\[1em]
        0, & \tau \ge t,
    \end{cases}
\end{equation}
which satisfies the adjoint identities $(-\partial_\tau - \Delta_y)G = \delta(t-\tau)\,\delta(x-y)$ and $(\partial_t - \Delta_x)G = \delta$.

We begin by applying Green's second identity on $\Omega_\tau$ for each fixed $\tau \in (0,t)$:
\begin{equation}
    \int_{\Omega_\tau} (u\,\Delta G - G\,\Delta u)\,\mathrm{d}y
    = \int_{\Gamma_\tau} \left(u\,\frac{\partial G}{\partial \vec{n}} - G\,\frac{\partial u}{\partial \vec{n}}\right)\mathrm{d}\sigma_y.
\end{equation}
Let $u$ satisfy \oldtext{$(\partial_\tau - \Delta)u = -f$}\,\changedtext{$(\partial_\tau - \Delta)u = f$}, so that \oldtext{$\Delta u = \partial_\tau u + f$}\,\changedtext{$\Delta u = \partial_\tau u - f$}, while $\Delta G = -\partial_\tau G$ away from the singularity (the delta at $\tau=t$ is handled by the time integration below). Substituting,
\begin{equation}
    u\,\Delta G - G\,\Delta u = u(-\partial_\tau G) - G(\partial_\tau u \oldtext{+}\changedtext{-} f) = -\partial_\tau(uG) \oldtext{-}\changedtext{+} Gf,
\end{equation}
yielding
\begin{equation}
    \label{equ:green_step}
    -\int_{\Omega_\tau} \partial_\tau(uG)\,\mathrm{d}y
    = \int_{\Gamma_\tau} \left(u\,\frac{\partial G}{\partial \vec{n}} - G\,\frac{\partial u}{\partial \vec{n}}\right)\mathrm{d}\sigma_y
      \;\oldtext{+}\changedtext{-}\; \int_{\Omega_\tau} G f\,\mathrm{d}y.
\end{equation}

Because $\Omega_\tau$ is moving, the time derivative cannot be pulled outside the integral. The Reynolds transport theorem gives
\begin{equation}
    \frac{\mathrm{d}}{\mathrm{d}\tau} \int_{\Omega_\tau} f\,\mathrm{d}y
    = \int_{\Omega_\tau} \partial_\tau f\,\mathrm{d}y
    + \int_{\Gamma_\tau} f\,\langle V, \vec{n}\rangle\,\mathrm{d}\sigma_y.
\end{equation}
Applying this to $f = uG$ and substituting into~\eqref{equ:green_step}:
\begin{equation}
    \begin{aligned}
    -\frac{\mathrm{d}}{\mathrm{d}\tau} \int_{\Omega_\tau} uG\,\mathrm{d}y
    &+ \int_{\Gamma_\tau} uG\,\langle V, \vec{n}\rangle\,\mathrm{d}\sigma_y \\
    &= \int_{\Gamma_\tau} \left(u\,\frac{\partial G}{\partial \vec{n}} - G\,\frac{\partial u}{\partial \vec{n}}\right)\mathrm{d}\sigma_y
      \oldtext{+}\changedtext{-} \int_{\Omega_\tau} G f\,\mathrm{d}y.
    \end{aligned}
\end{equation}
Rearranging isolates the total derivative:
\begin{equation}
    \frac{\mathrm{d}}{\mathrm{d}\tau} \int_{\Omega_\tau} uG\,\mathrm{d}y
    = \int_{\Gamma_\tau} \left(
        G\,\frac{\partial u}{\partial \vec{n}}
        - u\,\frac{\partial G}{\partial \vec{n}}
        + uG\,\langle V, \vec{n}\rangle
    \right)\mathrm{d}\sigma_y
      \;+\; \int_{\Omega_\tau} G f\,\mathrm{d}y.
\end{equation}
This is the central identity. The $\langle V, \vec{n}\rangle$ term, which is absent for stationary domains, encodes the geometric effect of the moving boundary, while the volume integral captures the source term $f$.

Integrating over $\tau \in (0,t)$, the left-hand side telescopes:
\begin{equation}
    \int_0^t \frac{\mathrm{d}}{\mathrm{d}\tau} \int_{\Omega_\tau} uG\,\mathrm{d}y\,\mathrm{d}\tau
    = \Bigl[\,\int_{\Omega_\tau} uG\,\mathrm{d}y \Bigr]_{\tau=0}^{\tau = t^-}.
\end{equation}
At $\tau = 0$, the integral evaluates to $\int_{\Omega_0} G(0,y)\,u(0,y)\,\mathrm{d}y$. Writing $u(0,\cdot) = g_0$ for the initial data, this becomes $\int_{\Omega_0} G(0,y)\,g_0(y)\,\mathrm{d}y$. At $\tau \to t^-$, the heat kernel concentrates. $\lim_{\tau \to t^-} G(\tau, y) = \delta(x-y)$, giving
\begin{equation}
    \lim_{\tau \to t^-} \int_{\Omega_\tau} u(\tau, y)\,G(\tau, y)\,\mathrm{d}y = \alpha(x)\,u(t, x),
\end{equation}
where $\alpha(x) = 1$ for interior points and $\alpha(x) = \tfrac{1}{2}$ on smooth boundaries.
Assembling the pieces yields the non-cylindrical boundary integral representation:
\begin{equation}
    \label{equ:bie_noncylindrical_derived}
    \;
    \begin{aligned}
    \alpha(x)\,u(t, x)
    &= \int_{\Omega_0} G(0,y)\,g_0(y)\,\mathrm{d}y
        + \int_0^t \!\!\int_{\Omega_\tau} G\,f\,\mathrm{d}y\,\mathrm{d}\tau \\
    &+ \int_0^t \!\!\int_{\Gamma_\tau} G\,\frac{\partial u}{\partial \vec{n}}\,\mathrm{d}\sigma_y\mathrm{d}\tau
        - \int_0^t \!\!\int_{\Gamma_\tau} \frac{\partial G}{\partial \vec{n}}\,u\,\mathrm{d}\sigma_y\mathrm{d}\tau \\
    &+ \int_0^t \!\!\int_{\Gamma_\tau} G\,\langle V, \vec{n}\rangle\,u\,\mathrm{d}\sigma_y\mathrm{d}\tau
    \;
    \end{aligned}
\end{equation}

The five terms are, respectively, the initial-condition term ($\int_{\Omega_0} G\,g_0$), the volume source integral ($\iint G\,f$), the single-layer potential ($\int G\,\partial u / \partial \vec{n}$), the double-layer potential ($-\int \partial G / \partial \vec{n}\,u$), and the moving-boundary correction ($\int G\langle V,\vec{n}\rangle u$). When $\Omega_\tau \equiv \Omega$ is stationary ($V \equiv 0$) with zero source and initial data, the correction, source, and initial terms vanish and~\eqref{equ:bie_noncylindrical_derived} reduces to the classical cylindrical representation~\eqref{equ:bie_cylinder}.


\section{Sampling Subroutine Details}
\label{app:sampling}

\subsection{Directional Sampling}
\label{app:sampling_directional}

\oldtext{We provide the complete derivation of the geometric cancellation that enables exact importance sampling of the double-layer kernel $-\partial G / \partial \vec{n}$ on $\partial\mathcal{H}^\star$ via the product distribution $u \sim \Gamma(\frac{n}{2}+1, \frac{2}{n}) \times \omega \sim \mathrm{Uniform}(S^{n-1})$.}
\changedtext{We first derive the time density on the heat sphere, then on the Neumann wall. We then show how to sample $\Gamma_\star(c,d)$ of Proposition~\ref{prop:directional_sampling} from one ordinary Gamma draw.}

Define dimensionless coordinates that unify the temporal and spatial degrees of freedom:
\begin{equation}
    \oldtext{u}\changedtext{\mu} = \frac{|x|^2}{2ns}, \qquad \omega = \frac{x}{|x|} \in S^{n-1}.
\end{equation}
Introducing the radial parameter $\rho \in (0, c]$ as in Section~\ref{sec:heat_ball}, the level-set condition $\Phi = \tau(c)$ on the heat-sphere surface corresponds to $\rho = c$, which enforces $\oldtext{u}\changedtext{\mu} = \log(c/s)$. This yields the parameterization of the surface:
\begin{equation}
    \label{equ:heat_param_app}
    s = \rho\,e^{-\oldtext{u}\changedtext{\mu}}, \qquad |x| = \sqrt{2n\,\rho\,\oldtext{u}\changedtext{\mu}\,e^{-\oldtext{u}\changedtext{\mu}}} = \sqrt{2n\,\oldtext{u}\changedtext{\mu}\,s}.
\end{equation}
For fixed $(\oldtext{u}\changedtext{\mu}, \omega)$, varying $\rho$ sweeps a radial path in space-time, which we call a \emph{space-time radial trace}. At $\rho = c$ the trace reaches the heat-sphere surface.

In a restricted domain, the boundary measure on $\partial\mathcal{H}^\star$ generalizes to
\begin{equation}
    -\frac{\partial G}{\partial \vec{n}}(x, s) \;\mathrm{d}\sigma \; \mathrm{d}s
    \;=\; \frac{x \cdot \vec{n}}{2s} \, \Phi(x, s) \;\mathrm{d}\sigma \; \mathrm{d}s.
\end{equation}
Change variables to $(\omega, \oldtext{u}\changedtext{\mu})$, keeping the fixed radius $c$ in the coefficients. The solid-angle identity $\mathrm{d}\omega = \frac{x \cdot \vec{n}}{|x|^n} \mathrm{d}\sigma$ eliminates the boundary orientation $\vec{n}$. On the heat-sphere surface, \oldtext{$\mathrm{d}s = c\,e^{-u}\,\mathrm{d}u$.}\,\changedtext{the Jacobian magnitude is $|\mathrm{d}s|=c e^{-\mu}\,\mathrm{d}\mu$.} Substituting both:
\begin{align}
    -\frac{\partial G}{\partial \vec{n}} \;\mathrm{d}\sigma \; \mathrm{d}s
    &= \frac{|x|^n}{2s} \, \Phi(x, s) \; c\,e^{-\oldtext{u}\changedtext{\mu}} \;\mathrm{d}\omega \; \mathrm{d}\oldtext{u}\changedtext{\mu} \nonumber\\
    &= \frac{1}{2} (2\pi)^{-n/2} \, n^{\,n/2} \, \oldtext{u}\changedtext{\mu}^{\,n/2} \, \exp\!\left(-\frac{n}{2}\oldtext{u}\changedtext{\mu}\right) \;\mathrm{d}\omega \; \mathrm{d}\oldtext{u}\changedtext{\mu} \nonumber\\
    &= \frac{1}{|S^{n-1}|} \; \Gamma\!\left(\oldtext{u}\changedtext{\mu};\, \frac{n}{2}+1,\; \frac{2}{n}\right) \;\mathrm{d}\omega \; \mathrm{d}\oldtext{u}\changedtext{\mu}.
\end{align}

\oldtext{Thus the double-layer kernel factorizes exactly into independent Gamma and uniform components. The directional marginal $p(\omega) = 1/|S^{n-1}|$ is constant, so uniform sampling on $S^{n-1}$ is exact with importance weight $w \equiv 1$. When the ray strikes $\partial\Omega_N$ at distance $\rho$ before the heat-sphere surface, the inversion $c_{\mathrm{wall}} = \rho^2 / (2n u e^{-u})$ maps to the wall component of the caloric measure with no additional correction.}
\changedtext{On the heat sphere, the calculation above gives the density $\Gamma(\mu;\frac n2+1,\frac 2n)$ in $\mu$ for a fixed direction $\omega$. If the radial trace meets $\partial\Omega_N$ before the heat sphere, the radius is fixed at $d=d_{\mathrm{wall}}(\omega)$, so $s=d^2/(2n\mu)$, the solid-angle identity reads $x\cdot\vec{n}\,\mathrm{d}\sigma=d^n\,\mathrm{d}\omega$, and the Jacobian magnitude is $|\mathrm{d}s|=s\,\mathrm{d}\mu/\mu$. On the wall the kernel evaluates to $\Phi=(n\mu/(2\pi d^2))^{n/2}e^{-n\mu/2}$. The same computation then gives
\begin{align}
\label{eq:directional_wall_density}
 -\frac{\partial G}{\partial \vec{n}} \;\mathrm{d}\sigma \; \mathrm{d}s
 &= \frac{d^n}{2s} \, \Phi(x, s) \; \frac{s}{\mu} \;\mathrm{d}\omega \; \mathrm{d}\mu \nonumber\\
 &= \frac{1}{|S^{n-1}|} \; \Gamma\!\left(\mu;\, \frac{n}{2},\; \frac{2}{n}\right) \;\mathrm{d}\omega \; \mathrm{d}\mu \nonumber\\
 &= \frac{1}{\mu}\cdot\frac{1}{|S^{n-1}|} \; \Gamma\!\left(\mu;\, \frac{n}{2}+1,\; \frac{2}{n}\right) \;\mathrm{d}\omega \; \mathrm{d}\mu ,
\end{align}
where the last form isolates the factor $1/\mu$ contributed by the wall Jacobian. The two pieces are selected by comparing $r_{\mathrm{sphere}}(\mu)$ with $d$: the trace reaches the heat sphere where $r_{\mathrm{sphere}}(\mu)\le d$ and meets the wall where $r_{\mathrm{sphere}}(\mu)>d$. Since $\mu e^{-\mu}$ is unimodal with its maximum at $\mu=1$ and vanishes at both ends of $(0,c)$, the wall branch occupies exactly the interval $(\mu_-,\mu_+)$ on which $r_{\mathrm{sphere}}(\mu)>d$, whose endpoints satisfy $r_{\mathrm{sphere}}(\mu_\pm)=d$. Together the two pieces form the piecewise density $\Gamma_\star(\cdot;c,d)$ of Proposition~\ref{prop:directional_sampling}.

Because the two pieces come from different changes of variables, we check that the combination is still a probability density. The piecewise density keeps the sphere piece outside $(\mu_-,\mu_+)$ and the wall piece inside, so its total mass is one exactly when the two pieces carry equal mass on $(\mu_-,\mu_+)$. Integrating the wall density minus the sphere density and applying the Newton--Leibniz formula gives
\begin{equation}
\label{eq:directional_wall_mass_difference}
\begin{aligned}
 &\int_{\mu_-}^{\mu_+}
 \left[\Gamma\!\left(v;\tfrac n2,\tfrac 2n\right)
      -\Gamma\!\left(v;\tfrac n2+1,\tfrac 2n\right)\right]\mathrm dv =\frac{1}{\Gamma(\frac n2+1)}
   \left[\left(\frac{nv}{2}e^{-v}\right)^{n/2}\right]_{\mu_-}^{\mu_+}.
\end{aligned}
\end{equation}
The endpoints satisfy $r_{\mathrm{sphere}}(\mu_\pm)=d$, which gives
\begin{equation}
 \label{eq:directional_wall_endpoints}
 \frac{n}{2}\,\mu_\pm e^{-\mu_\pm}=\frac{d^2}{4c}.
\end{equation}
Thus the bracket in~\eqref{eq:directional_wall_mass_difference} has the same value at both endpoints, the two pieces have equal mass on $(\mu_-,\mu_+)$, and $\Gamma_\star$ is normalized.

To sample $\Gamma_\star$ we draw $\mu$ from the sphere density and apply the branch test. Outside $(\mu_-,\mu_+)$ the trace reaches the heat sphere, and $\mu$ already has the required density. Inside the interval the trace hits the wall, and $\mu_N$ is defined by matching cumulative probabilities from the lower endpoint:
\begin{equation}
\label{eq:directional_wall_cdf_match}
 \int_{\mu_-}^{\mu}\!\Gamma(v;\frac n2+1,\frac 2n)\,\mathrm dv
 =\int_{\mu_-}^{\mu_N}\!\Gamma(v;\frac n2,\frac 2n)\,\mathrm dv .
\end{equation}
Since both densities are continuous and positive on $(\mu_-,\mu_+)$, this is the unique increasing map that turns the conditional sphere density into the conditional wall density on that interval.
To evaluate~\eqref{eq:directional_wall_cdf_match}, consider a Gamma random variable $X$ with shape $a>0$ and scale one. Its probability of falling below $z\geq 0$ is
\begin{equation}
\label{eq:directional_gamma_cdf}
 P(a,z)=\Pr(X\leq z)
       =\frac{1}{\Gamma(a)}\int_0^z v^{a-1}e^{-v}\,\mathrm dv .
\end{equation}
This cumulative probability is called the \emph{incomplete Gamma function} where the denominator $\Gamma(a)$ is the same integral over $[0,\infty)$. For the scale $2/n$ used here, the probability of a draw at or below $\mu$ is $P(a,n\mu/2)$.
\changedtext{Both $P(a,z)$ and its inverse in the second argument are available in standard numerical libraries, and we evaluate them with standard incomplete Gamma routines~\cite{boostmath,didonato1986computation}.} Evaluating each integral in~\eqref{eq:directional_wall_cdf_match} by subtracting its cumulative probability at $\mu_-$ gives
\begin{equation}
 \label{eq:directional_wall_cdf_match_p}
 \begin{aligned}
 &P\!\left(\frac n2+1,\frac{n\mu}{2}\right)
  -P\!\left(\frac n2+1,\frac{n\mu_-}{2}\right)=P\!\left(\frac n2,\frac{n\mu_N}{2}\right)
  -P\!\left(\frac n2,\frac{n\mu_-}{2}\right).
 \end{aligned}
\end{equation}

\begin{proposition}
\label{prop:gamma_recurrence}
For $a>0$ and $z\geq 0$,
\begin{equation}
 \label{eq:directional_gamma_recurrence}
 P(a,z)-P(a+1,z)=\frac{z^a e^{-z}}{\Gamma(a+1)}.
\end{equation}
\end{proposition}
The identity follows by integrating $v^a e^{-v}$ by parts in the definition of $P(a+1,z)$ and using $\Gamma(a+1)=a\Gamma(a)$. Rearranging~\eqref{eq:directional_wall_cdf_match_p} to group the two terms evaluated at $\mu_-$ gives
\begin{equation}
\label{eq:directional_wall_cdf_offset}
\begin{aligned}
 &P\!\left(\frac n2,\frac{n\mu_N}{2}\right)
       -P\!\left(\frac n2+1,\frac{n\mu}{2}\right)\\
 &\quad =P\!\left(\frac n2,\frac{n\mu_-}{2}\right)
       -P\!\left(\frac n2+1,\frac{n\mu_-}{2}\right)
       =\frac{(d^2/(4c))^{n/2}}{\Gamma(\frac n2+1)},
\end{aligned}
\end{equation}
where the last equality uses~\eqref{eq:directional_gamma_recurrence} at $z=n\mu_-/2$ and the endpoint identity~\eqref{eq:directional_wall_endpoints}. Inverting $P$ in its second argument yields the wall coordinate
\begin{equation}
 \label{eq:directional_wall_transform}
 \mu_N=\frac{2}{n}\,P^{-1}\!\left(\frac n2,\;
    P\!\left(\frac n2+1,\frac{n\mu}{2}\right)
    +\frac{(d^2/(4c))^{n/2}}{\Gamma(\frac n2+1)}\right).
\end{equation}
The resulting sample, $\mu$ on a sphere hit and $\mu_N$ on a wall hit, therefore has the piecewise density $\Gamma_\star(\cdot;c,d)$ of Proposition~\ref{prop:directional_sampling}.}

\subsection{Source Term Sampling}
\label{app:sampling_source}

We provide the derivation of Proposition~\ref{prop:source_sampling} and the associated sampling procedure for the volume source integral $I_f = \int_{\mathcal{H}^\star} G(y,s) \, f(y, t-s) \;\mathrm{d}y\,\mathrm{d}s$.

Define the coordinates
\begin{equation}
\label{equ:heat_spherical_app}
s = \rho e^{-\oldtext{u}\changedtext{\mu}}, \qquad |y| = \sqrt{2n\,\rho\,\oldtext{u}\changedtext{\mu}\,e^{-\oldtext{u}\changedtext{\mu}}}, \qquad \omega = y/|y| \in S^{n-1},
\end{equation}
with $0 < \rho < c$ and $\oldtext{u}\changedtext{\mu} > 0$. Substituting into the heat kernel gives $|y|^2/(4s) = n \oldtext{u}\changedtext{\mu} / 2$, hence
\begin{equation}
\begin{aligned}
\Phi(y, s) &= (4\pi s)^{-n/2} \exp\!\left(-\frac{|y|^2}{4s}\right) \\
&= (4\pi \rho e^{-\oldtext{u}\changedtext{\mu}})^{-n/2} e^{-n \oldtext{u}\changedtext{\mu}/2} \\
&= (4\pi \rho)^{-n/2} = \tau(\rho).
\end{aligned}
\end{equation}
The Green's function therefore simplifies to $G(y,s) = \tau(\rho) - \tau(c)$, which depends \emph{only on $\rho$} and is independent of $\oldtext{u}\changedtext{\mu}$.

\oldtext{Writing $\mathrm{d}y = \rho^{n-1}\,\mathrm{d}\rho\,\mathrm{d}\omega$ with $\rho = |y|$, the volume element is $\mathrm{d}y\,\mathrm{d}s = \rho^{\,n-1}\,|\det J|\,\mathrm{d}\rho\,\mathrm{d}u\,\mathrm{d}\omega$, where $J$ is the Jacobian of the map $(\rho,u) \mapsto (\rho, s)$}\,\changedtext{Write $r=|y|$ for the spatial radius, distinct from the heat-spherical coordinate $\rho$. Then $\mathrm{d}y=r^{n-1}\,\mathrm{d}r\,\mathrm{d}\omega$, and $\mathrm{d}y\,\mathrm{d}s=r^{n-1}|\det J|\,\mathrm{d}\rho\,\mathrm{d}\mu\,\mathrm{d}\omega$, where $J$ is the Jacobian of $(\rho,\mu)\mapsto(r,s)$}. A direct computation yields
\begin{equation}
|\det J| = \frac{\sqrt{2n\rho}}{2\sqrt{\oldtext{u}\changedtext{\mu}}}\,e^{-3\oldtext{u}\changedtext{\mu}/2},
\end{equation}
and assembling the factors gives
\begin{equation}
\label{equ:volume_element_app}
\mathrm{d}y\,\mathrm{d}s
= \frac{1}{2}\, (2n)^{n/2}\, \rho^{\,n/2}\, \oldtext{u}\changedtext{\mu}^{\,n/2-1}\, e^{-(n/2+1)\oldtext{u}\changedtext{\mu}} \;\mathrm{d}\rho\,\mathrm{d}\oldtext{u}\changedtext{\mu}\,\mathrm{d}\omega.
\end{equation}

Multiplying by $G(y,s)$ and normalizing yields the factorized sampling density stated in Proposition~\ref{prop:source_sampling}:
\begin{align}
p_\omega(\omega) &= \frac{1}{|S^{n-1}|}, \\
\oldtext{p_u}\changedtext{p_\mu}(\oldtext{u}\changedtext{\mu}) &= \frac{(n/2+1)^{\,n/2}}{\Gamma(n/2)}\, \oldtext{u}\changedtext{\mu}^{\,n/2-1}\, e^{-(n/2+1)\oldtext{u}\changedtext{\mu}}
= \Gamma\!\left(\oldtext{u}\changedtext{\mu};\, \frac{n}{2},\; \frac{1}{n/2+1}\right), \\
p_\rho(\rho) &= \frac{n+2}{c n}\left[1 - \left(\frac{\rho}{c}\right)^{\!n/2}\right], \qquad 0 < \rho < c.
\end{align}

\smallskip\noindent
\textit{Normalization constant.}
Integrating $G$ over the heat ball:
\begin{equation}
Z = \int_{B(c)} G(y,s)\,\mathrm{d}y\,\mathrm{d}s = c\left(\frac{n}{n+2}\right)^{\!n/2+1}.
\end{equation}
For $n=2$ this gives $Z = c/4$. For $n=3$, $Z = c\,(3/5)^{5/2} \approx 0.279\,c$.

The radial coordinate $\rho$ is sampled by rejection. Draw $\rho \sim \mathrm{Uniform}(0,c)$ and accept with probability $1 - (\rho/c)^{\,n/2}$. The acceptance rate is $n/(n+2) \ge 50\%$ for all $n \ge 2$. The temporal coordinate $\oldtext{u}\changedtext{\mu}$ is drawn directly from the Gamma distribution. The direction $\omega$ is reused from the directional sampling step (Section~\ref{sec:directional_sampling}), requiring no additional random variate. The point $(y, s)$ is then mapped via~\eqref{equ:heat_spherical_app}. If $y \notin \mathcal{H}^\star$ the contribution is zero. The estimator is $\widehat{I}_f = Z \cdot f(y, t-s)$, with no explicit evaluation of $G$ required.

\section{Gradient Estimator Derivation}
\label{app:gradient_derivation}

We provide the complete derivation of the gradient formula in Section~\ref{sec:gradient}, \oldtext{beginning with the familiar Laplace case as motivation}\changedtext{beginning with the Laplace gradient estimator used by~\citet{sawhney2020monte} as motivation} and then proceeding to the heat equation.

\subsection{Laplace Case (Motivation)}

For a harmonic function $\Delta u = 0$ on $\Omega \subset \mathbb{R}^n$, the volume mean value property on a ball $B(x,R) \subset \Omega$ states
\begin{equation}
    u(x) = \frac{1}{|B(x,R)|} \int_{B(x,R)} u(y) \; \mathrm{d}y,
\end{equation}
where $|B| = |S^{n-1}| R^n / n$. Changing variables $y = x + z$ and differentiating with respect to $x$ gives
\begin{equation}
    \nabla_x u(x) = \frac{1}{|B|} \int_{B(x,R)} \nabla_y u(y) \; \mathrm{d}y.
\end{equation}
Applying the divergence theorem with $F(y) = u(y)\,e_j$ converts this to a boundary integral:
\begin{equation}
    \int_{B(x,R)} \nabla u(y) \; \mathrm{d}y = \int_{\partial B(x,R)} u(y) \; n(y) \; \mathrm{d}S_y,
\end{equation}
where $n(y) = (y-x)/R$. Using $|B| = R\,|\partial B|/n$ yields the familiar gradient formula
\begin{equation}
    \nabla u(x) = \frac{n}{R\;|\partial B(x,R)|} \int_{\partial B(x,R)} u(y) \; n(y) \; \mathrm{d}S_y.
\end{equation}
The gradient is expressed as a surface integral of $u$. The recursive variable is $u$, not $\nabla u$. Note that the volume integral of $\nabla u$ converts entirely to a boundary integral because the weight is constant ($w \equiv 1$, so $\nabla w = 0$).

\subsection{Heat Equation Case}

The boundary MVP~\eqref{equ:mean_value} integrates over the heat-sphere \emph{surface}. For the gradient we require the \emph{volumetric} heat MVP, which integrates over the heat-ball interior with weight $w(r) = r^2/(4s^2)$~\cite{watson2012introduction}:
\begin{equation}
    \label{equ:volume_mvp}
    u(x_0, t_0) = \tau(c) \int_0^c \int_{B(x_0,\,\rho(s))} u(y, t_0-s) \; \frac{|x_0-y|^2}{4 s^2} \; \mathrm{d}y \, \mathrm{d}s.
\end{equation}

Changing variables $\tilde{y} = y - x_0$ to fix the integration domain, only $u$ depends on $x_0$. Differentiating under the integral gives
\begin{equation}
    \nabla_{x_0} u = \tau(c) \int_0^c \int_{B(x_0,\rho)} \nabla_y u(y,t_0-s) \; w(r) \; \mathrm{d}y \, \mathrm{d}s, \qquad r = |x_0-y|.
\end{equation}

For each fixed $s$, apply the divergence theorem on $B_\rho = B(x_0, \rho(s))$:
\begin{equation}
    \int_{B_\rho} \nabla_y u \cdot w \; \mathrm{d}y
    = \oint_{\partial B_\rho} u \; n \cdot w \; \mathrm{d}S \;-\; \int_{B_\rho} u \; \nabla_y w \; \mathrm{d}y.
\end{equation}
With $n = \omega = (y-x_0)/\rho$, $w(\rho) = \rho^2/(4 s^2)$, and $\nabla_y w = (y-x_0)/(2 s^2)$:
\begin{equation}
    \int_{B_\rho} \nabla_y u \cdot w \; \mathrm{d}y
    = \frac{\rho^2}{4 s^2} \oint_{\partial B_\rho} u \; \omega \; \mathrm{d}S
      \;-\; \frac{1}{2 s^2} \int_{B_\rho} u \; (y-x_0) \; \mathrm{d}y.
\end{equation}
Multiplying by $\tau(c)$ and integrating over $s \in (0,c)$ yields the gradient formula of Proposition~\ref{prop:gradient_formula}. The critical difference from the Laplace case is that $w(r)$ is non-constant, so $\nabla_y w \neq 0$ and the volume term survives.

With $\omega_i = (y_i - x_{0i})/r$, the surface and volume terms separate as
\begin{align}
    S_i &= \tau(c) \int_0^c \frac{r(s)^{n+1}}{4 s^2} \int_{S^{n-1}} u(x_0+r(s)\omega, t_0-s) \; \omega_i \; \mathrm{d}\omega \, \mathrm{d}s, \\
    V_i &= -\tau(c) \int_0^c \frac{1}{2 s^2} \int_0^{r(s)} r^n \int_{S^{n-1}} u(x_0+r\omega, t_0-s) \; \omega_i \; \mathrm{d}\omega \, \mathrm{d}r \, \mathrm{d}s,
\end{align}
with $\nabla_i u = S_i + V_i$.

Both terms share the $s$-marginal $p_S(s) \propto r(s)^{n+1}/s^2$. The constants are computed by evaluating $\int_0^c r(s)^{n+1}/s^2\,\mathrm{d}s$ via the substitution $\oldtext{u_g}\changedtext{\mu_g} = \log(c/s)$:
\begin{align}
    \int_0^c \frac{r(s)^{n+1}}{s^2}\,\mathrm{d}s
    &= (2n)^{\frac{n+1}{2}} c^{\frac{n-1}{2}} \int_0^\infty e^{-\frac{n-1}{2}\oldtext{u_g}\changedtext{\mu_g}}\, \oldtext{u_g}\changedtext{\mu_g}^{\frac{n+1}{2}}\,\mathrm{d}\oldtext{u_g}\changedtext{\mu_g} \nonumber\\
    &= (2n)^{\frac{n+1}{2}} c^{\frac{n-1}{2}}\,
      \frac{\Gamma\!(\frac{n+3}{2})}{(\frac{n-1}{2})^{\frac{n+3}{2}}}.
\end{align}
Multiplying by the prefactors gives $Z_S$ and $Z_V = \frac{2}{n+1}Z_S$ as stated in Section~\ref{sec:gradient}.


\end{document}